\documentclass[twocolumn]{aastex631}

\usepackage{soul}
\usepackage{amsmath}
\usepackage{booktabs}

\begin{document}

\title{Inferring Planet and Disk Parameters from Protoplanetary Disk Images Using a Variational Autoencoder}

\author[0000-0000-0000-0000]{Sayed Shafaat Mahmud}
\affiliation{Colgate University, 13 Oak Drive, Hamilton, NY 13346, USA}

\author[0000-0003-3784-8913]{Sayantan Auddy}
\affiliation{Fields Institute, 222 College St, Toronto, ON M5T 3J1, Canada}
\affiliation{Jet Propulsion Laboratory, California Institute of Technology, 4800 Oak Grove Drive, Pasadena, CA 91109, USA}

\author[0000-0001-8292-1943]{Neal Turner}
\affiliation{Jet Propulsion Laboratory, California Institute of Technology, 4800 Oak Grove Drive, Pasadena, CA 91109, USA}

\author[0000-0001-8642-5867]{Jeffrey S. Bary}
\affiliation{Colgate University, 13 Oak Drive, Hamilton, NY 13346, USA}

\begin{abstract}
Dust-continuum observations of many protoplanetary disks reveal rings and gaps that are widely interpreted as evidence of ongoing planet formation.  Here we present the first framework for inferring planet and disk parameters from such images using variational autoencoder (VAE) based generative machine learning (ML).  The new framework is called VADER (Variational Autoencoder for Disks with Embedded Rings).  We train VADER on synthetic images of dust continuum emission, generated from \texttt{FARGO3D} hydrodynamic simulations post-processed with Monte Carlo radiative transfer calculations.  VADER infers the masses of up to three embedded planets as well as the disk parameters viscous $\alpha$, dust-to-gas ratio, Stokes number, and flaring index.  VADER returns a full posterior distribution for each of these quantities.  We demonstrate that VADER reconstructs disk morphologies with high structural similarity (index $>$ 0.99), accurately recovers planet parameters with $R^2 > 0.9$ across planet masses, and reliably predicts disk parameters.  Applied to ALMA dust continuum images of 23 protoplanetary disks, our model returns mass estimates for embedded planets of 0.3-2~$M_{\mathrm{Jup}}$ that agree to within $1\sigma$ of published values in most cases, and infers disk parameters consistent with current literature.  Once trained, the VAE performs full posterior parameter inference in a matter of minutes, offering statistical rigor with enough computational speed for application to large-scale ALMA surveys.  These results establish VAE-based models as powerful tools for inferring from disk structure the masses of embedded planets and the global disk parameters, with their associated uncertainties.
\end{abstract}


\section{Introduction} \label{sec:intro}

Protoplanetary disks (PPDs) are the cradles of planet formation, providing the essential conditions for the early stages of planetary growth. Observations with the Atacama Large Millimeter/submillimeter Array (ALMA) \citep{brogan20152014} have revealed a diverse array of substructures in these disks, including concentric rings, gaps \citep{andrews2018scaling, isella2018signatures,huang2020large, fedele2018alma}, and spirals \citep{bae2022structured, cieza2021ophiuchus}. 
The origin of these substructures, although debated, is broadly attributed to disk physics and chemistry such as planet-disk interaction. Disk-specific mechanisms include zonal flows driven by magnetorotational turbulence \citep{johansen2009zonal, dittrich2013gravoturbulent, flock2015gaps}, secular gravitational instabilities \citep{youdin2011formation, takahashi2014two}, dust evolution around snowlines due to molecular condensation and fragmentation \citep{zhang2019systematic, pinilla2017dust}, self-induced dust pile-ups \citep{gonzalez2017self}, gap formation by large-scale vortices \citep{barge2017gaps}, dust sintering fronts \citep{okuzumi2016sintering}, and other radiation hydrodynamic instabilities \citep{ziampras2025dusty}.

However, the most popular explanation, and the one that is of interest to us, interprets these substructures as signatures of young, embedded planets, shaped by their dynamical interactions with the disk \citep{quanz2013gaps, winters2003gap, zhu2019inclined, pinilla2015sequential, akiyama2016planetary}. These planet-induced substructures may be used to reveal the properties of the unseen planets and the dynamical processes governing the evolution of the disk \citep{kanagawa2015mass, teague2018kinematical, pinilla2012ring, dullemond2018disk}. This is intriguing, since direct detection of these embedded protoplanets in the early stages of formation remains difficult \citep{wolf2002detecting}, owing to the complexity of the environment as well as the limitations in the current planet search techniques \citep{fis14,lee18}.



Traditional approaches rely on hydrodynamical and radiative transfer modeling of disks with embedded planets to predict the development of disk structure, which is then compared against observations.  This method involves sweeping a large parameter space to match specific observations \citep{toci2019long, veronesi2019multiwavelength, veronesi2021dynamical, teague2018kinematical} and thus presents significant computational challenges \citep{haworth2016grand, rosotti2023empirical, dominik2024bouncing, jiang2024grain}. Moreover, degeneracies mean multiple sets of parameter choices can produce similar observational signatures. The computational challenge has been addressed by using hydrodynamical modeling to develop empirical relations between gap width and planet mass \citep{kanagawa2016mass}. Additional observable disk features may carry information we can use to distinguish among models with similar gap widths but differing turbulence strength and dust properties, parameters that are quite uncertain and may vary substantially from one disk to another \citep{dipierro2015dust, lodato2019newborn, zhang2018disk, long2020dual}.

Machine Learning (ML) models trained on large datasets of synthetic disk images \citep{aud20,auddy2022using, ribas2020modeling} have emerged as potentially powerful tools for analyzing protoplanetary disk observations to infer both disk parameters and the properties of embedded planets. In particular, Convolutional Neural Network (CNN)-based methods \citep{auddy2021dpnnet, zhang2022pgnets, ruzza2024dbnets} have demonstrated success in predicting planet masses from dust-continuum images of PPDs.
The DPNNet-3.0 model (\textit{Praharaj et al.\ 2025, under review}) extends this approach to multi-planet systems. The DBNets model \citep{ruzza2024dbnets} utilizes an ensemble of CNNs to estimate planet masses while accounting in part for observational uncertainties
by adding noise to an ensemble of realizations and convolving each image with suitable telescope beam patterns, to evaluate how much the noise changes the inferred values of the parameters. A recent advance, DBNets2.0 \citep{ruzza2025dbnets2}, develops this framework by combining CNNs with normalizing flows \citep{papamakarios2021normalizing}. The resulting model infers the joint posterior distribution over planet mass, Stokes number, disk aspect ratio, and viscosity. This approach improves uncertainty quantification and enables a deeper exploration of degeneracies among physical parameters. However, DBNets2.0 is trained on \texttt{FARGO3D} simulations post-processed with simplified approximate radiative transfer to estimate disk surface brightness maps. This approach is valid only for disks viewed face-on.

Here we introduce VADER (Variational Autoencoder for Disks with Embedded Rings), a novel framework based on Variational Autoencoders (VAEs) \citep{pinheiro2021variational}, to jointly infer planetary and disk parameters. A VAE is a class of generative neural network model that encodes input data into a lower-dimensional latent space. The encoder maps the latent space variables probabilistically, associating a statistical significance to each value. Variations in key disk parameters correspond to smooth and physically meaningful changes in the latent space. Unlike CNN-based models \citep{auddy2021dpnnet,ruzza2024dbnets, zhang2022pgnets}, which perform direct regression from images to planetary properties, VAEs can use the structured latent space encoded with the underlying characteristics of the disk morphology to infer the masses of the embedded planets as well as key disk properties with uncertainties. In addition, the generative nature of the VAE model allows it to synthesize new models within the parameter space of the models used in training. This work represents the first application of generative ML models to the inference of planetary and disk parameters from dust continuum images of PPDs. In the context of protoplanetary disks, generative ML models have been used to generate disk surface density, radial velocity, and azimuthal velocity maps given $\alpha$-viscosities, disk aspect ratios, and planet-to-star mass ratios \citep{mao2023ppdonet, mao2024disk2planet}. They have also been used to generate edge-on images of protoplanetary disks \citep{telkamp2022machine}. 

Although novel to protoplanetary disk studies, VAEs have previously been applied in the context of astronomy. They have been applied to perform parameter inference on gravitational waves \citep{gabbard2022bayesian}, unsupervised classification and image generation of galaxies \citep{spindler2021astrovader}, to reduce complexity of chemical networks in astrochemical models \citep{grassi2022reducing}, to predict stellar mass and redshift of galaxies in the nearby universe \citep{gagliano2023physics}, and to reduce dimensionality of SDSS spectra \citep{portillo2020dimensionality}.

The rest of this paper is structured as follows. In Section~\ref{sec:modeling}, we describe the model planet-disk systems and how the synthetic images are constructed from them. 
Section~\ref{sec: NN Architecture} introduces VADER and its core architecture, the VAE, along with the two feedforward neural networks (FNN) used for planetary and disk parameter inference.
In Section~\ref{sec: Modelperform}, we present both quantitative and qualitative results of VADER's performance on simulated data while in Section~\ref{sec: Resultobs} we compare VADER's predictions with observational and modeling studies of systems including HD~163296 and HL~Tau. In Section~\ref{sec: Discussions}, we analyze trends in the model’s estimated uncertainties and examine situations where the model may encounter difficulties during prediction. In section~\ref{sec:Conclusions} we discuss the future prospects for using generative AI models to infer disk-planet parameters. 

\section{Modeling Disk-Planet Interaction and Postprocessing} \label{sec:modeling}

We employ \texttt{FARGO3D} \citep{benitez2016fargo3d} to simulate the dynamical evolution of dust and gas in protoplanetary disks perturbed by embedded planets. Extending prior work \citep{auddy2022using}, our simulations replicate disk systems hosting up to three embedded planets, enabling more realistic representations of multi-planetary disk morphologies. These simulations provide the foundational dataset for generating the synthetic observations used to train our neural network model.

\begin{deluxetable}{cc}
\tablecaption{Parameter space for hydrodynamic simulations \label{tab:params}}
\tablehead{
\colhead{Parameter} & \colhead{Range}
}
\startdata
Number of Planets ($\mathrm{N_p}$) & {1, 2, 3} \\
Dust-to-gas ratio ($\epsilon_0$) & \{0.05, 0.025, 0.01\} \\
Flaring Index ($F$) & \{0.25, 0.075, 0.01\} \\
$\alpha$-viscosity & \{0.05, 0.01, 0.005, 0.001, 0.0001\} \\
Stokes Number ($\mathrm{St}$) & \{1.57, 0.0157, 0.00523, 0.00157\} \\
Aspect Ratio ($h_0$) & [0.025, 0.1] \\
$R_{p1}$ & [0.8, 1.2] $R_0$ \\
$R_{p2}$ & [1.8, 2.2] $R_0$ \\
$R_{p3}$ & [2.9, 3.5] $R_0$ \\
$M_{p1}$ & [8$M_\oplus$, 3$M_J$] \\
$M_{p2}$ & [24$M_\oplus$, 3$M_J$] \\
$M_{p3}$ & [30$M_\oplus$, 3$M_J$] \\
\enddata
\tablecomments{Square brackets [ ] indicate a continuous or sampled range while curly brackets \{ \} indicate discrete values.  While the \texttt{FARGO3D} calculations are dimensionless, \texttt{RADMC3D} requires units.  $R_0$ is equal to 30~au. The listed Stokes number, $\mathrm{St}$, is the midplane value at $R_0$.}
\end{deluxetable}

\subsection{Disk-Planet Hydrodynamic Modeling}

We model a locally isothermal, non-self-gravitating 2D vertically-averaged disk orbiting a solar-mass star. The simulation domain spans radius $R \in [0.4, 6.0]\,R_0$, radially (where $R_0=1$ in the dimensionless code unit), and $\phi \in [0, 2\pi]$, azimuthally, resolved on a $512 \times 512$ grid spaced logarithmically in radius. We apply wave-killing boundary damping as described in \citet{de2006comparative} to prevent artificial reflections of density waves. The planets follow fixed, circular Keplerian orbits without migration and their masses increase linearly over the first 100 orbits of the innermost planet to mitigate start-up disk transients in the disk response. 
Each simulation is allowed to evolve for 3000 orbits or about $\approx$ 1~Myr at 45~au around a 1 M$_\odot$ star. Planet masses are normalized to the central stellar mass. 



The initial gas surface density follows:
\begin{equation}
    \Sigma_g(R) = \Sigma_{g,0} \left( \frac{R}{R_0} \right)^{-\sigma},
\end{equation}
with $\Sigma_{g,0} = 10^{-4}$ (code units) and the power law index $\sigma = 1$, ensuring self-gravity can be neglected since the Toomre parameter $Q \geq 80$ at $R_0$. The disks are flared, with aspect ratio
\begin{equation}
    \frac{H}{R} = h_0 \left( \frac{R}{R_0} \right)^F,
\end{equation}
where $h_0$ is the scale height at $R_0$ and $F$ is the flaring index. 
Disk turbulence is parameterized via the $\alpha$-viscosity prescription of \citet{shakura197313}.

We model the dust as a single, pressureless fluid \citep{jacquet2011linear}, coupled to the gas through aerodynamic drag. The degree of coupling is quantified by the Stokes number $\mathrm{St} = t_s \Omega_K$, with $t_s$ being the stopping time and $\Omega_K$ is the Keplerian angular frequency. We fix the Stokes Number at $R_0$. For details on how $\mathrm{St}$ varies radially, see Equation A15 from \cite{auddy2022using}. We include dust back-reaction and set the initial dust surface density as \(\Sigma_{\mathrm{d}} = \epsilon_0 \, \Sigma_{\mathrm{g}}\), where \(\epsilon_0\) is the initial dust-to-gas mass ratio at $R_0$. Over time, the local dust-to-gas ratio \(\epsilon = \Sigma_{\mathrm{d}} / \Sigma_{\mathrm{g}}\) can deviate from \(\epsilon_0\) due to radial drift, turbulent diffusion, and planet-induced redistribution of material \citep[see Equation A19 in][]{auddy2022using}.

To optimize computational efficiency while probing as much of disk parameter space as possible, we discretize the parameter space across key disk and planetary variables. The full setup, including $\alpha$-viscosity, flaring index, Stokes number, dust-to-gas ratio, and planet masses and locations, is summarized in Table~\ref{tab:params}.

We use a stratified sampling method called Latin Hypercube Sampling (LHS) \citep{mckay2000comparison} to sample the parameter space, ensuring both efficient and uniform coverage, to initialize the \texttt{FARGO3D} simulations.
For each Stokes number (defined at $R_0$), we sample the remaining disk parameters listed in Table~\ref{tab:params} and sample the corresponding particle size from Table~\ref{tab:stokes_particles} using LHS. In each simulation, the particle size remains fixed, while the Stokes number and the gas density vary radially. 
We run 140 distinct simulations for each of the five Stokes numbers, yielding a total of 700 hydrodynamic simulations.

\begin{deluxetable}{cccc}
\tablecaption{Particle sizes and Stokes number values for RADMC-3D.  \label{tab:stokes_particles}}
\tablehead{
\colhead{St} & \colhead{Particle Size} & \colhead{St} & \colhead{Particle Size} \\
\colhead{} & \colhead{($\mu$m)} & \colhead{} & \colhead{($\mu$m)}
}
\startdata
1.57   & 1000   & 0.00523 & 100 \\
1.57   & 3000   & 0.00523 & 300 \\
1.57   & 10000  & 0.00523 & 1000 \\
0.0157 & 100    & 0.00523 & 3000 \\
0.0157 & 300    & 0.00157 & 100 \\
0.0157 & 1000   & 0.00157 & 300 \\
0.0157 & 3000   & 0.00157 & 1000 \\
0.0157 & 10000  & 0.000157 & 100 \\
\enddata
\tablecomments{The different particle sizes used corresponding to each Stokes number for radiative transfer modelling. The Stokes Number for a given disk is the mid plane value at $R_0$ where we set $R_0 = 30$AU .}
\end{deluxetable}

\subsection{Radiative Transfer and Disk Image Generation}

We convert the simulated surface density maps into 2D images of dust continuum emission by post-processing the \texttt{FARGO3D} outputs with \texttt{RADMC-3D} \citep{dullemond2012radmc}, a Monte Carlo radiative transfer tool. We model the spatial distribution of dust emission at 1.3~mm (corresponding to ALMA Band 6, 230 GHz), simulating a beam with a FWHM~$\sim 0.02^{\prime\prime}$ to match ALMA's extended baseline resolution.


\texttt{RADMC-3D} directly imports the gas and dust density structures, along with the Stokes number and dust-to-gas ratio. Since \texttt{RADMC-3D} requires a fixed grain size, we force \texttt{FARGO3D} to keep grain size fixed and allow the Stokes number to be variable. We assume a grain density of $3.7\,\mathrm{g~cm^{-3}}$, corresponding to amorphous olivine with 50\% Mg content \citep{dorschner1995steps}. For each Stokes number, we vary the particle sizes in order to decouple grain size from Stokes number, as detailed in Table~\ref{tab:stokes_particles}.

To construct 3D disk structures from 2D hydro outputs, we apply the following vertical gas density profile assuming hydrostatic equilibrium:
\begin{equation}\label{hydro_equi}
    \rho_{g,3D} = \frac{\Sigma_{g,2D}}{\sqrt{2\pi} H} \exp\left( -\frac{(z/r)^2}{2(H/r)^2} \right),
\end{equation}
and adopt the dust profile formulation from equation 19 of \citet{Fromang2009}, which incorporates turbulent diffusion and vertical settling. Since \texttt{RADMC-3D} requires the kinematic viscosity $\nu$ rather than $\alpha$ for kinematic heating, we use:
\begin{equation}
    \nu = \alpha \frac{c_s^2}{\Omega_K}
\end{equation}
where $c_s^2$ is the sound speed and $\Omega_K$ is the Keplerian angular frequency. The dust temperature profile is calculated using a Monte Carlo thermal solver based on \citet{bjorkman2001radiative}. 

\subsection{Image Augmentation \& Normalization}

We apply a three-fold augmentation process to each radiative transfer output to account for degeneracies in disk morphologies due to inclination and position angles as well as translation. The augmentation parameters include inclination along the line of sight: $i \in \{0^\circ, 15^\circ, 30^\circ, 45^\circ, 60^\circ\}$, rotation in the plane of the sky (position angle): $\mathrm{PA} \in \{0^\circ, 15^\circ, 30^\circ, 45^\circ, 60^\circ, 75^\circ\}$, and translation left to right across the image: $t \in \{0, 2.5, 5, 7.5, 10\}$ code units. The translation allows the orientation of the image to be anywhere within the observed frame without being cropped. Once the augmentation is performed we normalize the images such that the maximum value is unity and make sure to have a similar number of 1 planet, 2 planet, and 3 planet systems in our dataset.

Each simulation yields 150 synthetic images per dust configuration through this augmentation pipeline, resulting in a dataset with 105,000 images. These images, encompassing a wide variety of disk inclinations and orientations, form the final dataset for training and evaluating VADER.

\section{Network Architecture}\label{sec: NN Architecture}
VADER is a generative ML model built on a VAE based neural network. It is designed to predict the masses of embedded planets and infer key physical properties of protoplanetary disks from dust continuum images. VADER builds on our earlier model for estimating the mass of a single exoplanet from disk images \citep{mahmud2024vae_exoplanet_masses} by extending that approach to handle more complex disk structures formed by multiple planets.



\begin{figure*}[ht!]
    \centering
    \includegraphics[width=\textwidth]{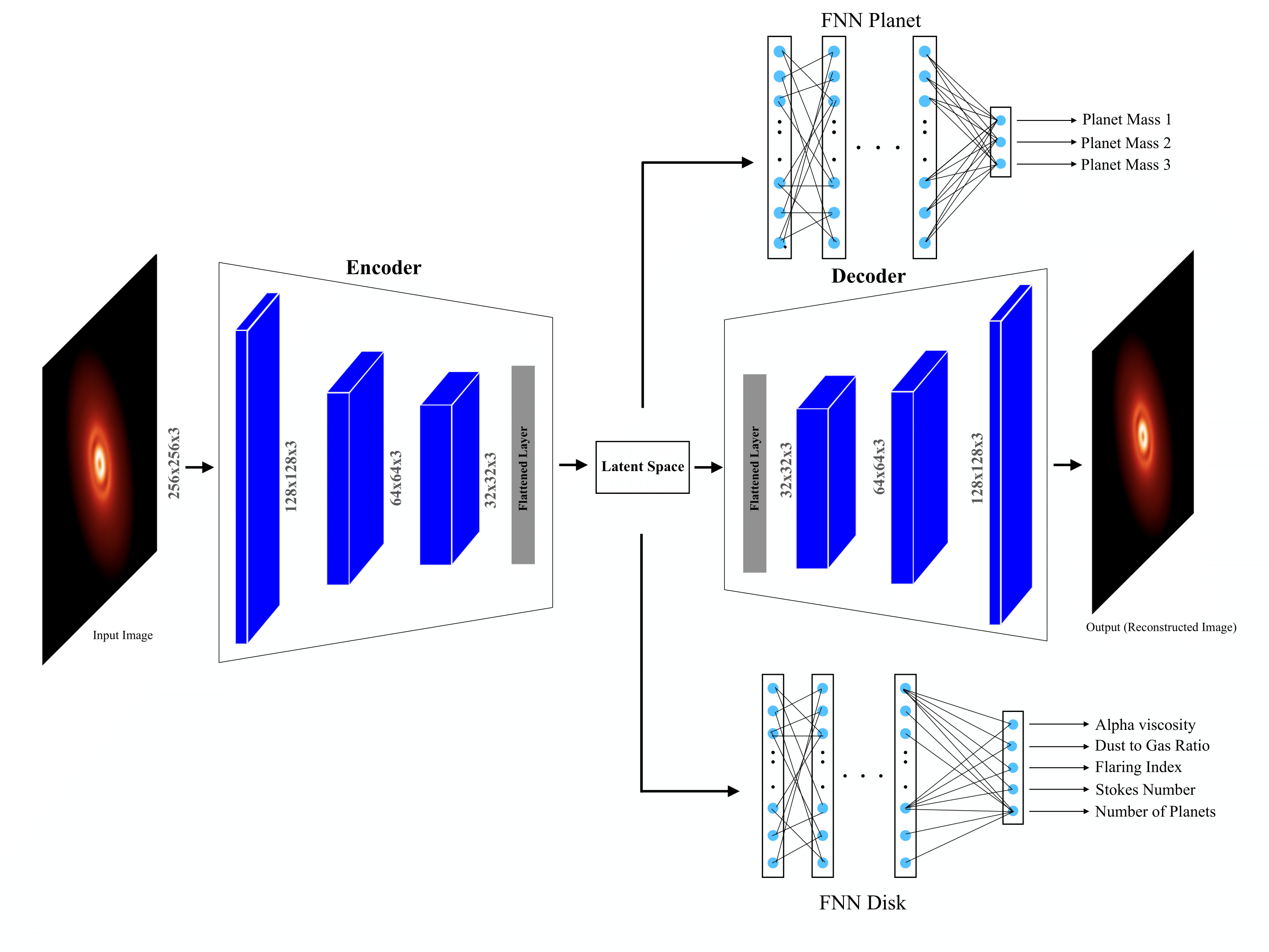}
    \caption{
A schematic of the variational autoencoder (VAE) architecture used in this work. The encoder is comprised of three convolutional layers (LEFT BLUE squares), compressing the input dust continuum image of a protoplanetary disk (\(256 \times 256 \times 3\)) into a 32-dimensional latent space. Two FNNs independently process the latent vector to infer planetary masses and disk parameters, respectively. The decoder (RIGHT BLUE squares) reconstructs the input image from the latent representation. 
}

    \label{fig:vae_ffnn}
\end{figure*}

We build VADER using a two-stage approach. First, we train a VAE to compress disk images into a lower-dimensional latent space that captures their essential structural features, as shown in Figure~\ref{fig:vae_ffnn}. Next, we use this latent representation as the input to two separate feedforward neural networks (FNNs): one for predicting planetary parameters (FNN-Planet) and the other for disk properties (FNN-Disk). In the following subsections, we provide an overview of the VADER network, including its training and optimization procedures. A breakdown of the convolutional operations and detailed architectural specifications are presented in Appendix~\ref{VADER_architecture}.

\subsection{Variational Autoencoder (VAE)}

A VAE is a type of generative ML model that learns to represent input data---in our case, synthetic images of protoplanetary disks---in a compressed form referred to as a latent space. In our application, this compression retains the most salient morphological features of the disk, such as rings, gaps, asymmetries, and flaring. This enables the model to focus on information most relevant for reconstructing disk geometry and inferring physical parameters.

The VAE consists of two main components: an \textit{encoder} and a \textit{decoder}. The encoder maps the input image into a lower-dimensional latent space parameterized by a multivariate Gaussian distribution. Latent vectors are then sampled from this distribution and passed to the decoder, which reconstructs the original input image. Figure~\ref{fig:vae_ffnn} illustrates the overall architecture of the encoder-decoder pipeline.


This structure allows the VAE to learn a compact, smooth, and continuous representation of disk morphologies. The latent space in VADER has 32 dimensions, a value which is chosen after experimenting with several latent dimensionalities. This size provides a good balance between the fidelity of the image reconstruction and parameter inference performance, while avoiding overfitting or degenerate encodings.


\begin{deluxetable*}{ll}[ht!]
\tablecaption{Categorized values used in predicting disk parameters.
\label{tab:disk_param_mapping}}
\tablehead{
\colhead{Parameter} & \colhead{Categorized Values}
}
\startdata
Alpha & {0.0001: 1, 0.001: 2, 0.005: 3, 0.01: 4, 0.05: 5} \\ 
Epsilon  & {0.01: 1, 0.025: 2, 0.05: 3} \\ 
Stokes & {0.000157: 1, 0.00157: 2, 0.00523: 3, 0.0157: 4, 1.57: 5} \\ 
Flaring Index & {0.01: 1, 0.075: 2, 0.25: 3} \\
\enddata
\tablecomments{Each numerical value corresponds to a categorical label, allowing for structured predictions. This table serves as a reference for comparing predicted categories with real-world physical values.}
\end{deluxetable*}

\subsection{Feedforward Neural Networks (FNNs)}

Once the VAE is trained, we sample 32 latent points from their respective distributions (defined by their mean, $\mu$, and variance) as input to the two dedicated feedforward neural networks. The FNN-Planet predicts the existence and masses of up to three embedded planets. Since these are continuous quantities, the network is trained as a regression model, minimizing the mean squared error (MSE) between predicted and actual values.

The FNN-Disk infers the disk parameters: $\alpha$-viscosity ($\alpha$), Stokes Number ($\mathrm{St}$) at $R_0$, Dust-to-Gas Ratio ($\mathrm{\epsilon}$), Flaring Index ($\mathrm{F}$), and Number of Planets ($\mathrm{N_p}$). 
Each of these parameters takes on values from a small set of categorical choices as given in Table~\ref{tab:params}. Rather than treating this as a classification problem, we map each parameter value to an integer label. For example, $\alpha = 0.0001$ maps to `1', $\alpha = 0.001$ maps to `2', $\alpha = 0.005$ maps to `3', $\alpha = 0.01$ maps to `4' and $\alpha = 0.05$ maps to `5'. The full mapping used in training is shown in Table~\ref{tab:disk_param_mapping}. While $\mathrm{N_p}$ is not strictly a disk parameter, we predict it using FNN-Disk because in our training dataset we have only three unique values for number of planets: 1, 2, and 3. The network is trained and treated as a multi-class regression problem, such that the output for each of the disk parameters takes on continuous values. This allows the network to benefit from the simplicity of regression and at the same time capture the corresponding uncertainty associated with predictions.

\subsection{Training and Optimization}
The VAE is trained to minimize a composite loss function, $\mathcal{L}_{\text{VAE}}$, consisting of a reconstruction term and a regularization term:

\begin{equation}
\begin{split}
     \mathcal{L}_{\text{VAE}} =\ 
     &\mathbb{E}_{q_\phi(z|x)}[\log p_\theta(x|z)] \\
     & - D_{\text{KL}}(q_\phi(z|x) \parallel p(z))
\end{split}
\end{equation}


\noindent
where $x$ is the input image and $z$ is a latent variable representing the encoded features. The encoder $q_\phi(z|x)$, parameterized by a neural network with parameter $\phi$, maps the input image to a probability distribution in the latent space, typically modeled as a Gaussian with learnable mean and log-variance. The first term, $\mathbb{E}_{q_\phi(z|x)}[\log p_\theta(x|z)]$, is the reconstruction term. It measures how well the decoder (with parameter $\theta$) can reconstruct the input $x$ from the latent vector $z$. This term encourages the model to generate outputs that are close to the original inputs. The second term, $D_{\text{KL}}(q_\phi(z|x) \parallel p(z))$, is the regularization term (also called the Kullback–Leibler (KL) divergence), which penalizes the divergence between the learned latent distribution and a prior distribution $p(z)$. The prior distribution $p(z)$ is usually chosen as a standard normal $\mathcal{N}(0, I)$, where \( \textit{I} \) denotes the identity matrix, implying independent unit-variance noise in each latent dimension. The double vertical bar ``\(\| \)'' indicates that the divergence is computed from \( q_\phi(z \mid x) \) relative to \( p(z) \). This regularization term ensures that the latent space is continuous and allows for meaningful interpolation between points. This also avoids overfitting to isolated regions of the latent space. Together, these two components balance reconstruction quality with latent space regularization, enabling the model to generate coherent and generalizable representations.

We use 80\% of the total images after image augmentation followed by normalization for training and the remaining images are split equally for validation and testing. The training process works iteratively, where we monitor the evolution of $\mathcal{L}_{\text{VAE}}$ after each epoch.
The reconstruction loss is computed as the MSE between the input image and the decoder's output. This term quantifies how accurately the model reproduces the original image from its compressed latent representation. By minimizing this loss, the model learns to capture and preserve essential visual and structural features of the disk. As training progresses, the decoder becomes increasingly effective at mapping latent vectors back to realistic disk images, ensuring high-fidelity reconstructions. Details regarding the code implementation of the VAE architecture are provided in Appendix~\ref{appendix:vae}

\begin{figure*}[ht!]
    \centering
    \includegraphics[width=\textwidth]{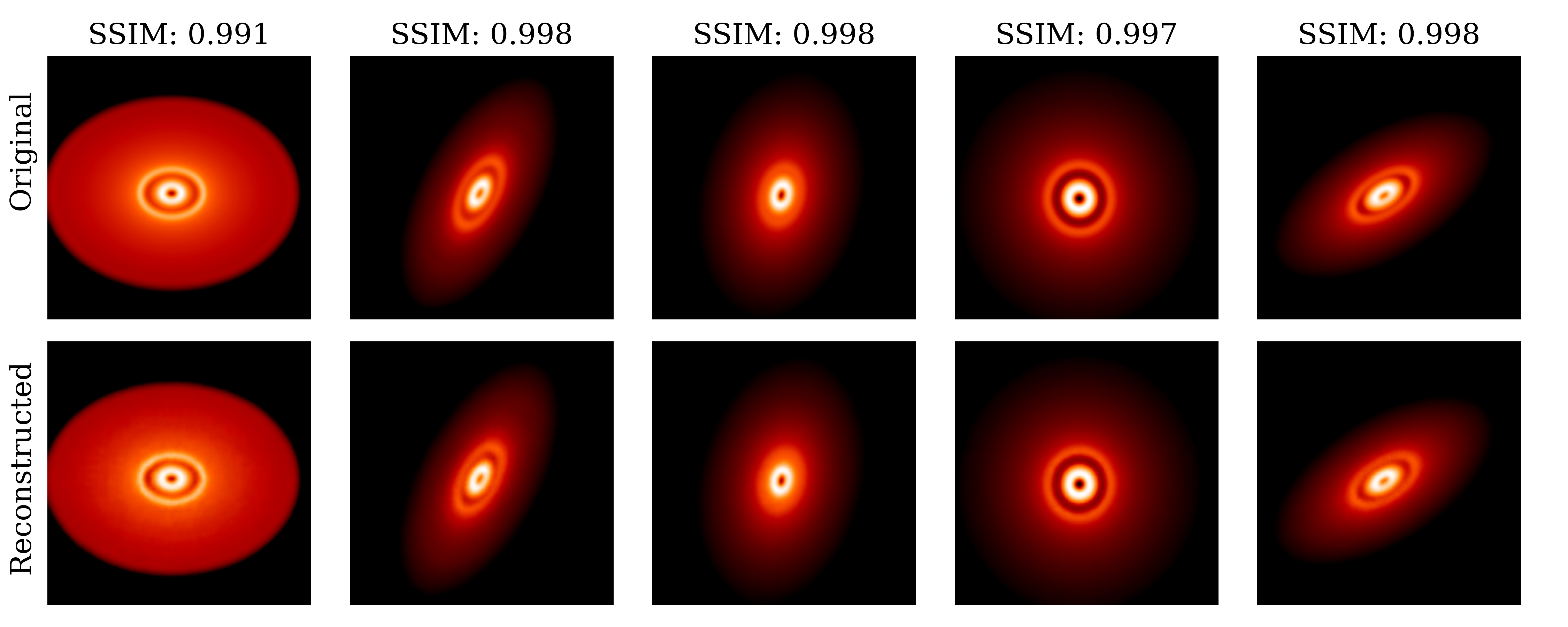}
    \caption{Comparison of original (top row) and reconstructed (bottom row) images from the test dataset using our VADER. Each image was randomly selected, and the structural similarity index between the original and reconstructed images exceeds 99\%.}
    \label{fig:vae_comparison}
\end{figure*}

After the VAE is trained, we independently train two FNNs to predict physical parameters from the learned latent space. The first network, FNN-Planet, is optimized to predict embedded planet masses by minimizing the MSE between the predicted and true values. For FNN-Disk, the loss is defined as the MSE between the true and the predicted disk parameters. The FNNs take in values sampled from the latent space for a given image as input and optimizes its weights to predicts parameters. Both networks are trained using the Adam optimizer \citep{kingma2014adam} with a learning rate of $10^{-3}$ and a batch size of 32 over 150 training epochs. Adam is a standard optimization method that adjusts how much the model updates its weights during training, combining ideas from momentum and adaptive step sizes to help the network converge more quickly and smoothly. This two-stage approach—consisting of the VAE, followed by FNN-Planet and FNN-Disk—forms the complete VADER framework.

\section{Evaluation on Synthetic Images}\label{sec: Modelperform}

To evaluate VADER's performance, we deploy it on 10,500 (10\% of the total dataset) synthetic test images. We assess both how well it can reconstruct the input images and how accurately it predicts the parameters of protoplanetary disks and embedded planets.

\begin{figure*}[ht!]
    \centering
    \includegraphics[width=\textwidth]{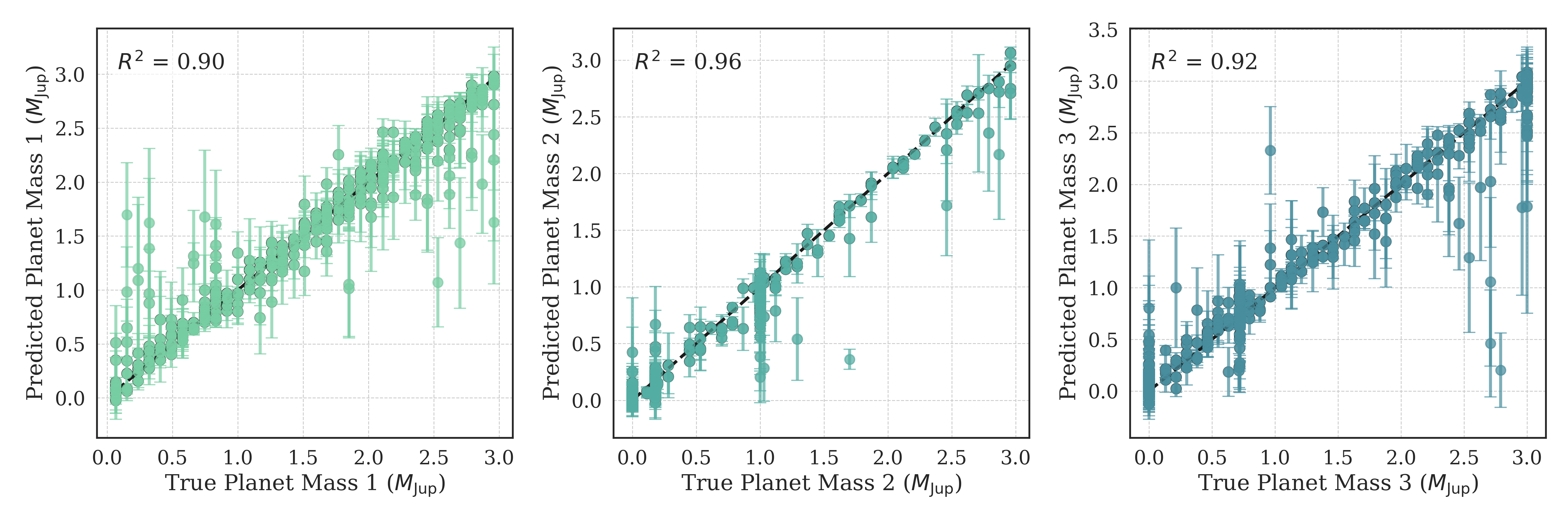}
    \caption{True vs.\ predicted masses of the up to three embedded planets, with error bars indicating VADER's uncertainty estimates. We find the uncertainty by sampling observations multiple times from the mean and log-variance in the latent space for a given image and passing these observations to the FNN iteratively. For all cases, the coefficient of determination $R^2$ is at least 0.9, indicating strong predictive performance. The uncertainties are in most cases comparable to the deviations from the true values.
    \label{fig:planet_mass_radii_performance}}
\end{figure*}

\subsection{Image Reconstruction}
Figure~\ref{fig:vae_comparison} shows five randomly selected examples from the test dataset. The original disk images are in the top row, and the corresponding VADER reconstructions are in the bottom row. We quantify reconstruction quality using the structural similarity index (SSIM) \citep{wang2004image}, defined as:
\begin{equation}
    \text{SSIM}(x, y) = \frac{(2\mu_x \mu_y + C_1)(2\sigma_{xy} + C_2)}{(\mu_x^2 + \mu_y^2 + C_1)(\sigma_x^2 + \sigma_y^2 + C_2)},
\end{equation}
where $\mu_x$ and $\mu_y$ are the mean pixel values of the two images $x$ and $y$, $\sigma_x^2$ and $\sigma_y^2$ are their variances, $\sigma_{xy}$ is the sample covariance between the corresponding local regions (in our case, 7 by 7 pixels) of images $x$ and $y$, and $C_1$, $C_2$ are small constants (relative to the number of pixels) to stabilize the division. This metric reflects perceptual similarity by considering luminance, contrast, and structural information. As reflected from Figure~\ref{fig:vae_comparison}, the reconstructions preserve critical morphological structures such as rings, gaps, and eccentricities, with an average SSIM exceeding 99\% for the test dataset. This demonstrates that the VAE has learned a latent space that successfully captures the essential features of PPD structures. 

\subsection{Planet Mass Inference}
Figure~\ref{fig:planet_mass_radii_performance} illustrates VADER's performance in estimating planet masses. It captures the correlation between predicted and actual values for systems with up to three embedded planets from the test dataset. The VAE model achieves strong predictive accuracy in all cases, with the coefficient of determination $R^2$ exceeding 0.9. $R^2$ is computed using the standard regression-based formula:
\begin{equation}
    R^2 = 1 - \frac{\sum_{i=1}^{n} (y_i - \hat{y}_i)^2}{\sum_{i=1}^{n} (y_i - \bar{y})^2},
\end{equation}
where $y_i$ and $\hat{y}_i$ are the true and predicted values, respectively, and $\bar{y}$ is the mean of the true values. Moreover, the predicted uncertainties grow (as indicated by the widening error bars) as the predicted values deviate further from the true value. This suggests that VADER’s uncertainty estimates meaningfully reflect prediction confidence (For more details see subsection~\ref{subsec:Uncertainty_Quantification}).  

\begin{figure}
    \centering
    \includegraphics[width=0.5\textwidth]{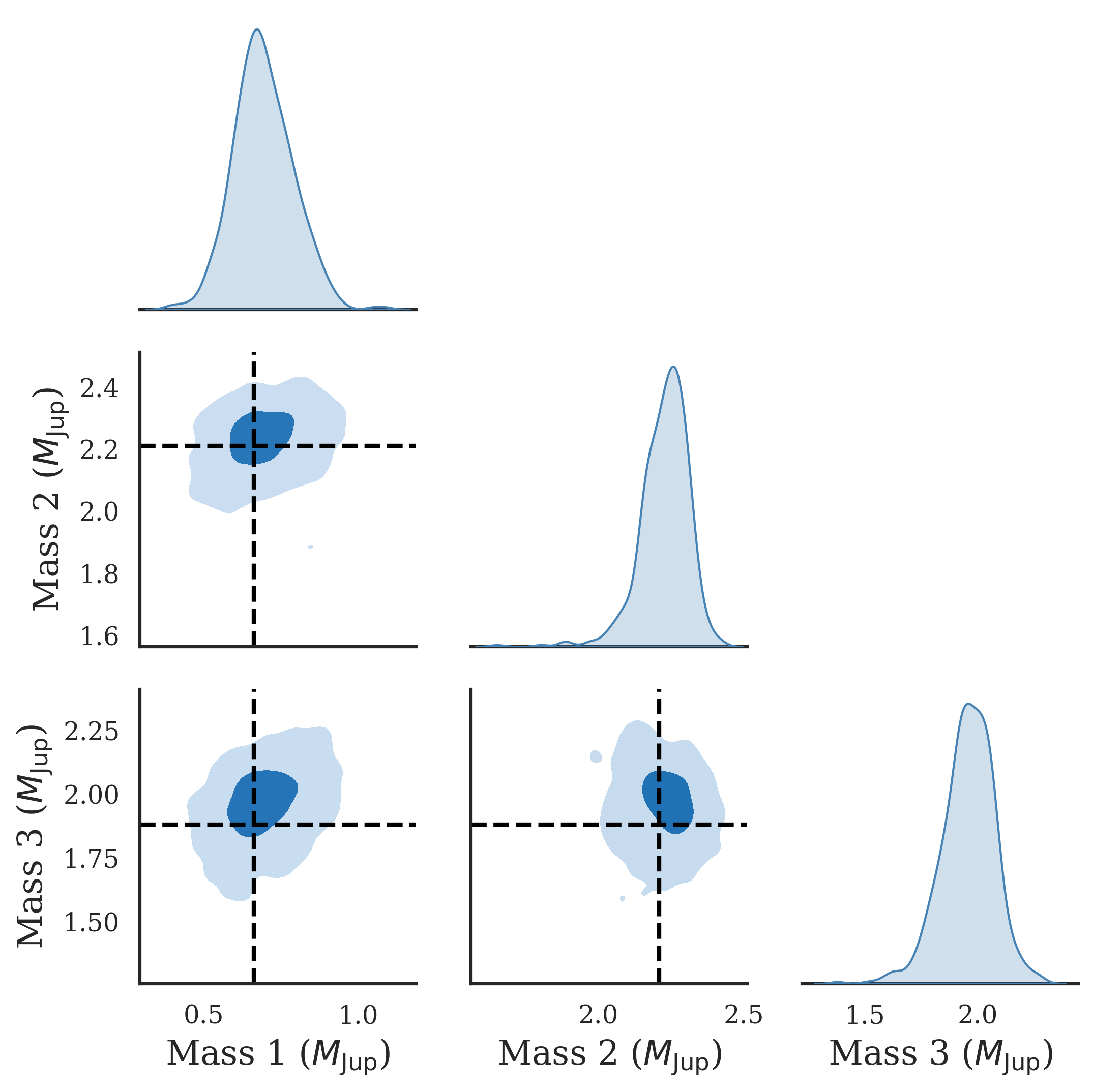}
    \caption{Corner plot showing the distribution of predicted planet masses (in Jupiter masses, $M_{\mathrm{Jup}}$) for a single randomly chosen example from the test dataset. The diagonal panels display the marginal distributions of each parameter, while the off-diagonal panels show joint distributions with confidence contours. The dark blue region corresponds to 1$\sigma$ confidence interval while the light blue region corresponds to 2$\sigma$. The vertical and horizontal dashed lines indicate the actual values for each mass.}
    \label{fig:corner_plot_planet_predictions}
\end{figure}

To further illustrate this capability, we present a corner plot in Figure~\ref{fig:corner_plot_planet_predictions} for a representative example from the test dataset, showing the marginal distributions and joint correlations between predicted planet masses. The vertical and horizontal dashed lines mark the true values. The tightly peaked distributions centered near the intersection of the dashed lines highlight the model’s ability to make accurate predictions and provide well-calibrated uncertainty estimates.


\begin{figure*}[ht!]
    \centering
    \includegraphics[width=\textwidth]{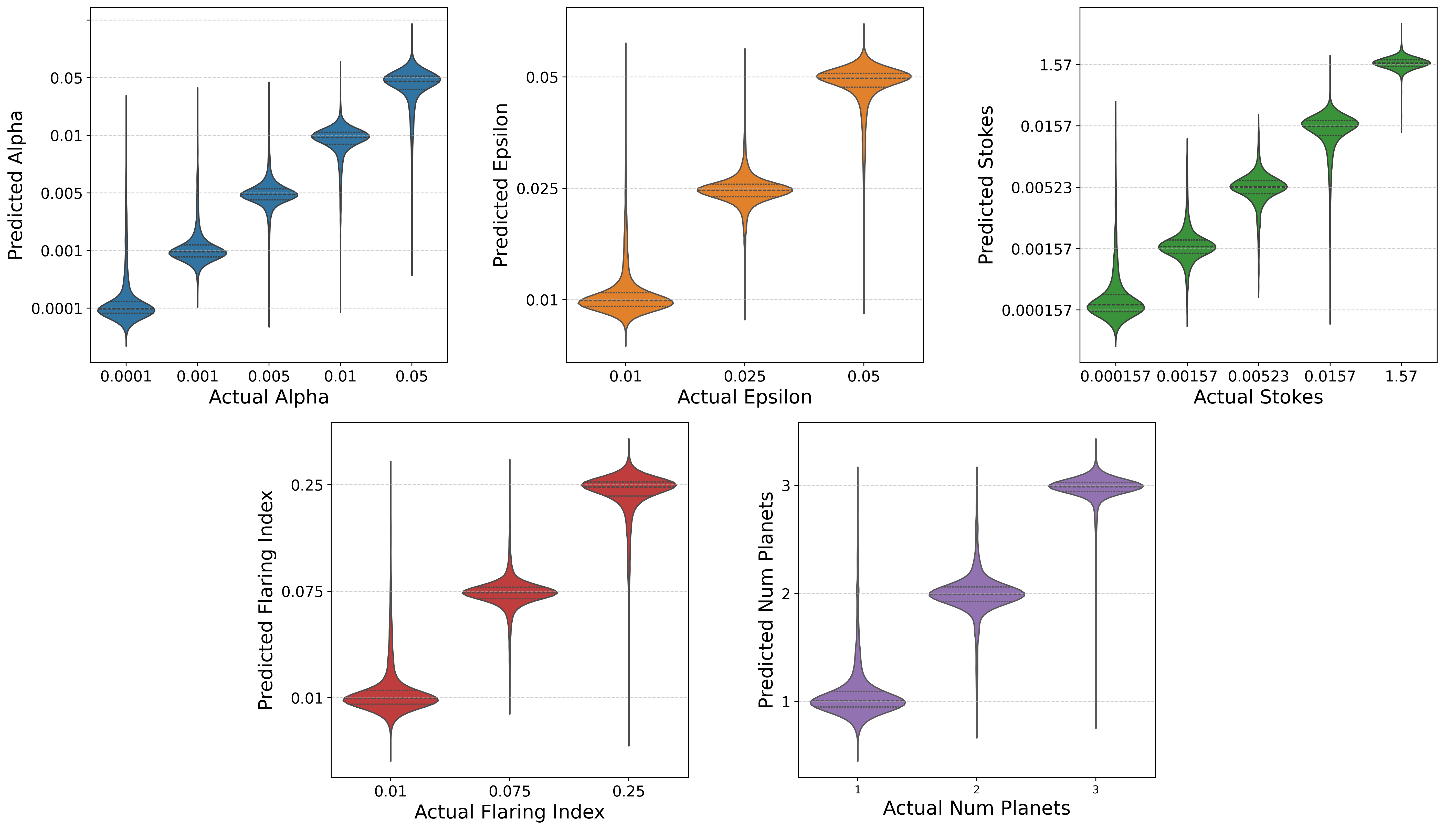} 
    \caption{Violin plots comparing predicted and actual parameter values for protoplanetary disks across five disk parameters in the test data: $\alpha$-viscosity (top left), dust-to-gas ratio (top center), Stokes number (top right), flaring index (bottom left), and number of planets (bottom right). For each parameter, the violin shows the distribution of corresponding predictions made by our model. The interior dashed lines within each violin denote the 25th, 50th (median), and 75th percentiles of the predictive distribution, illustrating both the central tendency and spread. Considering we only had discrete values for the disk parameters, the extremely narrow spread across all values for all the parameters show the predictive capability of the VAE framework for predicting disk parameters.}
    \label{fig:violin_plots}
\end{figure*}

\subsection{Disk Parameter Inference}
We next evaluate VADER's performance in predicting key disk parameters—$\alpha$-viscosity ($\alpha$), dust-to-gas ratio ($\epsilon$), Stokes number ($\mathrm{St}$), and flaring index ($\mathrm{F}$)—using the test dataset. Since these parameters were discretized into a small number of categories, we assess prediction quality by examining the distribution of predicted values conditioned on each true category. Specifically, we evaluate performance based on the concentration, alignment, and separation of predictions across categories. Figure~\ref{fig:violin_plots} shows that VADER produces predictions of disk parameters that are tightly clustered around the correct class, with minimal overlap between adjacent categories. On the 25th percentile, median, and 75th percentile of the predicted values for each actual class, we observe narrow distributions centered near the correct values and little to no spillover into neighboring categories. To compute accuracy of our predictions, we discretize model outputs by rounding each prediction to the nearest integer class using half–integer thresholds (i.e., bins centered at $k$ with edges at $k\pm0.5$). We then compare these binned predictions to the true labels and report exact–match accuracy for each parameter:
$\alpha$–viscosity: $91.2\%$, dust–to–gas ratio: $95.5\%$, Stokes number: $92.1\%$, flaring index: $95.6\%$, and number of planets: $95.1\%$.

\begin{figure*}[ht!]
    \centering
    \includegraphics[width=0.8\textwidth]{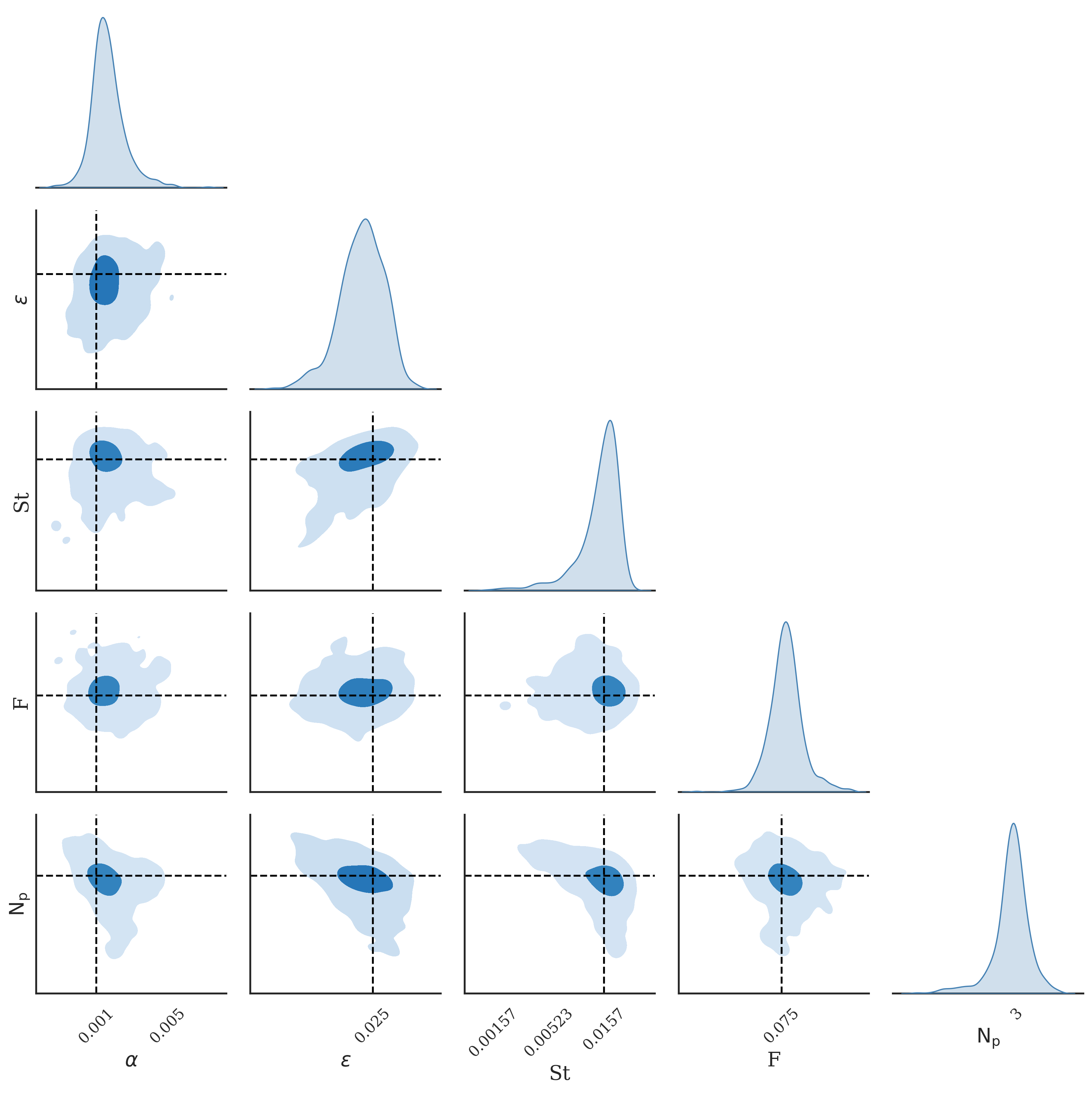}
    \caption{Corner plot illustrating the predicted disk parameters for a randomly selected planetary system from the test dataset. The diagonal panels display the marginal distributions of each parameter, while the off-diagonal panels show joint distributions with confidence contours. The vertical and horizontal dashed lines represent the actual values, allowing for direct comparison with model predictions. The spread in the distributions reflects the uncertainty in predictions, with tighter contours indicating higher confidence.}
    \label{fig:corner_plot_disk_params}
\end{figure*}

Figure~\ref{fig:corner_plot_disk_params} further illustrates VADER's performance by showing a corner plot of predicted disk parameters for a representative test planetary system. This plot visualizes both the marginal distributions and pairwise relationships among the predicted parameters. The predicted distributions are tightly concentrated around the true values, all within 1$\sigma$. These results highlight VADER's ability to infer disk parameters directly from disk morphology.

\section{Application to Observational Data}\label{sec: Resultobs}

\begin{figure*}[ht!]
    \centering
    \includegraphics[width=0.85\textwidth]{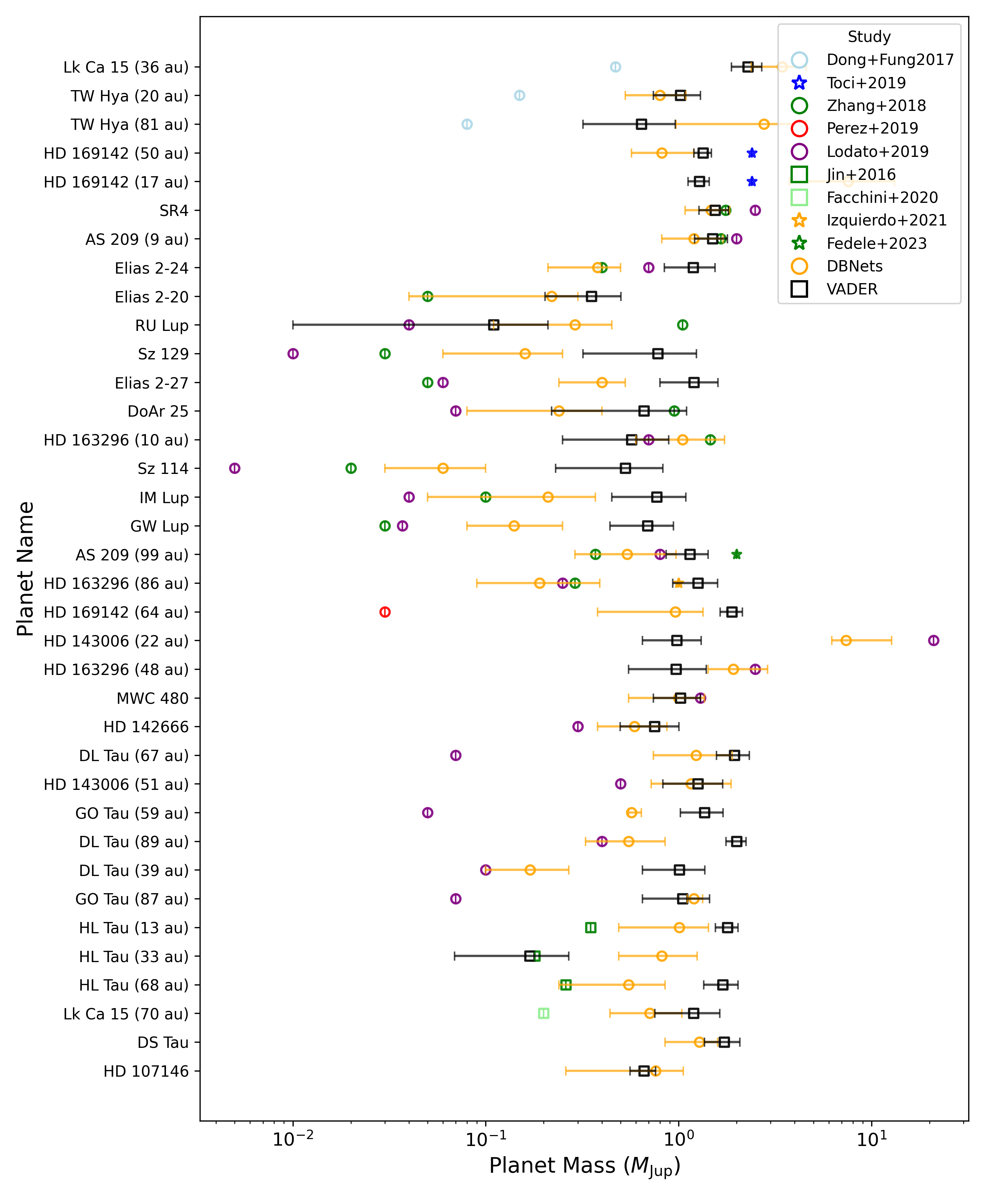}
    \caption{ Comparison of predicted planet masses using VADER with estimates from previous studies across different protoplanetary systems \citep{dong2017mass, toci2019long, zhang2018disk, perez2019dust, lodato2019newborn, jin2016modeling, facchini2020annular, izquierdo2022new, fedele2018alma, ruzza2024dbnets}. The error bars indicate the uncertainty in mass predictions, with VADER predictions shown in black squares. A key validation of our model is that many of its predicted masses are within 1$\sigma$ of the values reported in prior studies, demonstrating strong agreement with independent analyses. Additionally, the predictions from DBNets (orange circles) are shown for reference, further supporting the consistency of our results. }
    \label{fig:vae_planet_mass_comparison}
\end{figure*}

To assess the performance of VADER on real systems, we apply it to a sample of 23 protoplanetary disks, as reported in the DBNets study \citep{ruzza2024dbnets}. Of the 23 disks studied, 14 are from the DSHARP survey \citep{andrews2018disk} and the rest are from \cite{perez2019dust, pinte2019kinematic, toci2019long, facchini2020annular}. Figure~\ref{fig:vae_planet_mass_comparison} showcases the planet mass predictions using VADER (black squares) for these systems. For comparison, we include estimates from DBNets (orange circles) along with other independent values reported in the literature, derived using methods such as hydrodynamic modeling, gap fitting, and kinematic analysis. Table~\ref{tab:gap_masses} provides the mass and the uncertainty estimates for all 23 PPDs. The VADER predictions fall within $1\sigma$ of published values for a majority of systems. Notably, both DBNets and VADER yield similar mass estimates in well-characterized systems, such as AS~209, HD~143006, and TW~Hya. Some divergence in inferred mass is observed in targets where the disk is less well characterized, such as Sz~129. We compute the differences of mean planet mass prediction between VADER and DBNets for multi-planetary and single planet systems and find no statistically significant difference (p-value$=0.73>0.05$) using the Welch's t-test \citep{welch1947generalization}.

In addition to planet masses, VADER provides predictions for disk parameters across all 23 systems. These predictions are summarized in Table~\ref{tab:disk-param-ranges}. In the table, we show the physical ranges of inferred predictions rather than integer-mapped values predicted by VADER. We compute the conversion as follows: VADER predicts each parameter on a discrete, integer scale (see Table~\ref{tab:disk_param_mapping}). For each disk we convert the model’s \emph{mean} predicted category to a physical range by bracketing the value between the two nearest integer categories. Thus, if the predicted category lies between $k$ and $k\!+\!1$ (e.g., 1.6), we report the interval spanned by those two levels in physical units (for $\alpha$, $10^{-4}$–$10^{-3}$). We do not interpolate within bins because the simulations exist only at these discrete grid points; the ranges therefore reflect the resolution of the training grid as well along with the uncertainty of VADER's predictions. The inferred ranges (on the physical grid) are consistent with typical estimates for protoplanetary disks. For example, the $\alpha$-viscosity values in the range of $10^{-4}$--$10^{-2}$ with a peak around $\approx 10^{-3}$ agree well with both theoretical predictions and observational constraints from ALMA turbulence measurements \citep{rafikov2017protoplanetary, hughes2011empirical}. Since $\alpha$-viscosity in our sample space was uniformly sampled from ($10^{-4}-0.05$), we can say that VADER is genuinely constraining $\alpha$-viscosity instead of just predicitng within the same limits of training data with the same distribution. The predicted dust-to-gas ratios of 0.01--0.05 align with the canonical interstellar medium value of about 1\% and with locally dust-enriched regions that are thought to promote planetesimal formation \citep{andrews2009protoplanetary, birnstiel2012simple}. Similarly, the Stokes numbers of $10^{-4}$--$10^{-2}$ fall within the expected regimes for millimeter- to centimeter-sized grains that are only partially coupled to the gas \citep{birnstiel2016dust}. For flaring index, we infer a range between 0.01 to 0.25 for all protoplanetary disks.

In the next subsection, we discuss in detail the inference from VADER applied to two of the most well-studied systems, HD~163296 and HL~Tau. We compare our model predictions for planetary masses,  $\alpha$, $\mathrm{St}$, and $\epsilon$ with established results from the literature.



\begin{deluxetable*}{llccrccclll}
\tabletypesize{\scriptsize}
\tablecaption{Gap Properties and Inferred Planet Masses\label{tab:gap_masses}}
\tablewidth{0pt}
\tablehead{
\colhead{Object Name} & \colhead{$M_\ast$} & \colhead{Distance} & \colhead{Inclination} & \colhead{PA} & \colhead{Resolution} & \colhead{Band} & \colhead{Gap} & \colhead{DBNets} & \colhead{VADER} & \colhead{Data Ref.}\\
\colhead{} & \colhead{($M_\odot$)} & \colhead{(pc)} & \colhead{($^\circ$)} & \colhead{($^\circ$)} & \colhead{(arcsec)} & \colhead{} & \colhead{(au)} & \colhead{$M_p$($M_J$)} & \colhead{$M_p$($M_J$)}
}
\startdata
AS 209 & 0.83 & 121 & 35 & 86 & 0.22 & 6 & 9 & $1.20^{+0.50}_{-0.38}$ & $1.501^{+0.29}_{-0.29}$ & a\\
AS 209 & 0.83 & 121 & 35 & 86 & 0.02 & 6 & 99 & $0.54^{+0.43}_{-0.25}$ & $1.143^{+0.28}_{-0.28}$ & a \\
Elias 2-24 & 0.78 & 136 & 29 & 46 & 0.04 & 6 & 57 & $0.38 \pm 0.17$ & $1.192 \pm 0.35$ & a\\
Elias 2-27 & 0.49 & 140 & 56 & 117 & 0.04 & 6 & 69 & $0.40^{+0.23}_{-0.12}$ & $1.2 \pm 0.4$ & a\\
GW Lup & 0.46 & 155 & 39 & 38 & 0.04 & 6 & 74 & $0.14^{+0.06}_{-0.04}$ & $0.69 \pm 0.25$ & a\\
HD 142666 & 1.58 & 148 & 62 & 162 & 0.13 & 6 & 16 & $0.59^{+0.28}_{-0.21}$ & $0.751 \pm 0.253$ & a\\
HD 143006 & 1.78 & 165 & 19 & 169 & 0.15 & 6 & 22 & $7.38^{+2.56}_{-2.35}$ & $0.98 \pm 0.33$ & a\\
HD 143006 & 1.78 & 165 & 19 & 169 & 0.06 & 6 & 51 & $1.16^{+0.71}_{-0.41}$ & $1.26 \pm 0.43$ & a \\
HD 163296 & 2.04 & 101 & 48 & 133 & 0.21 & 6 & 10 & $1.05^{+0.48}_{-0.45}$ & $0.57 \pm 0.32$ & a \\
HD 163296 & 2.04 & 101 & 48 & 133 & 0.04 & 6 & 48 & $1.92^{+0.97}_{-0.19}$ & $0.97 \pm 0.42$ & a \\
HD 163296 & 2.04 & 101 & 48 & 133 & 0.02 & 6 & 86 & $1.0^{+0.10}_{-0.10}$ & $1.26 \pm 0.33$ & a\\
SR 4 & 0.68 & 134 & 22 & 18 & 0.18 & 6 & 11 & $1.47^{+0.39}_{-0.37}$ & $1.55 \pm 0.27$ & a\\
DoAr 25 & 0.95 & 138 & 67 & 11 & 0.02 & 6 & 111 & $0.24^{+0.16}_{-0.15}$ & $0.66 \pm 0.44$ & a\\
Elias 2-20 & 0.48 & 138 & 49 & 153 & 0.08 & 6 & 25 & $0.22^{+0.11}_{-0.08}$ & $0.35 \pm 0.15$ & a\\
RU Lup & 0.63 & 154 & 19 & 121 & 0.06 & 6 & 29 & $0.29^{+0.16}_{-0.10}$ & $0.11 \pm 0.1$ & a\\
IM Lup & 0.89 & 158 & 48 & 143 & 0.03 & 6 & 117 & $0.21^{+0.16}_{-0.06}$ & $0.77 \pm 0.32$ & a\\
Sz 114 & 0.17 & 162 & 21 & 165 & 0.19 & 6 & 24 & $0.06^{+0.03}_{-0.02}$ & $0.53 \pm 0.3$ & a\\
Sz 129 & 0.83 & 161 & 34 & 151 & 0.07 & 6 & 41 & $0.16^{+0.09}_{-0.06}$ & $0.78 \pm 0.46$ & a\\
HL Tau & 1.0 & 140 & 47 & 138 & 0.14 & 6 & 13.1 & $1.01^{+0.42}_{-0.32}$ & $1.79 \pm 0.24$ & b\\
HL Tau & 1.0 & 140 & 47 & 138 & 0.05 & 6 & 33 & $0.82^{+0.43}_{-0.31}$ & $0.169 \pm 0.1$ & b\\
HL Tau & 1.0 & 140 & 47 & 138 &  0.03 & 6 & 68.6 & $0.55^{+0.30}_{-0.21}$ & $1.69 \pm 0.34$ & b\\
HD 169142 & 1.65 & 117 & 5 & 13 & 0.08 & 6 & 17 & $7.57^{+5.59}_{-3.82}$ & $1.28 \pm 0.16$ & c\\
HD 169142 & 1.65 & 117 & 5 & 13 & 0.03 & 6 & 50 & $0.63^{+0.36}_{-0.29}$ & $1.34 \pm 0.14$ & c\\
HD 169142 & 1.65 & 117 & 5 & 13 & 0.02 & 6 & 64 & $0.95^{+0.65}_{-0.35}$ & $1.89 \pm 0.25$ & c\\
Lk Ca 15 & 1.25 & 159 & 50 & 62 & 0.11 & 6 & 36 & $3.45^{+1.83}_{-0.91}$ & $2.29 \pm 0.41$ & d\\
Lk Ca 15 & 1.25 & 159 & 50 & 62 & 0.06 & 6 & 70 & $0.71^{+0.27}_{-0.23}$ & $1.193 \pm 0.44$ & d\\
TW Hya & 0.8 & 56 & 56 & -8 & 0.03 & 7 & 20 & $0.80^{+0.39}_{-0.27}$ & $1.02 \pm 0.28$ & e\\
TW Hya & 0.8 & 56 & 56 & -8 & 0.01 & 7 & 81 & $2.77^{+1.59}_{-0.88}$ & $0.64 \pm 0.32$ & e\\
DS Tau & 0.83 & 159 & 65 & -19 & 0.29 & 3 & 33 & $1.28^{+0.43}_{-0.35}$ & $1.72 \pm 0.36$ & f\\
DL Tau & 0.98 & 159 & 45 & 52 & 0.24 & 6 & 39 & $0.55^{+0.22}_{-0.20}$ & $1.01 \pm 0.36$ & f\\
DL Tau & 0.98 & 159 & 45 & 52 & 0.14 & 6 & 67 & $1.23^{+0.67}_{-0.49}$ & $1.95 \pm 0.38$ & f\\
DL Tau & 0.98 & 159 & 45 & 52 & 0.11 & 6 & 89 & $0.17^{+0.10}_{-0.07}$ & $2 \pm 0.24$ & f\\
GO Tau & 0.36 & 144 & 54 & 21 & 0.15 & 6 & 59 & $0.22^{+0.14}_{-0.08}$ & $1.36 \pm 0.34$ & f\\
GO Tau & 0.36 & 144 & 54 & 21 & 0.10 & 6 & 87 & $0.12^{+0.06}_{-0.05}$ & $1.05 \pm 0.4$ & f\\
HD 107146 & 1.0 & 27.5 & 19 & 153 & 0.12 & 6 & 80 & $0.76^{+0.43}_{-0.34}$ & $0.66 \pm 0.1$ & g\\
MWC 480 & 1.91 & 161 & 36.5 & 147.5 & 0.16 & 6 & 73 & $1.06^{+0.45}_{-0.36}$ & $1.02 \pm 0.28$ & f\\
\enddata
\tablecomments{The stellar mass, inclination angle, spatial resolution, and gap data are adopted from \cite{ruzza2024dbnets}. \textit{Data References:} (a)\href{https://almascience.eso.org/almadata/lp/DSHARP/}{DSHARP Data Release}, (b)\href{https://almascience.eso.org/alma-data/science-verification}{ALMA Science Verification Data} (c)\cite{perez2019dust}, (d)\cite{facchini2020annular}(e)\cite{andrews2016ringed}, (f)\cite{long2018gaps}, (g)\cite{marino2021constraining}. For a couple of observations, we use bands other than default ALMA band 6. We check their mass predictions with other studies (Figure \ref{fig:vae_planet_mass_comparison}) and find planet mass predictions to lie within 1$\sigma$ for both observations.}
\end{deluxetable*}

\begin{table*}[t]
\centering
\caption{Disk parameter ranges for 23 protoplanetary disk systems as predicted by VADER}

\label{tab:disk-param-ranges}
\small
\setlength{\tabcolsep}{6pt}
\begin{tabular*}{\textwidth}{@{\extracolsep{\fill}} lcccc}
\toprule
Disk System & Viscosity ($\alpha$) & Dust to Gas Ratio ($\epsilon$) & Stokes Number (St) & Flaring Index (F)\\
\midrule
AS~209   & $10^{-3}\,\text{--}\,5\times10^{-3}$ & $0.01\,\text{--}\,0.025$ & $1.57\times10^{-3}\,\text{--}\,5.23\times10^{-3}$ & $0.01\,\text{--}\,0.075$ \\
DS~Tau   & $10^{-3}\,\text{--}\,5\times10^{-3}$ & $0.01\,\text{--}\,0.025$ & $5.23\times10^{-3}\,\text{--}\,1.57\times10^{-2}$ & $0.01\,\text{--}\,0.075$ \\
DoAr~25  & $10^{-3}\,\text{--}\,5\times10^{-3}$ & $0.025\,\text{--}\,0.05$ & $1.57\times10^{-4}\,\text{--}\,1.57\times10^{-3}$ & $0.075\,\text{--}\,0.25$ \\
HD~169142  & $5\times10^{-3}\,\text{--}\,10^{-2}$ & $0.01\,\text{--}\,0.025$ & $1.57\times10^{-2}\,\text{--}\,1.57$ & $0.01\,\text{--}\,0.075$ \\
Elias~20 & $5\times10^{-3}\,\text{--}\,10^{-2}$ & $0.025\,\text{--}\,0.05$ & $1.57\times10^{-4}\,\text{--}\,1.57\times10^{-3}$ & $0.075\,\text{--}\,0.25$ \\
Elias~24 & $5\times10^{-3}\,\text{--}\,10^{-2}$ & $0.01\,\text{--}\,0.025$ & $1.57\times10^{-3}\,\text{--}\,5.23\times10^{-3}$ & $0.01\,\text{--}\,0.075$ \\
Elias~27 & $10^{-2}\,\text{--}\,5\times10^{-2}$ & $0.01\,\text{--}\,0.025$ & $1.57\times10^{-4}\,\text{--}\,1.57\times10^{-3}$ & $0.01\,\text{--}\,0.075$ \\
GW~Lup   & $5\times10^{-3}\,\text{--}\,10^{-2}$ & $0.01\,\text{--}\,0.025$ & $1.57\times10^{-4}\,\text{--}\,1.57\times10^{-3}$ & $0.075\,\text{--}\,0.25$ \\
HD~142666& $10^{-3}\,\text{--}\,5\times10^{-3}$ & $0.01\,\text{--}\,0.025$ & $5.23\times10^{-3}\,\text{--}\,1.57\times10^{-2}$ & $0.01\,\text{--}\,0.075$ \\
HD~107146   & $5\times10^{-3}\,\text{--}\,10^{-2}$ & $0.01\,\text{--}\,0.025$ & $5.23\times10^{-3}\,\text{--}\,1.57\times10^{-2}$ & $0.01\,\text{--}\,0.075$ \\
HD~143006& $10^{-3}\,\text{--}\,5\times10^{-3}$ & $0.01\,\text{--}\,0.025$ & $1.57\times10^{-2}\,\text{--}\,1.57$ & $0.075\,\text{--}\,0.25$ \\
HD~163296& $10^{-3}\,\text{--}\,5\times10^{-3}$ & $0.01\,\text{--}\,0.025$ & $1.57\times10^{-3}\,\text{--}\,5.23\times10^{-3}$ & $0.075\,\text{--}\,0.25$ \\
HL~Tau& $10^{-3}\,\text{--}\,5\times10^{-3}$ & $0.025\,\text{--}\,0.05$ & $1.57\times10^{-4}\,\text{--}\,1.57\times10^{-3}$ & $0.075\,\text{--}\,0.25$ \\
HT~Lup   & $5\times10^{-3}\,\text{--}\,10^{-2}$ & $0.01\,\text{--}\,0.025$ & $5.23\times10^{-3}\,\text{--}\,1.57\times10^{-2}$ & $0.01\,\text{--}\,0.075$ \\
IM~Lup   & $5\times10^{-3}\,\text{--}\,10^{-2}$ & $0.01\,\text{--}\,0.025$ & $1.57\times10^{-3}\,\text{--}\,5.23\times10^{-3}$ & $0.01\,\text{--}\,0.075$ \\
TW~Hya   & $5\times10^{-3}\,\text{--}\,10^{-2}$ & $0.01\,\text{--}\,0.025$ & $1.57\times10^{-3}\,\text{--}\,5.23\times10^{-3}$ & $0.01\,\text{--}\,0.075$ \\
RU~Lup   & $10^{-3}\,\text{--}\,5\times10^{-3}$ & $0.01\,\text{--}\,0.025$ & $5.23\times10^{-3}\,\text{--}\,1.57\times10^{-2}$ & $0.01\,\text{--}\,0.075$ \\
SR~4     & $10^{-4}\,\text{--}\,10^{-3}$        & $0.025\,\text{--}\,0.05$ & $1.57\times10^{-2}\,\text{--}\,1.57$ & $0.075\,\text{--}\,0.25$ \\
Sz~114   & $5\times10^{-3}\,\text{--}\,10^{-2}$ & $0.01\,\text{--}\,0.025$ & $1.57\times10^{-4}\,\text{--}\,1.57\times10^{-3}$ & $0.01\,\text{--}\,0.075$ \\
Sz~129   & $5\times10^{-3}\,\text{--}\,10^{-2}$ & $0.025\,\text{--}\,0.05$ & $5.23\times10^{-3}\,\text{--}\,1.57\times10^{-2}$ & $0.01\,\text{--}\,0.075$ \\
Lk~Ca~15  & $10^{-2}\,\text{--}\,5\times10^{-2}$ & $0.01\,\text{--}\,0.025$ & $1.57\times10^{-4}\,\text{--}\,1.57\times10^{-3}$ & $0.01\,\text{--}\,0.075$ \\
GO~Tau   & $10^{-3}\,\text{--}\,5\times10^{-3}$ & $0.01\,\text{--}\,0.025$ & $1.57\times10^{-3}\,\text{--}\,5.23\times10^{-3}$ & $0.075\,\text{--}\,0.25$ \\
MWC~480   & $5\times10^{-3}\,\text{--}\,10^{-2}$ & $0.01\,\text{--}\,0.025$ & $1.57\times10^{-3}\,\text{--}\,5.23\times10^{-3}$ & $0.075\,\text{--}\,0.25$ \\
\bottomrule

\end{tabular*}
\tablecomments{VADER predicts each parameter on a discrete, integerized scale (see Table~\ref{tab:disk_param_mapping}). For each disk we convert the model’s \emph{mean} predicted category to a physical range by bracketing between the two nearest integer categories. Thus, if the predicted category lies between $k$ and $k\!+\!1$ (e.g., 1.6), we report the interval spanned by those two levels in physical units (for $\alpha$, $10^{-4}$–$10^{-3}$). We do not interpolate within bins because the simulations exist only at these discrete grid points; the quoted ranges therefore reflect the resolution of the training grid as well along with the uncertainty of VADER's predictions.}

\end{table*}


\begin{figure*}
\gridline{
  \fig{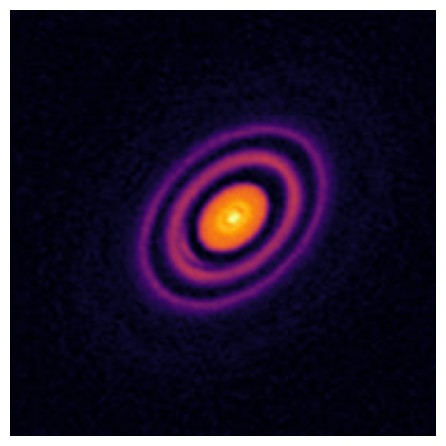}{0.48\textwidth}{}
  \hfill
  \fig{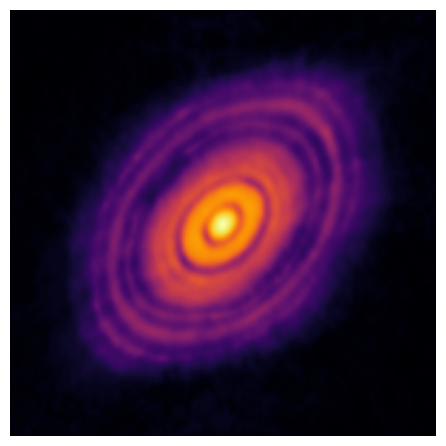}{0.48\textwidth}{}
}
\caption{ALMA observations at 1.3 mm of protoplanetary disks HD 163296 (left) and HL Tau (right).}
\label{fig:disk_comparison}
\end{figure*}

\subsection{HD 163296}\label{subsub hd163296}

HD~163296 has been extensively studied as a benchmark system for understanding planet-disk interactions and disk structures. The disk shows clear evidence of axisymmetric gaps in dust thermal emission (see Figure \ref{fig:disk_comparison}). Several studies \citep{isella2016ringed, isella2018signatures, teague2018kinematical}, including hydrodynamical simulations \citep{rodenkirch2021modeling}, suggest that the axisymmetric gaps are due to disk–planet interaction. 



\begin{figure*}
\gridline{
  \fig{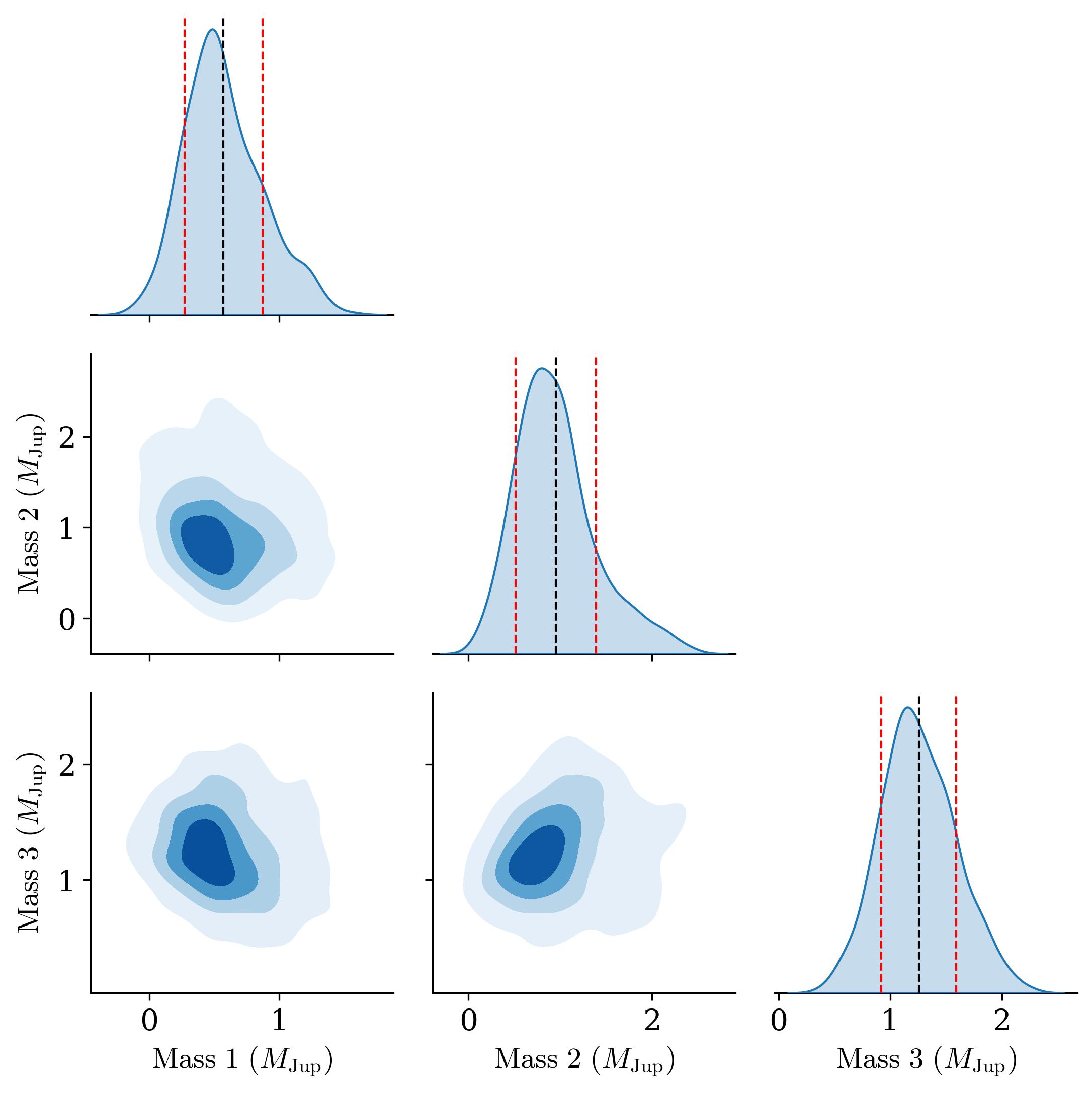}{0.48\textwidth}{}
  \hfill
  \fig{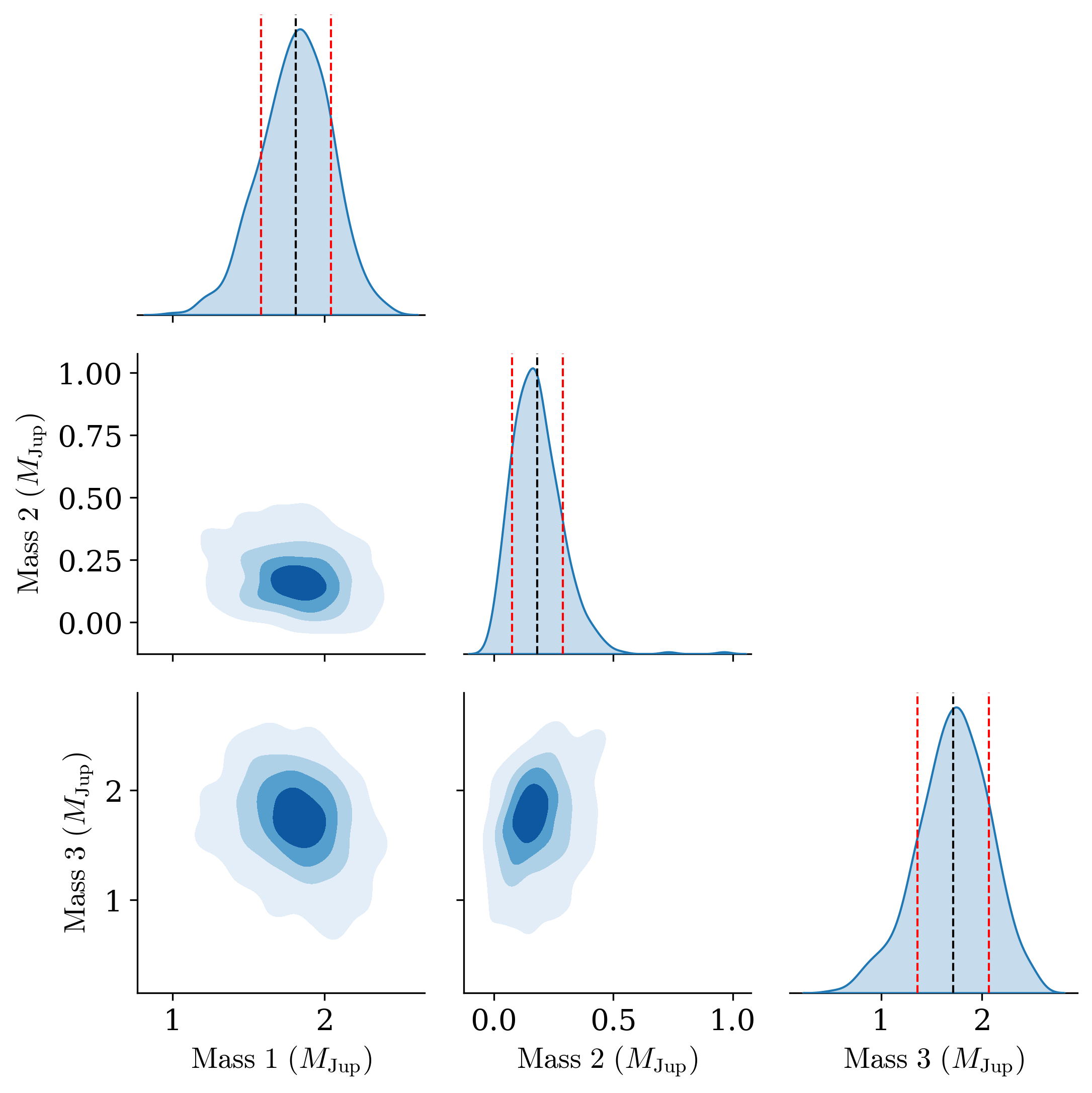}{0.48\textwidth}{}
}
\caption{Density corner plots for mass estimates of protoplanets embedded in HD~163296 (left) and HL~Tau (right).}
\label{fig:planet_mass_hltau_hd163296}
\end{figure*}

For the HD~163296 system, VADER predicts the presence of three planets with masses as shown in the right panel of Figure \ref{fig:planet_mass_hltau_hd163296}.
For ``Planet~1", our model predicts a mass of $0.57 \pm 0.30~M_{\text{Jup}}$, while the DBNets study predicts $1.05^{+0.68}_{-0.45}~M_{\text{Jup}}$. For comparison, \cite{zhang2018disk}, by using \texttt{FARGO} simulations estimated a mass range of $0.19 - 1.46~M_{\text{Jup}}$, depending on different combinations of initial conditions of $\alpha$, Stokes number, and dust density. For``Planet~2", VADER predicts $0.95 \pm 0.44~M_{\text{Jup}}$, while DBNets estimates $1.92^{+0.97}_{-0.50}~M_{\text{Jup}}$. \cite{zhang2018disk} suggested a range of $0.54 - 2.24~M_{\text{Jup}}$, depending on disk parameters such as $\alpha$ and dust-to-gas ratio. For ``Planet~3", VADER predicts $1.25 \pm 0.33~M_{\text{Jup}}$, while \cite{ruzza2024dbnets} predicts a significantly lower mass of $0.19^{+0.21}_{-0.10}~M_{\text{Jup}}$. \citet{zhang2018disk} estimated a range of $0.54 - 2.24~M_{\text{Jup}}$, again depending on different combinations of $\alpha$, Stokes number, and dust density. However, kinematic studies by \cite{izquierdo2022new} and \cite{teague2018kinematical} inferred a planetary mass close to $1~M_{\text{Jup}}$ from CO observations, aligning closely with our prediction.

\begin{figure*}[ht!]
    \centering
    \includegraphics[width=0.8\textwidth]{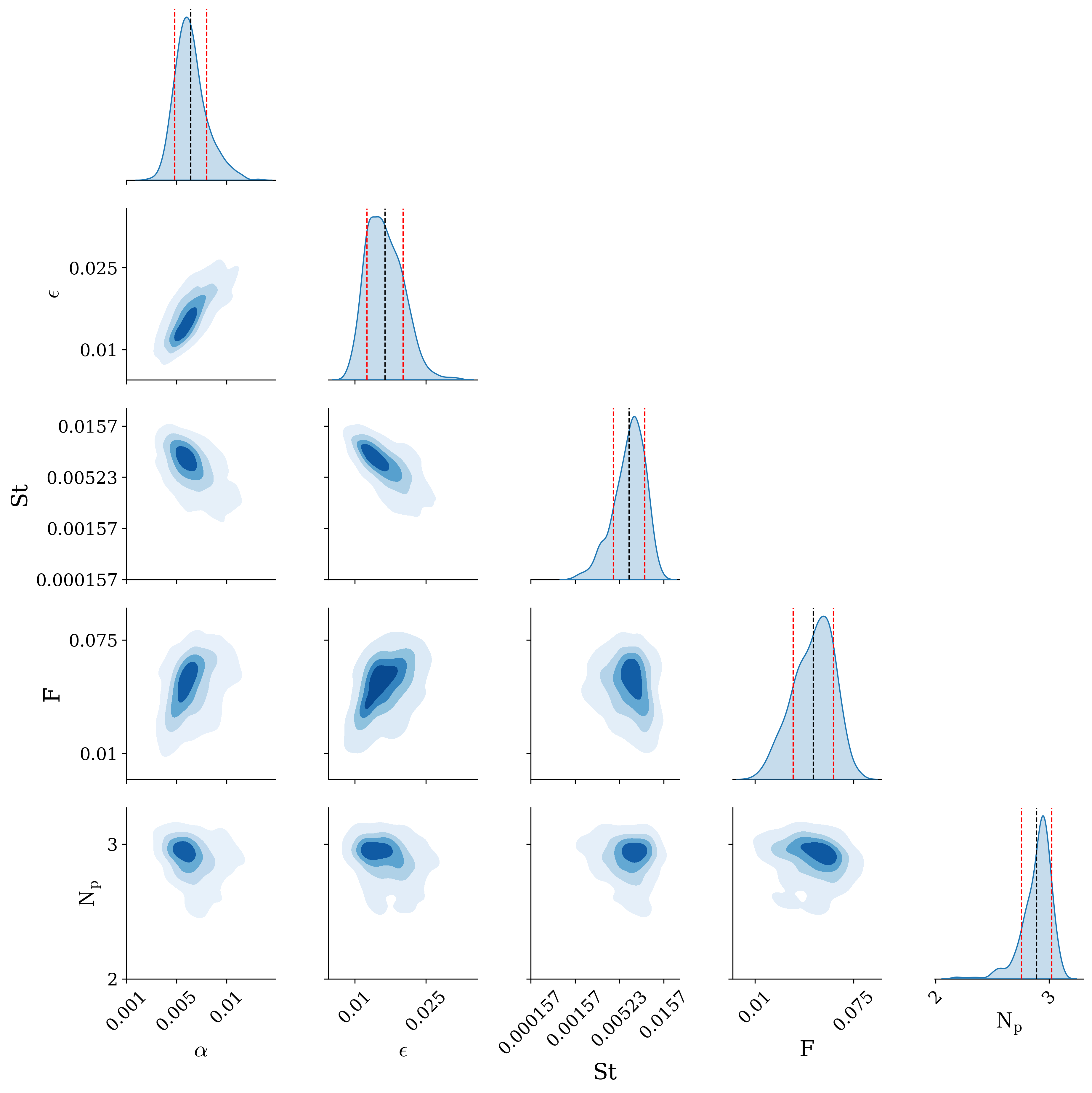}
    \caption{Corner plot showing the posterior distributions of inferred disk and planet parameters for HD~163296. Parameters include $\alpha$, $\mathrm{St}$, $\epsilon$, $\mathrm{F}$ and Number of Planets ($\mathrm{N_p}$). The vertical dashed lines in the marginalized distributions indicate the mean and 1$\sigma$ confidence bounds. The model identifies a three-planet configuration as the most probable explanation for the observed disk substructures in HD~163296.}
    \label{fig:diskparams_hd163296}
\end{figure*}

Figure~\ref{fig:diskparams_hd163296} shows the posterior distributions of the inferred disk parameters and the number of planets for HD~163296, as predicted by VADER. The predicted $\alpha$ values cluster around a few $\times 10^{-3}$, aligning well with estimates from both hydrodynamical simulations and observational studies. For example, \cite{isella2016ringed} estimated $10^{-3} < \alpha < 10^{-2}$, assuming a Saturn-mass planet influencing gas surface density. \cite{liu2018new} derived $\alpha \sim 7.5 \times 10^{-3}$ in the outer disk based on dust evolution modeling. \cite{muller2022emerging} assumed $3 \times 10^{-4} < \alpha < 10^{-3}$ to study planetary evolution in HD~163296 using the FLINTSTONE code developed by \cite{bitsch2019formation}. FLINTSTONE simulates planetary growth and orbital migration by coupling pebble and gas accretion with disk-driven migration in protoplanetary disks. \cite{rodenkirch2021modeling} suggested $\alpha < 2 \times 10^{-3}$, which is slightly lower but still within range of our predictions. 

Stokes number ($\mathrm{St}$) characterizes the aerodynamic coupling between dust grains and gas in the disk. VADER predicts $\mathrm{St} \sim 10^{-3} - 5 \times 10^{-3}$, which closely overlaps with values found in previous studies. \cite{rodenkirch2021modeling} suggested $\mathrm{St} < 0.036$ based on a 1.3~mm observation, supporting good dust-gas coupling, while \cite{jiang2024grain} found $\mathrm{St}$ ranging from $10^{-3}$ to $10^{-1}$ at ALMA observable wavelengths, depending on the nH$_2$O/nCO ratio. These findings suggest that HD~163296 exhibits moderate dust-gas coupling, where larger grains efficiently settle toward the midplane while continuing to experience radial drift. Such conditions are conducive to planetesimal formation and leads to the formation of gap.

Furthermore, the dust-to-gas mass ratio $\epsilon$ is a crucial metric for understanding planetesimal formation. VADER predicts $\epsilon$ from 1\% to $2.5\%$, which is consistent with past estimates. VADER is trained to recognize $\epsilon$ up to 5\% but does not find such signatures for this object. \cite{tilling2012gas} found $\epsilon$ ranging from 1\% to 11\%, depending on dust settling, while \cite{isella2016ringed} determined $\epsilon \sim 2.5\%$ within 10~au based on ALMA observations of HD~163296, closely matching our estimates. The consistency of our predictions with observational studies reinforces the idea that HD~163296 maintains an enhanced dust fraction relative to the canonical interstellar value (1\%).


\subsection{HL Tau} \label{subsub HLTau}

\begin{figure*}[ht!]
    \centering
    \includegraphics[width=0.8\textwidth]{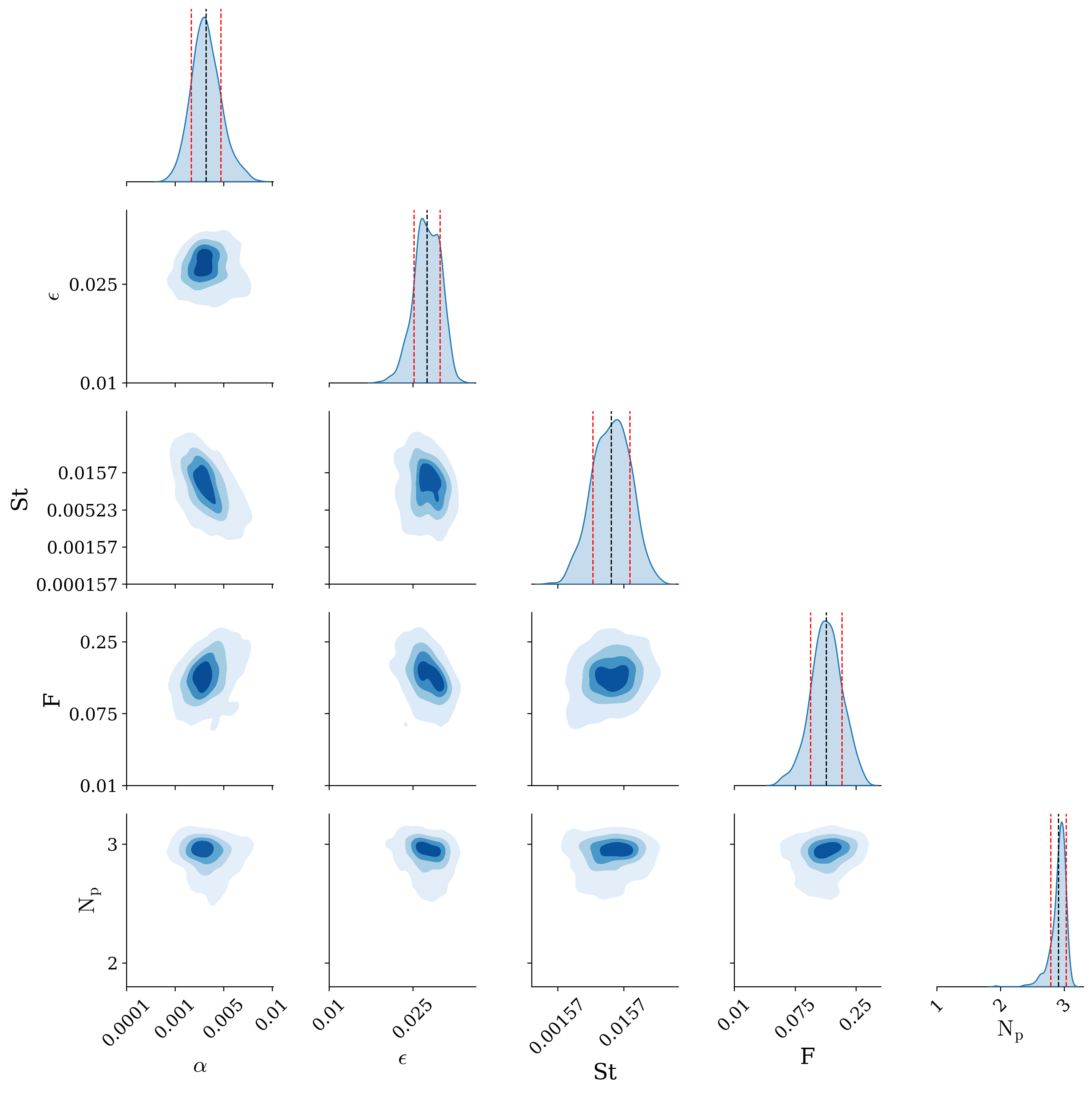}
    \caption{Corner plot showing the posterior distributions of inferred disk and planet parameters for HL~Tau. Parameters include the disk $\alpha$, $\mathrm{St}$, $\epsilon$, Flaring Index ($\mathrm{F}$), and the number of embedded planets ($\mathrm{N_p}$). The vertical dashed lines in the marginalized distributions indicate the mean and 1$\sigma$ confidence bounds. The model identifies a three-planet configuration as the most probable explanation for the observed disk substructures in HL~Tau.}
    \label{fig:diskparams_HLTau}
\end{figure*}

HL~Tau has been extensively studied due to its well-resolved disk structures observed with ALMA. VADER predicts the mass of three of the most massive planets within the disk, with estimated masses of \(1.81 \pm 0.23\, M_{\mathrm{Jup}}\)  \(0.18 \pm 0.11\, M_{\mathrm{Jup}}\) and \(1.72 \pm 0.35\, M_{\mathrm{Jup}}\). Figure~\ref{fig:planet_mass_hltau_hd163296} (left) displays these predictions, highlighting the spatial distribution and associated uncertainties. These predictions are broadly consistent with those from DBNets, which yield mass estimates of \(1.01^{+0.42}_{-0.52}\, M_{\mathrm{Jup}}\), \(0.82^{+0.43}_{-0.33}\, M_{\mathrm{Jup}}\), and \(0.55^{+0.30}_{-0.31}\, M_{\mathrm{Jup}}\). In contrast, \citet{jin2016modeling} estimated lower masses using the LA-COMPASS Code \citep{drazkowska2019including}, with estimates of \(0.35\, M_{\mathrm{Jup}}\), \(0.17\, M_{\mathrm{Jup}}\), and \(0.26\, M_{\mathrm{Jup}}\) respectively. LA-COMPASS is a forward hydrodynamics simulation code which enables planet mass predictions by simulating how a disk would look given a planet of known mass, and comparing those simulations to observed disk structures. The 2nd planet mass prediction from \cite{jin2016modeling} agrees remarkably well with our predictions. Despite variations across models, VADER predictions fall within ranges previously inferred for HL~Tau.

Figure~\ref{fig:diskparams_HLTau} shows the posterior distributions of the inferred disk parameters and the number of planets for HL~Tau, as predicted by VADER. Note that the system could have more planets \citep{tamayo2015dynamical, simbulan2017connecting} but VADER has been trained to recognize up to three. For the $\alpha$-viscosity parameter, VADER predicts values in the range of \(10^{-3}\) to \(5 \times 10^{-3}\). These estimates are consistent with previous studies. For example, \cite{kanagawa2015mass} inferred planet masses on the order of one Jupiter mass within HL~Tau and estimated a viscosity parameter of \(\alpha \approx 10^{-3}\), consistent with the observed gap widths having been carved by such planets. Their mass estimates were based on an empirical relationship between gap width and planet mass, derived by considering angular momentum exchange between the planet and the surrounding disk. Similarly, \cite{jiang2024grain}, using the DustPy framework \citep{stammler2022dustpy}, found that $\alpha$ varies from \(10^{-4}\) to the high \(10^{-3}\) range depending on the location within the disk. The agreement between VADER’s predicted $\alpha$ values and these independent results supports the conclusion that HL~Tau’s disk has relatively low viscosity, an important factor influencing dust dynamics and planet formation.

For Stokes number, VADER predicts values in the $10^{-3}$ to $10^{-2}$ range as shown in Figure \ref{fig:diskparams_HLTau}. This prediction is in line with \cite{jiang2024grain}, who estimated a few $10^{-3}$ to a few $10^{-2}$. The consistency of our results with established models suggests that dust particles in HL~Tau remain well-coupled to the gas.

For $\epsilon$, our model predicts ~2.5\%, which is higher than the canonical ISM value of 1\%. \cite{pinte2015dust} and \cite{wu2018physical} estimated that dust depletion in gaps leads to significantly high dust-to-gas ratio elsewhere, and our results support this conclusion.

Overall, VADER provides predictions that are largely consistent with other studies but suggest slightly high planetary masses and $\epsilon$. The ability of VADER to infer parameters rapidly from large observational datasets highlights its potential as a valuable tool for constraining disk parameters, the nature of disk structures and planet formation processes.

\section{Discussion} \label{sec: Discussions}

\subsection{Structure of Latent Space}
A key strength of VADER lies in the structure of the learned latent space. As shown in Appendix \ref{Apxc},  movement along latent dimensions, which correlate with a disk parameter, results in smooth and distinguishable changes in morphology of the decoder generated disk images. For example, varying the values in latent dimensions that directly correlate with the masses of the planets influences the widths of the gaps. Similarly, tuning the latent space coordinate related to the disk turbulence makes the overall disk structures more diffused.
This indicates that the model is not simply memorizing the data, but instead learning physically meaningful representations that capture the underlying processes shaping disk structures.

\subsection{Uncertainty Quantification}\label{subsec:Uncertainty_Quantification}

\begin{figure}
    \centering
    \includegraphics[width=0.5\textwidth]{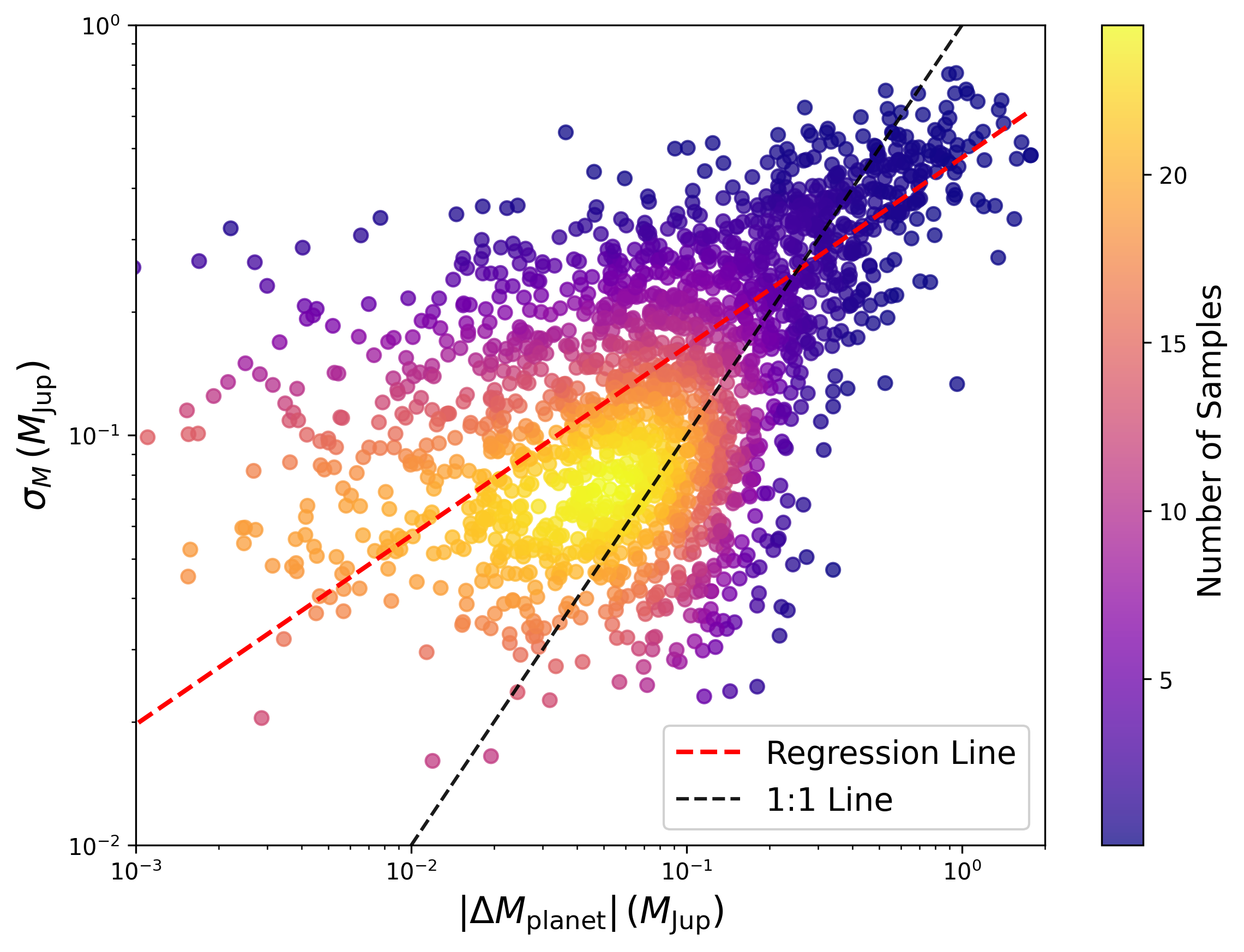}  
    \caption{ Standard deviation in planet mass prediction as a function of mass error, revealing confidence trends in the VAE model. As prediction significantly deviates from true mass, the uncertainty increases, highlighting it as a tool to quantify confidence in prediction. The color bar is used to show the relative concentration of points where higher values mean a larger number of observations per pixel in the image. The black dashed line represents the 1:1 correspondence between standard deviation of prediction and prediction error while the red line represents the best-fit line for standard deviation and planet mass prediction error.}
    \label{fig:std_vs_massdifference}
\end{figure}

The probabilistic nature of VADER enables uncertainty quantification at a per-prediction level. Figure~\ref{fig:std_vs_massdifference} illustrates how VADER increases its uncertainty as error in its prediction increases, providing a measure of confidence in its outputs. From the plot we can infer that for most cases VADER predictions have uncertainties between 0.01 and 0.1 $M_{\mathrm{Jup}}$. This slight difference in prediction is accounted for by the standard deviation of our predictions, thanks to the increasing nature of uncertainty with difference in predicted planet mass. For a difference in actual and predicted values of 0.01 to 0.1 $M_\mathrm{Jup}$, the standard deviation of planet mass predictions range from 0.05 to 0.2 $M_\mathrm{Jup}$. For differences larger than 0.1$M_\mathrm{Jup}$ standard deviations in predictions are usually $\geq0.1M_\mathrm{Jup}$.

\subsection{Predicted planet mass distribution}
 VADER predicts planet masses spanning $0.3\,M_\mathrm{Jup}$ to $2\,M_\mathrm{Jup}$ (Figure~\ref{fig:vae_planet_mass_comparison}) when applied to the   ALMA observations. This range is consistent with other estimates. For example, gaps observed by ALMA in AS~209 \citep{fedele2018alma} and HD~163296  \citep{muller2022emerging} have been linked to Jupiter-mass planets. Simulations by \citet{elbakyan2021gap} show that while gap formation can begin at $\sim$0.1\,$M_\mathrm{Jup}$, more prominent structures typically require masses above 0.5\,$M_\mathrm{Jup}$. The VADER results are consistent with predictions \citep{gonzalez2012planet} that Jupiter-mass planets are effective at producing observable gaps.

\begin{figure*}
\gridline{
  \fig{normalized_mass_distribution_3.png}{0.55\textwidth}{}
  \hfill
  \fig{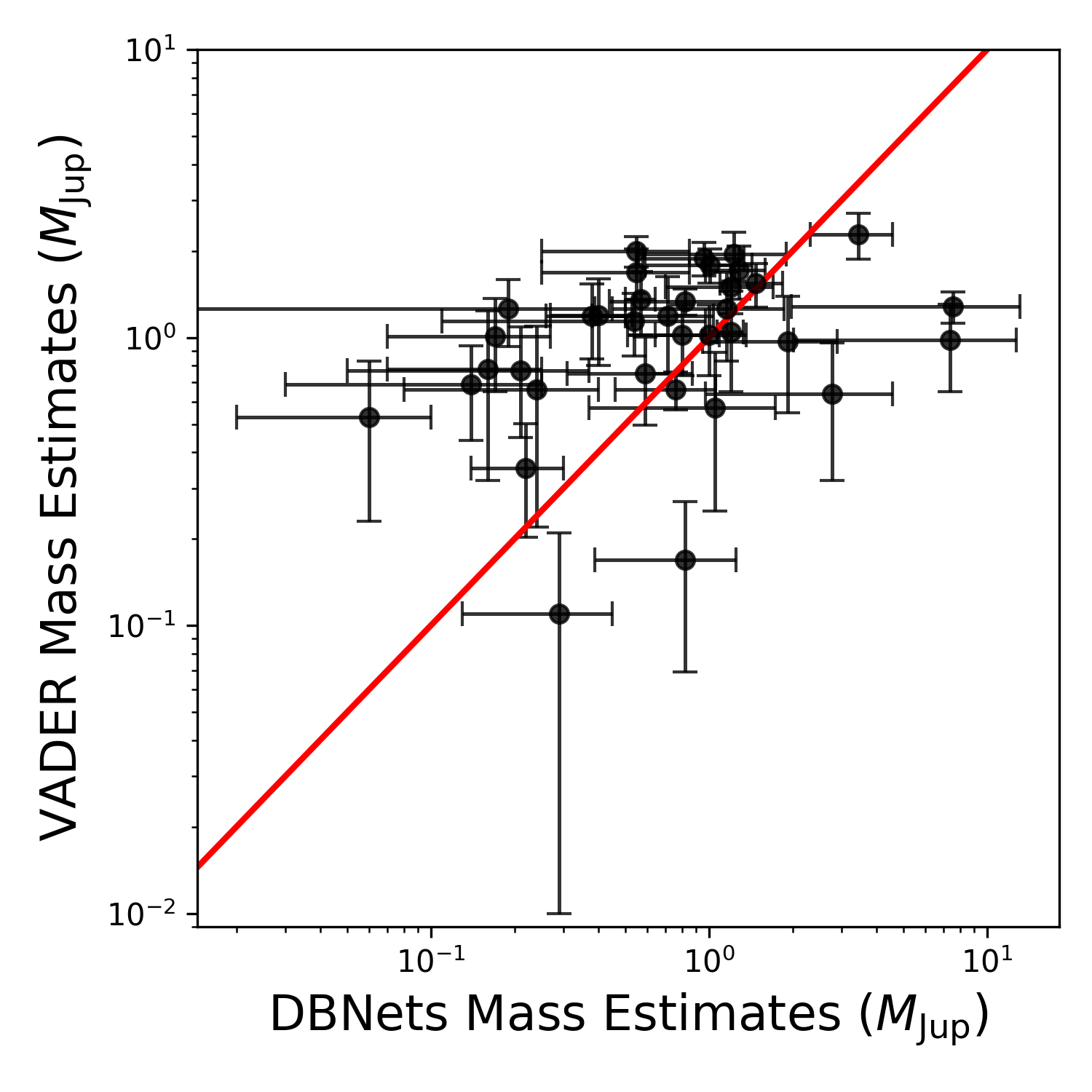}{0.43\textwidth}{}
}
\caption{(Left) Normalized distributions of predicted planet masses from 23 ALMA disk images. The results from VADER are shown alongside those from DBNets for comparison. This highlights both the overlap and slight differences in mass predictions between the two models. While largely similar, VADER, on average, predicts slightly higher masses than DBNets. (Right) Scatter plot comparing VADER mass predictions with DBNets. The red line marks the points in the plot where VADER and DBNets agree 1:1.}
\label{fig:vae_vs_dbnets}
\end{figure*}

\begin{figure}
    \centering
    \includegraphics[width=0.5\textwidth]{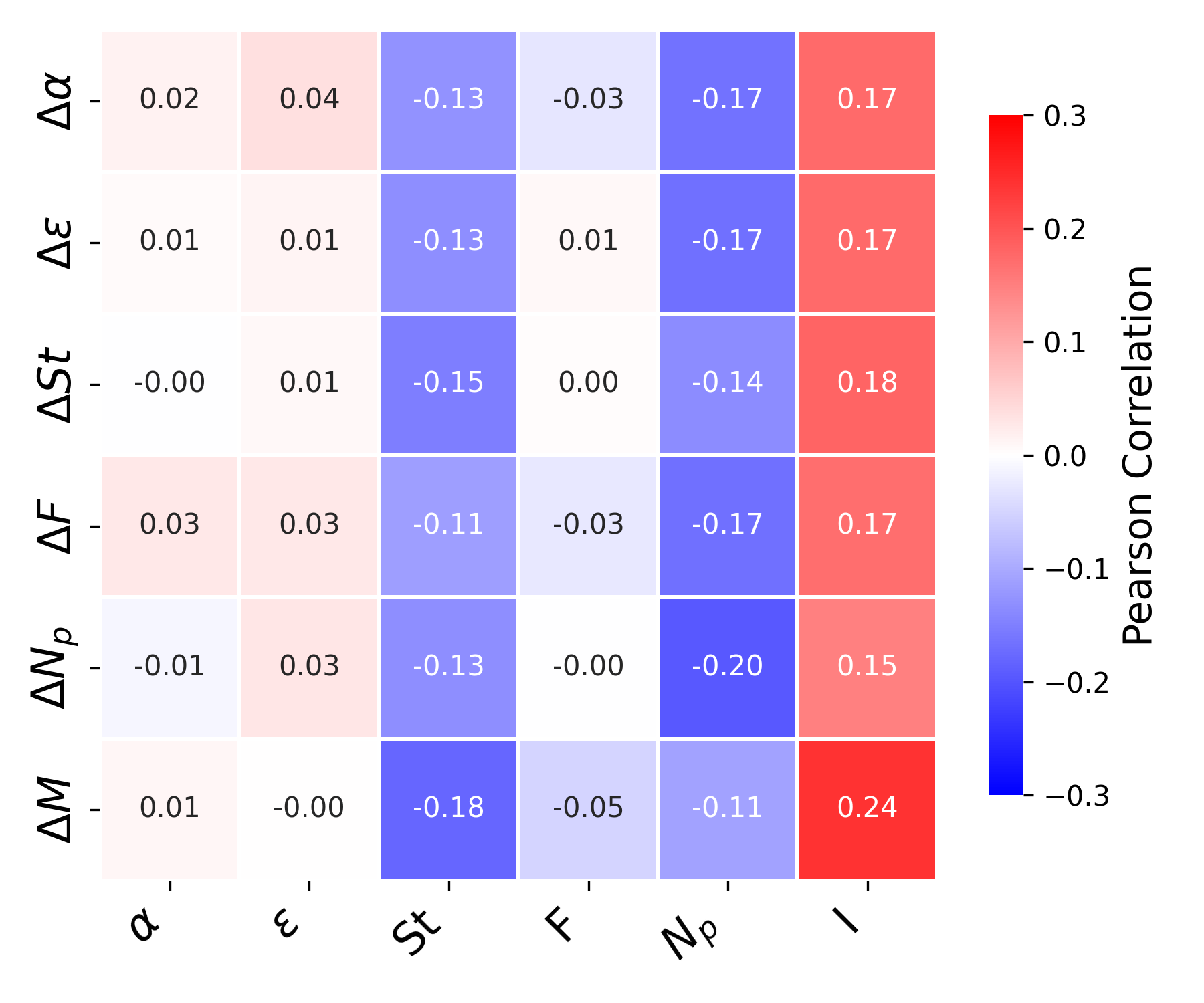}  
    \caption{Correlation heatmap showing the relationship between true disk parameters (x-axis) and the prediction errors (y-axis) for corresponding inferred quantities. Each row represents the difference between a predicted parameter and its true value (e.g., $\mathrm{\Delta} \alpha$ is the error in $\alpha$), while each column corresponds to a physical property of the disk, including $\alpha$, $\epsilon$, St, $F$, $N_p$, and inclination $I$. The values inside the cells indicate the Pearson correlation coefficients, with darker shades representing stronger correlations. A higher absolute correlation means that extreme values of the associated disk parameter tend to increase the model's prediction error for the corresponding quantity.}
    \label{fig:combined_plots}
\end{figure}



Furthermore, in Figure~\ref{fig:vae_vs_dbnets}, we compare the normalized distribution of the predicted planet masses from the 23 ALMA disks using VADER and DBNets.
Both models favor the presence of sub-Jovian planets, but exhibit slightly different mass distributions. DBNets yields a narrow peak around 0.1--0.3\,$M_\mathrm{Jup}$, while VADER shows a broader tail extending beyond 2\,$M_\mathrm{Jup}$ with a mode near 1\,$M_\mathrm{Jup}$. In fact, the secondary peak in planet mass distribution as predicted by DBNets directly overlaps with the peak of mass distribution as predicted by VADER. As shown in the right panel of Figure~\ref{fig:vae_vs_dbnets}, the mass predictions of both models overlap significantly within their respective uncertainty intervals, particularly in the mass range of approximately $0.5\text{--}1.5\, M_\mathrm{Jup}$, which captures the core population consistently predicted across both frameworks.

The differences in uncertainty quantification between VADER and DBNets stem from the differing architectures and distinct approaches to modeling and constraining predictive uncertainties.
DBNets relies on an ensemble of deterministic CNN-based networks to capture uncertainty through variations in predictions across multiple realizations. By contrast, VADER employs a fully generative VAE framework that accounts for both model and data-driven uncertainties. This enables VADER to capture degeneracies arising from shallow or ambiguous gap structures. A per-target comparison of mass estimates and uncertainties is provided in Table~\ref{tab:gap_masses}. The values in the table further illustrate the consistency and complementarity of the two approaches, with 19 out of 35 planets (approximately 54\%) showing overlapping mass estimates within their respective $1\sigma$ uncertainties.

\subsection{Limitations}


Although VADER is effective at inferring planet masses from ALMA disk images, several limitations remain.
These are captured in Figure~\ref{fig:combined_plots}, which shows a correlation heatmap between model prediction error and disk parameters. For instance, the error in mass prediction increases modestly with disk inclination, with a Pearson correlation coefficient of 0.24 between $\Delta M$ and inclination $I$.  Similar trends appear for other predicted parameters, all of which show increased errors with higher inclination.  This is due to projection effects. The 3D model disks are constructed from 2D hydrodynamic outputs under the assumptions of vertical hydrostatic equilibrium (Eq. \ref{hydro_equi}) and perfect gas-dust mixing.  The settling towards the midplane of the larger, higher-Stokes-number particles is not included.  In the more inclined disk images, the particles' vertical extent enables the rings to obscure the gaps, reducing the model's ability to accurately constrain planet masses and disk properties.  This is a limitation of the training set, rather than of the autoencoder.

A similar trend is seen with the Stokes number. The negative Pearson correlation coefficient of –0.18 between $\Delta M$ and the Stokes number indicates increased mass prediction error in low-$St$ regimes. Since in the low Stokes number regimes ($\mathrm{St} < 0.001$), dust remains tightly coupled to the gas, it suppresses the formation of sharp rings.  This leads to more diffuse and less distinguishable disk structures, hindering the model's ability to extract precise constraints.

Furthermore, systems that exhibit single gaps—often indicative of a single planet—provide fewer morphological constraints. This is once again captured vividly in Figure~\ref{fig:combined_plots}, where all the predicted parameters have weak negative correlation with the number of planets.  As a result, global parameter inference in these cases tends to be more degenerate, resulting in increased prediction error.

\subsection{Expanding the Mass–Semimajor-Axis Landscape}

\begin{figure}
    \centering
    \includegraphics[width=0.5\textwidth]{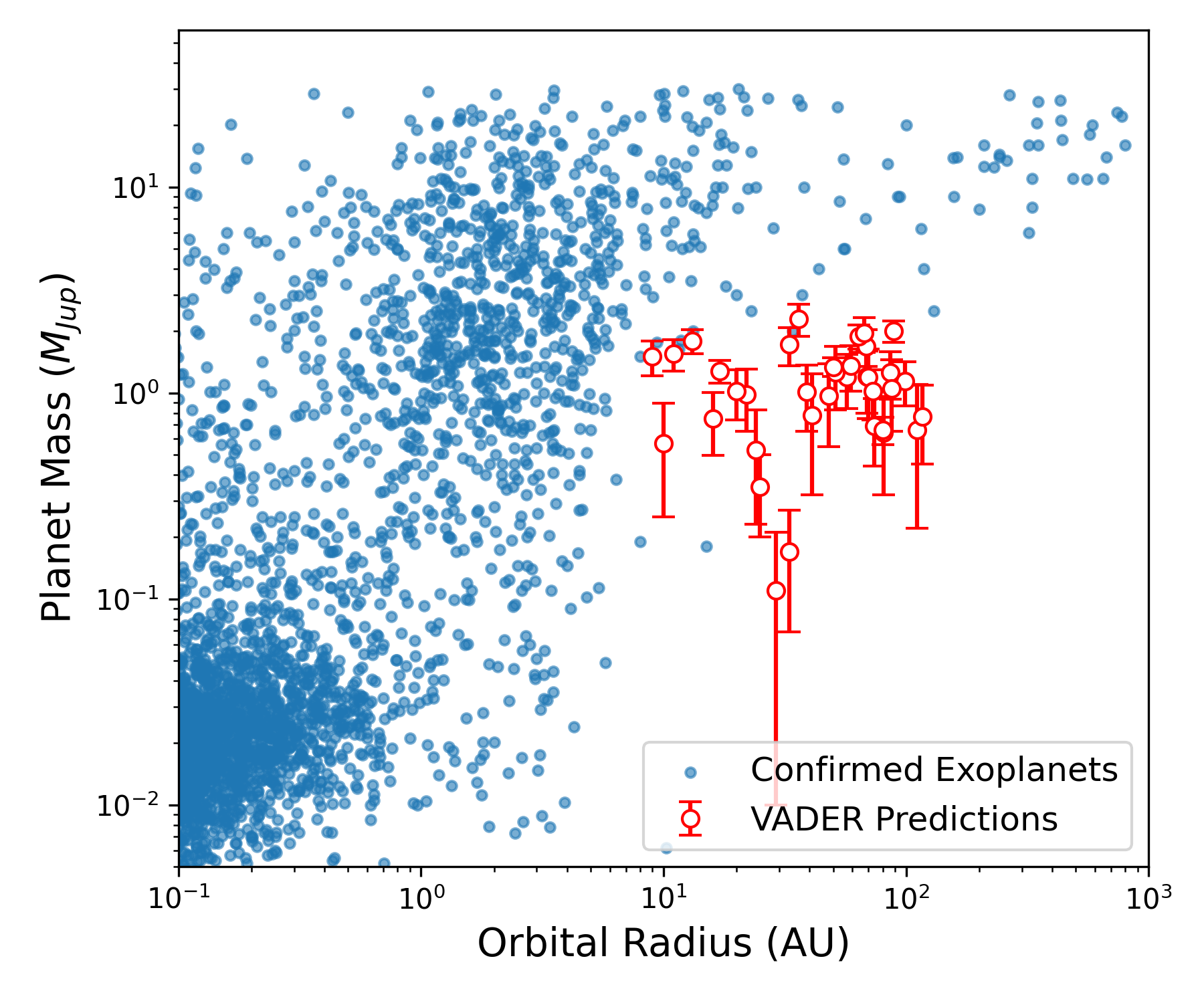}
    \caption{Comparison between confirmed exoplanets (blue) from the NASA Exoplanet Archive \citep{akeson2013nasa} and planet mass predictions (red, with error bars) from VADER. Both axes are shown in logarithmic scale. The VAE predictions are well-positioned in the parameter space occupied by observed gas giants, indicating physical plausibility.}
    \label{fig:vae_nasa_comparison}
\end{figure}

Our model expands upon the known distribution of exoplanets in mass–semi-major axis space. Figure~\ref{fig:vae_nasa_comparison} illustrates that the young planets inferred with VADER and listed in table~\ref{tab:gap_masses} constitute a population of sub-Jovian to Jovian-mass planets at wide separations (10-100~au) that are currently underrepresented in transit and radial velocity surveys of mature planetary systems due to selection biases. This suggests that ALMA-based parameter inference based on disk morphology can complement traditional planet detection methods by probing an otherwise inaccessible regime of planet formation.


These results also align with the goals of upcoming microlensing surveys, such as the Roman Space Telescope, which is expected to detect a substantial population of wide-separation planets \citep{penny2019predictions}. The synergy between morphology based parameter inference from ALMA and direct detection from Roman may open a new window into understanding the demographics of planets massive enough to clear gaps but beyond the detection limits of transit and radial velocity methods.

 While it remains premature to draw definitive conclusions about the dominant mechanism at play, our results suggest that ALMA-based imaging paired with machine learning techniques may guide observational studies and help uncover systems where traditional detection methods fall short, thereby contributing new empirical constraints to theoretical models of planet formation.

\section{Conclusions}\label{sec:Conclusions}
We have introduced VADER, a VAE-based framework for joint inference of planetary and disk parameters from ALMA dust continuum observations of protoplanetary systems.  The results show VAEs' promise for advancing the characterization of planet-forming environments.

When applied to ALMA 1.3-mm dust continuum images of 23~protoplanetary disks, VADER recovered planet masses and disk parameters broadly consistent with literature values.  The inferred planets lie in a relatively uncharted region of parameter space --- young, Jovian-mass planets orbiting at tens of AU --- highlighting that disk observations probe aspects of planetary system architecture inaccessible to transit and radial velocity surveys.

A key advantage of VADER is its interpretable latent space.  The algorithm learns a compact representation of disk structure in which the latent dimensions correlate with familiar disk and planet properties.  Controlled perturbations of the latent vectors yield smooth, meaningful changes in the morphology of the synthetic disk images generated by the decoder.  Training the algorithm thus lends insight into the astrophysical processes underlying the image features.  Model performance degrades somewhat in regimes including highly-inclined disks, weak dust density contrasts, and disks with a single gap.  The tests we have performed show the method's uncertainty estimates generally increase by about as much as needed to reflect the degradation.

Any machine learning method is only as good as the training data. For this initial exploration, we assumed a star of solar mass. We plan to extend our models to span a range of masses and explore the relationship between stellar mass and disk properties. Our training set furthermore was developed from forward models that include a subset of the effects likely to be important.  Not treated for example are dust grains' coagulation and fragmentation and the sublimation and freezeout of water or other ices.  Further effort will be needed to determine whether these significantly affect the distribution of the grains responsible for the millimeter-wave thermal emission. Other extensions will allow for the fact that ring-like substructures in protoplanetary disks are not uniquely diagnostic of embedded planets but may come also from disks' zonal flows \citep{flock2015gaps}, dust sintering fronts \citep{okuzumi2016sintering}, and magnetohydrodynamic instabilities \citep{suriano2019formation}.  Since our training set comprised exclusively planet-induced features, the model may over-interpret some structures as planetary in origin.  Expanding training sets to include the proposed mechanisms for forming gaps without planets will be necessary to ensure robustness.

VADER can also be augmented to account for the noise in ALMA and other observational data. Future study will involve injecting noise into the training data and adapting the VAE framework to capture disks' morphological features in the latent space despite the added noise. At the same time, we plan to compare planet mass and disk parameter measurements against those from other inference models such as DBNets2.0.

Beyond its current application, the VAE architecture introduced here is inherently well-suited for multi-modal astronomy. While this study focused on dust continuum maps, the framework can be naturally extended to integrate multiple observational modalities, including CO and other millimeter-wave molecular line channel maps, infrared thermal emission and spectral diagnostics, and scattered starlight.  Each of these provides complementary constraints: thermal emission traces dust morphology; millimeter lines encode gas kinematics; infrared line emission informs temperature, chemical composition, and snow lines; and scattered light reveals the distribution of dust stirred by turbulence.  Merging these diverse inputs into a shared latent space has the potential to break degeneracies, enable recovering further parameters, and allow for a more holistic understanding of disk environments.

In summary, this work has demonstrated the potential of generative machine learning and particularly variational autoencoders as powerful tools for recovering both planetary and disk parameters from astronomical images.  VAEs bring together morphological reconstruction, latent space interpretability, uncertainty quantification, and capacity for multi-modal data fusion, offering a flexible and extensible framework for data-driven exploration of planet formation.  With future enhancements that incorporate a broader range of physical processes and heterogeneous input channels, VADER can serve as a foundational tool for next-generation parameter inference across the protoplanetary disk population.

\begin{acknowledgments}
This work was carried out in part at the Jet Propulsion Laboratory, California Institute of Technology, under contract with NASA and supported by the Exoplanets Research Program through grant 21-XRP21-0088. Support for SSH was provided by a student wage grant awarded by the Research Council at Colgate University. This paper makes use of the following ALMA data: ADS/JAO.ALMA2011.0.000015.SV ALMA is a partnership of ESO (representing its member states), NSF (USA) and NINS (Japan), together with NRC (Canada), NSTC and ASIAA (Taiwan), and KASI (Republic of Korea), in cooperation with the Republic of Chile. The Joint ALMA Observatory is operated by ESO, AUI/NRAO and NAOJ.
\end{acknowledgments}


\vspace*{5mm}
\facilities{ALMA}

\noindent \textbf{Github code:} \href{https://github.com/sauddy/VADER}{\texttt{https://github.com/sauddy/VADER}}

A frozen, citable release for the data is archived at \texttt{Zenodo} \citep{auddy_sayantan_2022_7332423}

\software{pytorch \citep{paszke2019pytorch}, astropy \citep{price2022astropy, price2018astropy, 2013A&A...558A..33A}}



\appendix

\section{Architecture of the VADER VAE}\label{VADER_architecture}


The VAE consists of two main components: an encoder and a decoder, together designed to learn compact latent representations of protoplanetary disk images with three (RGB) color channels. The conversion from ALMA images to RGB is done using the standard RADMC3D heatmap color scale. Although the VAE contains three channels, the channels contain the same information for dust continuum maps. This, however, will prove to be very useful once we incorporate other maps such as gas velocity maps as it will convey a more robust and rich information about the protoplanetary disk in the latent space. 

\subsection{Encoder}

The encoder comprises three convolutional layers \citep{o2015introduction}, applied sequentially to extract hierarchical spatial features from input images. We chose this three-layer configuration empirically after two-layer networks failed to capture sufficient morphological complexity, while deeper models showed diminishing returns in performance and slower convergence.

Each convolutional layer detects structures such as edges, rings, and gaps by applying small, trainable filters. The first layer uses 64 filters of size 4$\times$4, with stride 2 and padding 1, followed by a ReLU activation \citep{agarap2018deep}. This stride reduces the spatial resolution by half, and padding ensures boundary features are preserved. The second and third layers increase the filter count to 128 and 256, respectively, maintaining the same filter size, stride, and padding.

As the image passes through deeper layers, its spatial resolution decreases while its feature depth increases, thanks to the progressively increasing filter count, enabling the network to learn increasingly abstract representations of disk morphology. After the final convolution, the resulting feature map is flattened into a vector of size \(256 \times 32 \times 32\), where $256$ is the filter count and $32 \times 32$ is the downsampled image dimension. This yields a high-dimensional embedding of the input image. This vector is then linearly projected onto a final layer of vector size 64 corresponding to the \textit{mean} and \textit{log-variance} of a 32-dimensional Gaussian distribution in the latent space.  

\subsection{Decoder}

The decoder mirrors the encoder in reverse but does not share weights with it; instead, it has its own set of independently learnable parameters. The 32-dimensional latent vector is first mapped to the original flattened size via a fully connected layer, reshaped, and then passed through three deconvolutional layers. These layers progressively upsample the feature maps to reconstruct the image. Channel dimensions are reduced from 256 \(\rightarrow\) 128 \(\rightarrow\) 64 \(\rightarrow\) 3, as the image size is upscaled from 32$\times$32 \(\rightarrow\) 64$\times$64 \(\rightarrow\) 128$\times$128 \(\rightarrow\) 256$\times$256. ReLU activations are applied in the first two layers and a sigmoid activation \citep{sharma2017activation} in the final layer to constrain pixel values to the (0, 1) range. The encoder and decoder working as a pair ensures the output matches the original image size while preserving learned morphological structures.

\subsection{Latent-Space Prediction Networks}

Two feedforward neural networks (FNNs) predict physical parameters from the 32-dimensional latent vector:

FNN-Planet maps the latent vector to the masses of up to three planets using seven fully connected hidden layers, each followed by batch normalization and a ReLU activation. Batch normalization stabilizes training by normalizing the layer outputs with a mean of 0 and unit standard deviation. 

FNN-Disk has a similar architecture (the only difference is having eight layers instead of seven) but outputs five disk properties: \(\alpha\)-viscosity, Stokes number, dust-to-gas ratio, flaring index, and the number of planets.

The depth and width of these networks were selected empirically to balance expressiveness and generalization. Shallower or narrower networks underperformed, while larger networks showed no significant gains and risked overfitting.

\subsection{Computational Breakdown of the Encoder and Decoder}
\label{appendix:computation}

To better understand the scale of computation in our architecture, we analyze the first encoder layer as a representative example. The first layer receives RGB images of shape \([3, 256, 256]\) and applies the following two-dimensional convolution operation:
\[
\texttt{nn.Conv2d(in\_channels=3, out\_channels=64, kernel\_size=4, stride=2, padding=1)}
\]
\noindent
Each of the 64 filters spans all three channels each with \(4 \times 4\) weights, giving \(3 \times 4 \times 4 = 48\) parameters per filter, plus one bias term, for a total of \(49\). With 64 filters, this layer has:
\[
64 \times 49 = 3{,}136 \text{ trainable parameters.}
\]
\noindent
The stride of 2 reduces the spatial dimensions by half, while padding ensures boundary preservation. The output shape becomes:
\[
\left\lfloor \frac{256 - 4 + 2(1)}{2} \right\rfloor + 1 = 128,
\quad \text{giving } [64, 128, 128]
\]
\noindent
This breakdown illustrates the architectural flow for a single convolution layer. For a visual illustration of how these operations apply to protoplanetary disk images, see Figure~1 of \citet{auddy2021dpnnet}.

\section{VAE ALGORITHM} \label{appendix:vae}
In this section, we give the details of the VAE algorithm and the implementation of the training loss:

\subsection{Probabilistic Encoding and the Latent Space}

Given an input image $x$, the encoder learns to produce the parameters of a multivariate Gaussian distribution: a mean vector $\mu(x)$ and a standard deviation vector $\sigma(x)$. Instead of mapping $x$ to a single latent point, the VAE maps it to a distribution:
\begin{equation}
z \sim \mathcal{N}(\mu(x), \sigma^2(x))
\end{equation}
where $z$ is the latent vector sampled from a normal distribution represented by $\mathcal{N}$.

This formulation enables the model to capture uncertainty and variability in the observed disk images. In our case, this means that disks with similar morphologies (e.g., rings or gaps) are clustered together in latent space, but small variations—like planet mass or disk inclination—still allow for distinguishable representations where image quality allows. 

\subsection{Reparameterization Trick}

Sampling from a distribution is a non-differentiable operation, which would typically prevent backpropagation during training. To overcome this, VAEs use a technique known as the reparameterization trick. Rather than sampling $z$ directly, we write:
\begin{equation}
    z = \mu(x) + \sigma(x) \odot \epsilon, \quad \epsilon \sim \mathcal{N}(0, I)
\end{equation}
In this formulation, \( \epsilon \sim \mathcal{N}(0, \mathbf{I}) \) represents a random noise vector drawn from a standard multivariate normal distribution with zero mean and identity covariance. The symbol \( \mathcal{N}(0, \mathbf{I}) \) denotes the normal distribution, where each dimension of \( \epsilon \) is sampled independently from \( \mathcal{N}(0, 1) \). This noise is then scaled by the encoder-predicted standard deviation \( \sigma(x) \) and shifted by the mean \( \mu(x) \), with \( \odot \) indicating element-wise multiplication. This makes the sampling process differentiable and gradients can now flow from encoder to decoder through $z$ during training.

\subsection{Loss Function: Balancing Fidelity and Structure}

\begin{figure*}
    \centering
    \includegraphics[width=0.98\textwidth]{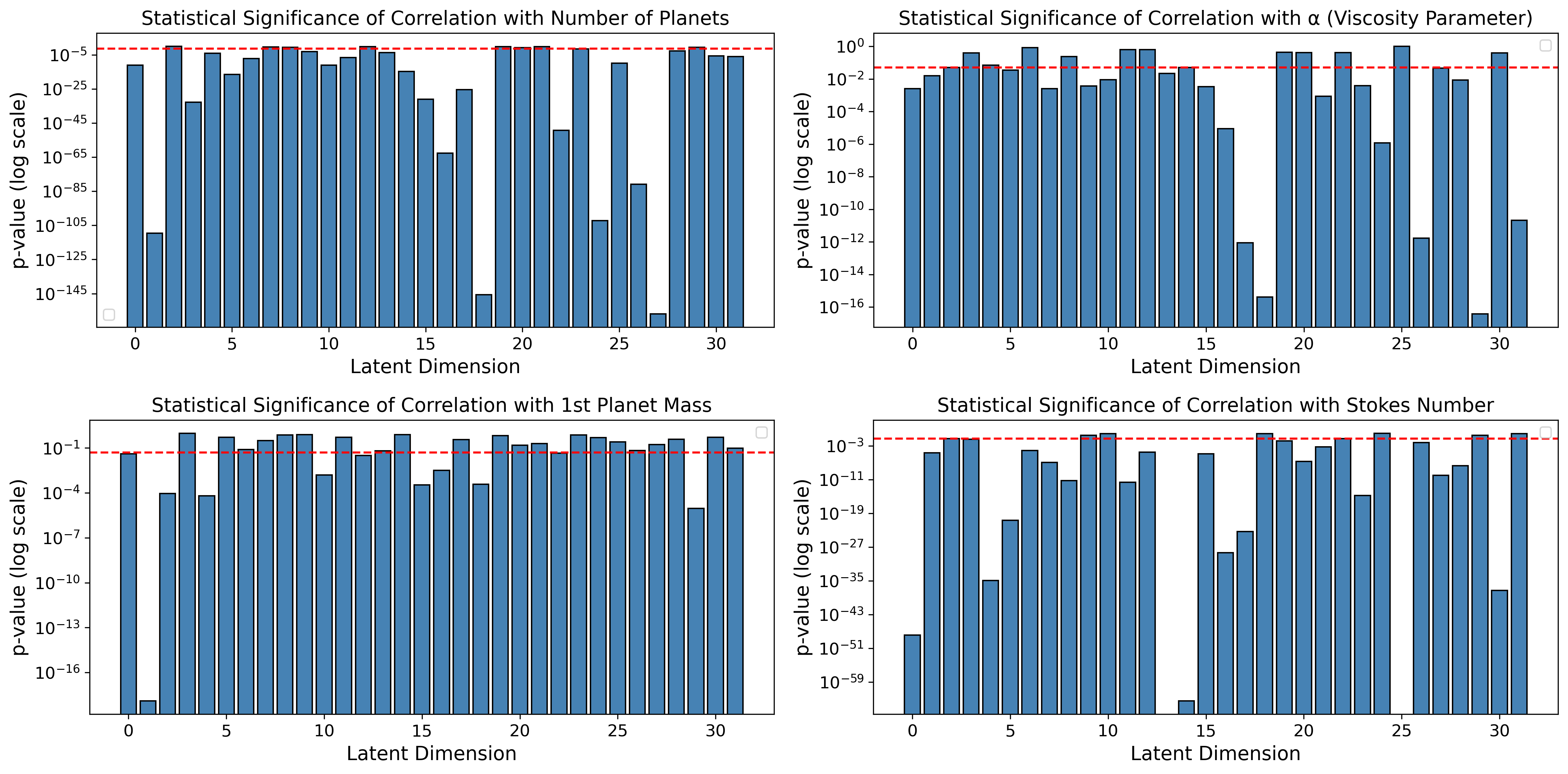}
    \caption{
    Statistical significance (p-values) of Pearson correlations between each latent dimension of the VAE and key physical parameters of the disk: number of planets, viscosity parameter ($\alpha$), first planet mass, and St. Each subplot shows the p-value (in log scale) for the correlation between one of these physical quantities and the 32-dimensional latent space. 
    A lower p-value (especially $p < 0.05$, marked by the red dashed line) indicates statistically significant evidence of correlation. These results confirm that specific latent dimensions capture morphological features that correlate with key astrophysical properties, such as the number of formed planets or turbulence (via $\alpha$).
    }
    \label{fig:vae_correlation_significance}
\end{figure*}

Training a VAE involves minimizing a loss function composed of two terms:

\begin{enumerate}
    \item Reconstruction Loss ($\mathcal{L}_{\text{recon}}$): Measures how closely the decoder can recreate the input image from the latent vector. For image data, this is often the mean squared error (MSE) or binary cross-entropy:
    \begin{equation}
         \mathcal{L}_{\text{recon}} = \| x - \hat{x} \|^2
    \end{equation}

    where, \( x \) represents the original input image, and \( \hat{x} \) is the reconstructed image produced by the decoder.

    \item KL Divergence ($D_{\text{KL}}(q(z|x) \, \| \, p(z))$): This term encourages the learned latent distributions to remain close to a standard normal distribution, while ensuring that the latent space is continuous and enabling smooth interpolation between samples:
    \begin{equation}
        D_{\text{KL}}(q(z|x) \, \| \, p(z)) = \frac{1}{2} \sum_i \left( \mu_i^2 + \sigma_i^2 - \log \sigma_i^2 - 1 \right)    
    \end{equation}
    
    Each component of this expression contributes to the regularization of the latent space in a meaningful way. The term $\mu_i^2$ penalizes the deviation of the learned latent means from zero, encouraging the encoder to center the latent distribution around the origin. The term $\sigma_i^2$ discourages excessively large variances, thereby keeping the latent space compact and preventing overdispersion. On the other hand, the term $-\log \sigma_i^2$ penalizes extremely small variances, ensuring that the latent distribution does not collapse into a delta function and that samples remain diverse. Finally, the constant $-1$ normalizes the expression so that the KL divergence is exactly zero when $\mu = 0$ and $\sigma^2 = 1$ (i.e., when the learned posterior exactly matches the standard normal prior).

\end{enumerate}

The total loss is the combination of both terms:
\begin{equation}
\mathcal{L}_{\text{VAE}} = \mathcal{L}_{\text{recon}} + \beta \, D_{\text{KL}}
\end{equation}
where $\beta$ is a weighting factor (typically 1) that can be tuned to control the balance between reconstruction fidelity and latent space regularization. In our implementation, we set \( \beta = 1 \).
For further reading, see the original work by \citet{kingma2014autoencoding}, as well as the tutorial by \citet{doersch2016tutorial}, which provides an accessible introduction to VAEs for scientific applications.









\section{Visualizing the Latent Space}\label{Apxc}

\begin{figure*}[ht!]
    \centering
    \includegraphics[scale=0.90]{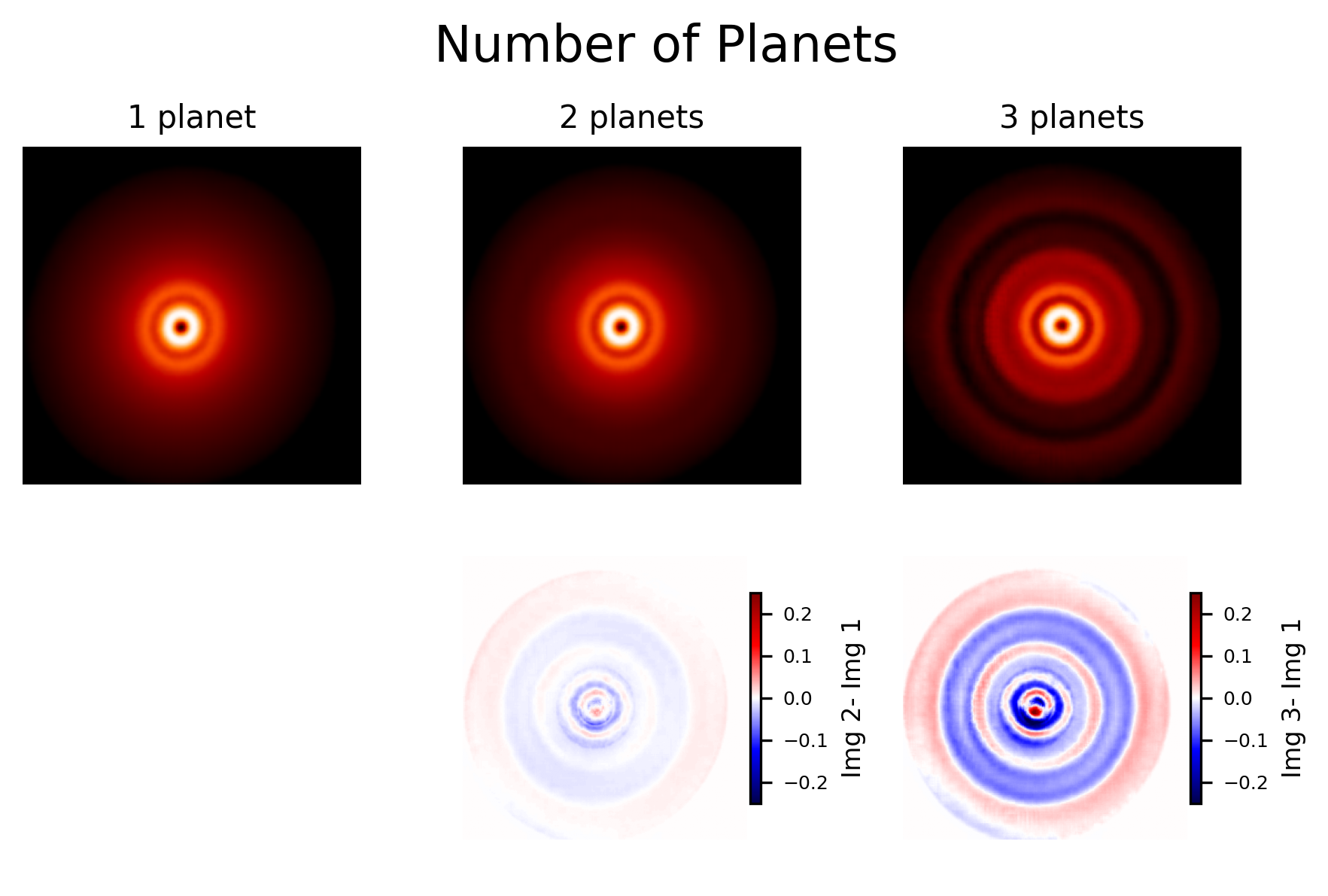}
    \caption{
        Morphological variation in protoplanetary disks as a function of the number of embedded planets. 
        The top row shows simulated disk images generated from a base latent representation, while progressively adjusting the few most important latent dimensions (latent dimension 1, 18, 27) correlated with the number of planets (1, 2, and 3) (Figure \ref{fig:vae_correlation_significance}a). The bottom row visualizes the corresponding difference maps (computed by subtracting the current image with the first image), revealing increasingly complex ring and gap structures with more planets. 
        All other parameters in the latent space are held constant, isolating the influence of the number of planets feature.
    }
    \label{fig:number_of_planets_latent}
\end{figure*}

\begin{figure}[ht!]
    \centering
    \includegraphics[width=\textwidth]{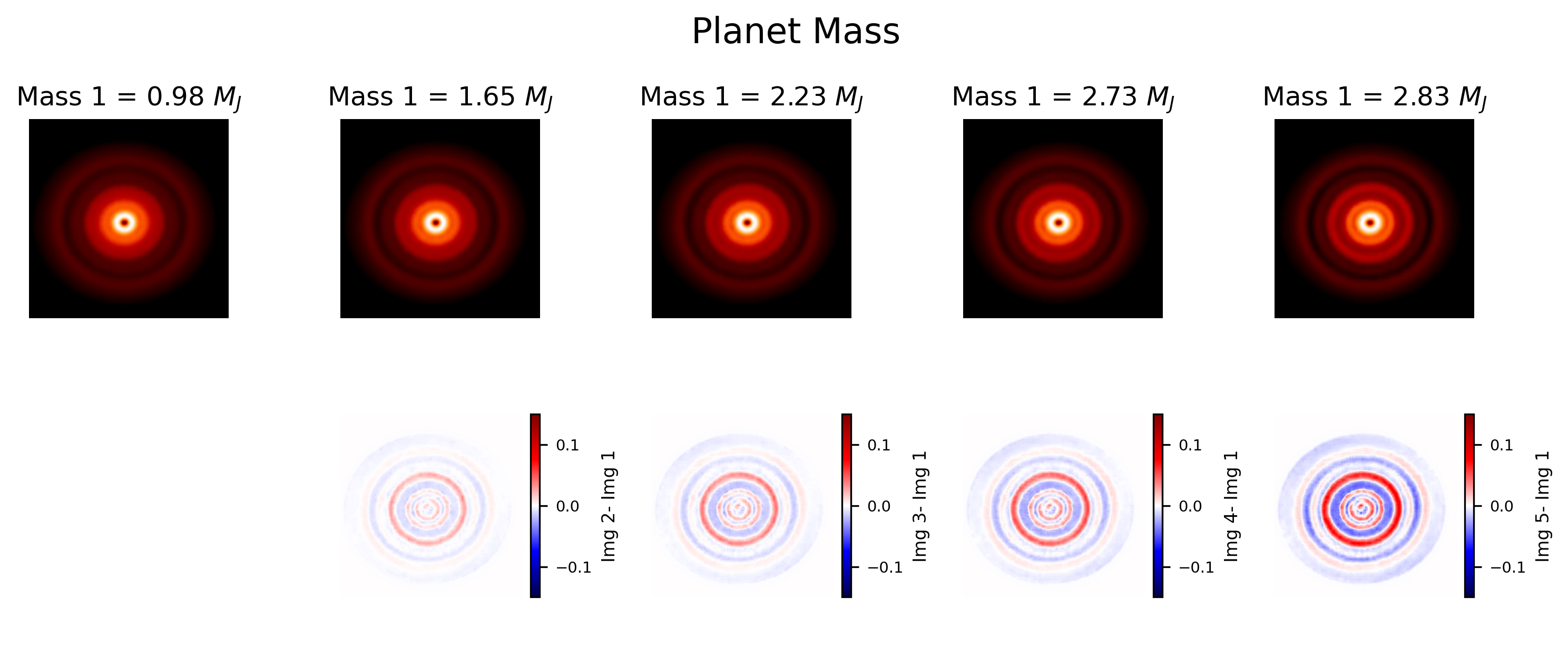}
    \caption{
        Morphological changes in protoplanetary disks as the mass of the innermost planet is varied. 
        The images are generated by manipulating the latent dimensions (latent dimension 1) most correlated with planet mass (Figure \ref{fig:vae_correlation_significance}c, while holding all other parameters fixed. 
        The top row shows synthesized disk images with increasing mass of the first planet, from $0.98\,M_J$ to $2.83\,M_J$. 
        The bottom row displays the residual maps (difference from the base case), illustrating the increasingly prominent gaps and ring structures induced by higher planetary mass.
    }
    \label{fig:planet_mass_latent}
\end{figure}

\begin{figure}
    \centering
    \includegraphics[scale=0.90]{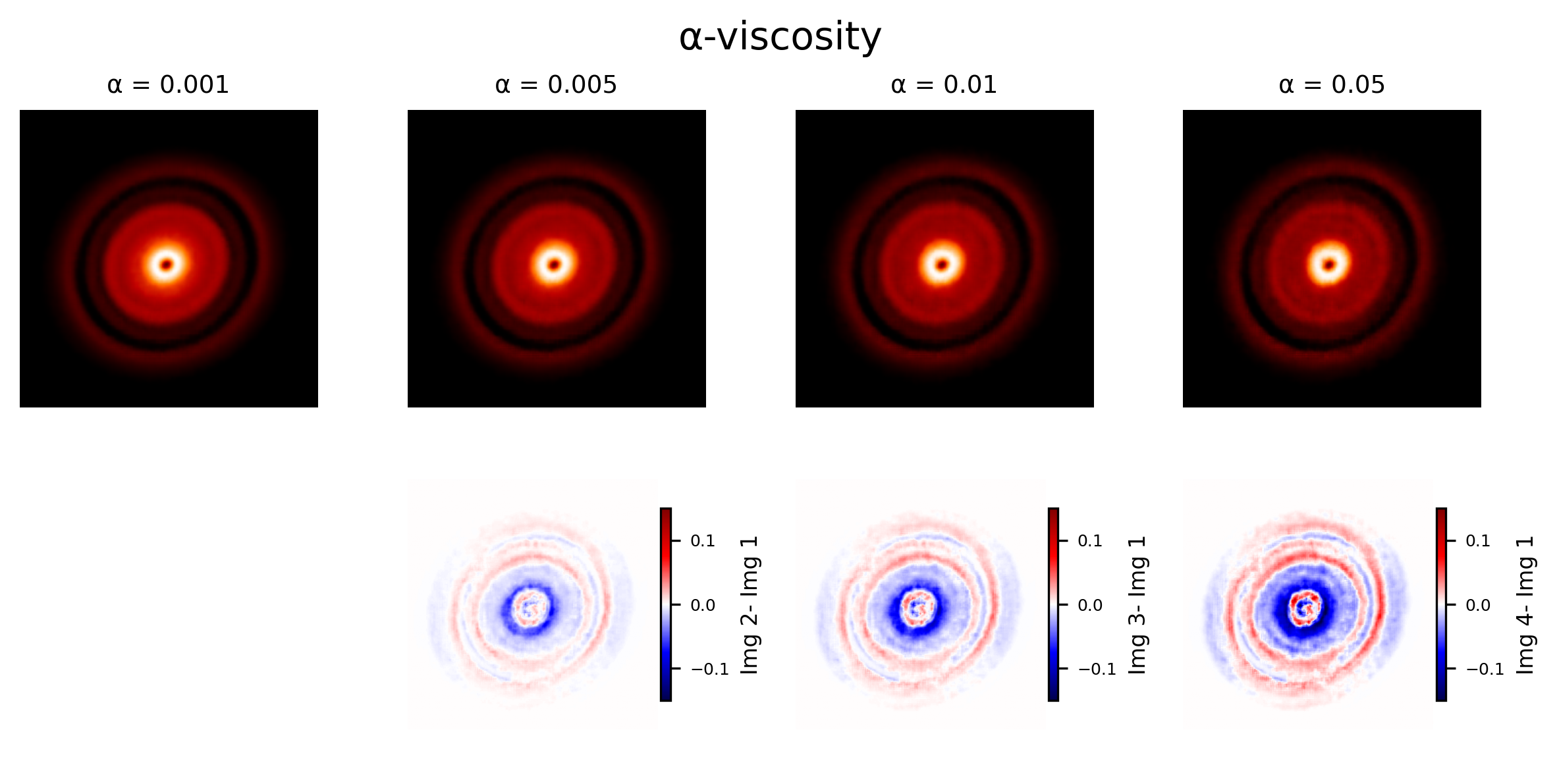}
    \caption{
        Impact of turbulent viscosity parameter $\alpha$ on disk morphology, as visualized through latent space traversal of latent dimension 29.
        The top row shows disk images generated by varying the most relevant latent dimension associated with $\alpha$-viscosity, while keeping all other latent variables fixed. 
        As $\alpha$ increases from 0.001 to 0.05, the resulting morphologies exhibit increasingly diffused gaps and smoother disk structures. 
        The bottom row shows residual maps compared to the lowest $\alpha$ case ($\alpha = 0.001$), highlighting the extent of smoothing and diffusion due to turbulence.
    }
    \label{fig:alpha_viscosity_latent}
\end{figure}

\begin{figure}
    \centering
    \includegraphics[width=\textwidth]{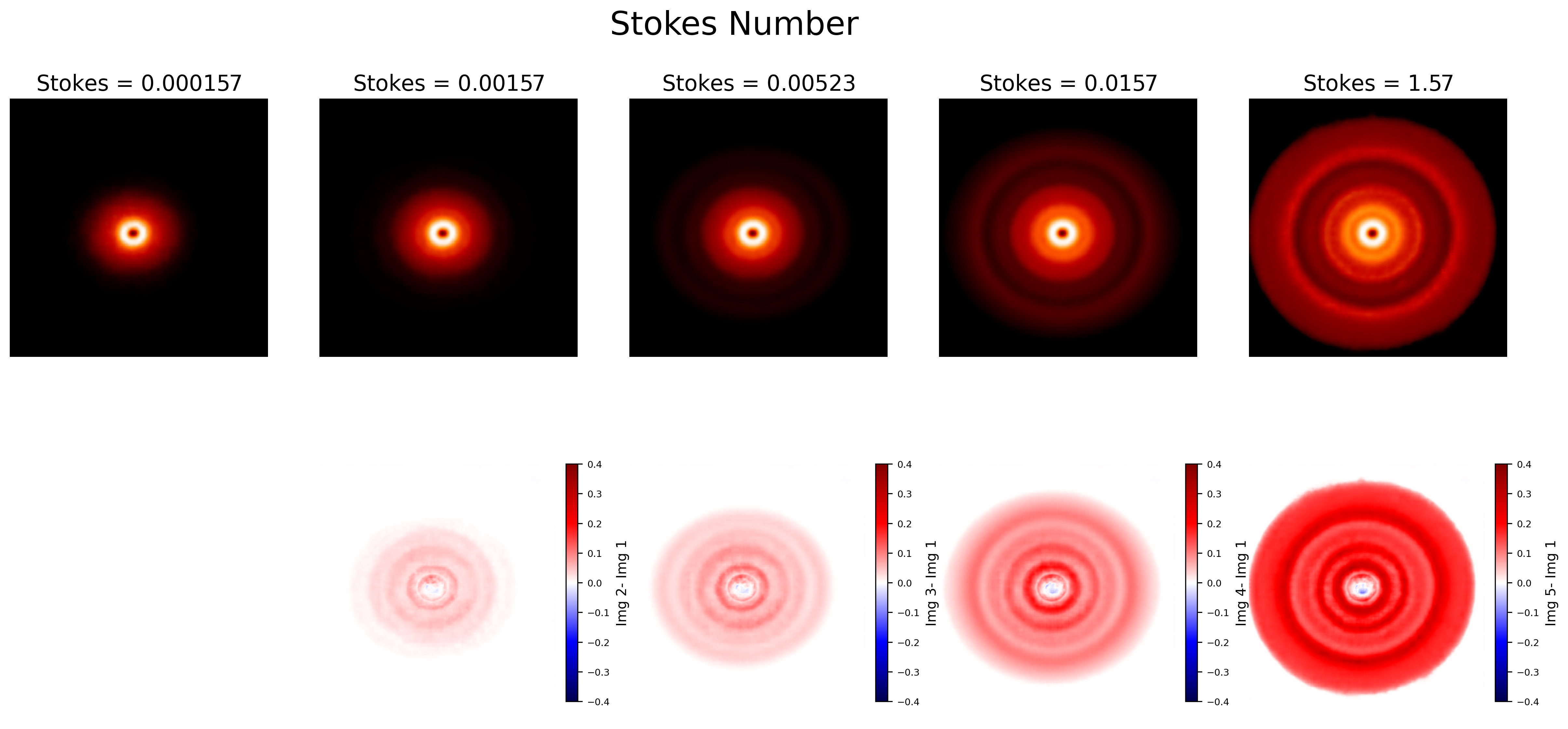}
    \caption{
        Effect of varying St on disk morphology, explored by varying latent dimensions 13 and 25. 
        The top row displays synthesized disk images where the latent vector dimension most correlated with St is varied from 0.000157 to 1.57, while keeping all other latent values fixed (Figure \ref{fig:vae_correlation_significance}d). As the Stokes number increases, dust trapping becomes more pronounced and the ring structures become sharper and more extended. 
        The bottom row shows residual maps relative to the lowest-Stokes case, revealing how dust concentration and ring prominence evolve with particle size coupling.
    }
    \label{fig:stokes_latent}
\end{figure}

One of the most compelling strengths of VAEs is their capacity to learn a low-dimensional, latent representation that is not only compact but also provides insight. Unlike CNNs, which often act as black-box, VAEs can disentangle the underlying latent space. Thus offering a mapping between complex image data and physical theory. This mapping between the disk morphology and the latent space is key to interpreting the disk physics.

To assess the degree of interpretability in the VAE's latent space, we perform a quantitative correlation analysis between each of the 32 latent dimensions and key physical parameters. Of the key physical parameters, we analyze in detail the number of embedded planets, the turbulent viscosity parameter, the mass of the innermost planet, and the Stokes number. As shown in Figure~\ref{fig:vae_correlation_significance}, we compute the statistical significance (p-values, plotted in log scale) of Pearson correlations between each latent variable and these physical quantities on the test dataset. Latent dimensions with low p-values—particularly those below $p < 0.05$, indicated by the red dashed line—demonstrate statistically robust correlations. The results reveal that distinct latent dimensions strongly encode specific image features governed by physical characteristics. Thus validating that the VAE has learned the latent space in a way that reflects astrophysical processes.

To qualitatively probe these correlations, we perform controlled latent space traversals using the most strongly correlated latent dimensions for each physical property. For each parameter, we fix all other latent dimensional values and systematically vary only the relevant dimension(s), generating synthetic disk images that isolate the morphological impact of a single physical parameter.

In Figure~\ref{fig:number_of_planets_latent}, we vary the latent dimensions most correlated with the number of embedded planets, namely latent dimensions 1, 18, and 27 (see Figure~\ref {fig:vae_correlation_significance}). The decoded disk images reveal a progressive increase in the number of concentric ring gaps, consistent with the expectation from our simulations that each planet carves out one distinct gap. The corresponding difference maps highlight how additional planets generate new pressure maxima and substructures in the disk.

In Figure~\ref{fig:planet_mass_latent}, we explore the effect of increasing the mass of the innermost planet by varying the most correlated vector: latent vector 1. As the planet mass increases from $0.98\,M_J$ to $2.83\,M_J$, the inner gap becomes significantly wider and deeper. The difference maps visually affirm that even modest changes in mass lead to nontrivial enhancements in ring contrast and radial clearing, capturing the nonlinear gravitational influence of a more massive planet.

The effect of $\alpha$-viscosity is shown in Figure~\ref{fig:alpha_viscosity_latent}. As the $\alpha$-parameter increases from 0.001 to 0.05, by varying the most influential latent dimension (dimension 29), the rate of change in brightness as we move away radially from the disk center decreases. As a result, the disk substructure becomes progressively more diffused. This aligns well with theoretical expectations: higher turbulence leads to enhanced dust diffusion, which in turn smears out otherwise sharp features. The residual maps further confirm that this smoothing primarily affects the contrast and sharpness of existing rings, particularly in the outer disk regions.

Finally, in Figure~\ref{fig:stokes_latent}, we examine the influence of St. Traversing the latent dimensions most correlated with St (latent vector 13 and 25) reveals increasingly sharp and well-defined ring structures. Since St quantifies the aerodynamic coupling between dust and gas, this behavior reflects enhanced dust trapping and reduced radial drift as grains become more decoupled from the gas. The residuals clearly show that dust ring contrast increases significantly with larger Stokes numbers, leading to narrower, brighter rings consistent with physical models of pressure bump trapping.

As for other physical parameters, we find that the effect of grain size is most directly captured by latent dimension which is exactly the most correlated feature with stokes number, reflecting the stokes number and grain size relationship. The effect of dust-to-gas ratio is most directly captured by latent dimension 20 with a p-value of $5*10^{-12}$. Lastly, the flaring index correlates the most with latent dimension 3 which has a p-value less than $10^{-12}$. 

Collectively, these visualizations—backed by the correlation analysis in Figure~\ref{fig:vae_correlation_significance}—demonstrate that the VAE does far more than compress disk images: it captures a morphologically aligned latent structure that encodes interpretable, physical features. These results highlight the utility of VAEs not just as generative models, but also as tools for physical inference, model interpretability, and hypothesis testing in astrophysical systems.

\bibliography{sample631}{}

@ARTICLE{aud20,
       author = {{Auddy}, Sayantan and {Lin}, Min-Kai},
        title = "{A Machine Learning Model to Infer Planet Masses from Gaps Observed in Protoplanetary Disks}",
      journal = {\apj},
         year = 2020,
        month = sep,
       volume = {900},
       number = {1},
          eid = {62},
        pages = {62},
          doi = {10.3847/1538-4357/aba95d},
archivePrefix = {arXiv},
       eprint = {2007.13779},
 primaryClass = {astro-ph.EP},
       adsurl = {https://ui.adsabs.harvard.edu/abs/2020ApJ...900...62A}
}

@ARTICLE{2013A&A...558A..33A,
       author = {{Astropy Collaboration} and {Robitaille}, Thomas P. and
         {Tollerud}, Erik J. and {Greenfield}, Perry and {Droettboom}, Michael and
         {Bray}, Erik and {Aldcroft}, Tom and {Davis}, Matt and
         {Ginsburg}, Adam and {Price-Whelan}, Adrian M. and
         {Kerzendorf}, Wolfgang E. and {Conley}, Alexander and {Crighton}, Neil and
         {Barbary}, Kyle and {Muna}, Demitri and {Ferguson}, Henry and
         {Grollier}, Fr{\'e}d{\'e}ric and {Parikh}, Madhura M. and
         {Nair}, Prasanth H. and {Unther}, Hans M. and {Deil}, Christoph and
         {Woillez}, Julien and {Conseil}, Simon and {Kramer}, Roban and
         {Turner}, James E.~H. and {Singer}, Leo and {Fox}, Ryan and
         {Weaver}, Benjamin A. and {Zabalza}, Victor and {Edwards}, Zachary I. and
         {Azalee Bostroem}, K. and {Burke}, D.~J. and {Casey}, Andrew R. and
         {Crawford}, Steven M. and {Dencheva}, Nadia and {Ely}, Justin and
         {Jenness}, Tim and {Labrie}, Kathleen and {Lim}, Pey Lian and
         {Pierfederici}, Francesco and {Pontzen}, Andrew and {Ptak}, Andy and
         {Refsdal}, Brian and {Servillat}, Mathieu and {Streicher}, Ole},
        title = "{Astropy: A community Python package for astronomy}",
      journal = {\aap},
         year = "2013",
        month = "Oct",
       volume = {558},
          eid = {A33},
        pages = {A33},
          doi = {10.1051/0004-6361/201322068},
archivePrefix = {arXiv},
       eprint = {1307.6212},
 primaryClass = {astro-ph.IM},
       adsurl = {https://ui.adsabs.harvard.edu/abs/2013A&A...558A..33A}
}

@article{auddy2022using,
  title={Using Bayesian deep learning to infer planet mass from gaps in protoplanetary disks},
  author={Auddy, Sayantan and Dey, Ramit and Lin, Min-Kai and Carrera, Daniel and Simon, Jacob B},
  journal={The Astrophysical Journal},
  volume={936},
  number={1},
  pages={93},
  year={2022},
  publisher={IOP Publishing}
}

@article{brogan20152014,
  title={The 2014 ALMA long baseline campaign: first results from high angular resolution observations toward the HL Tau region},
  author={Brogan, CL and P{\'e}rez, LM and Hunter, TR and Dent, WRF and Hales, AS and Hills, RE and Corder, S and Fomalont, EB and Vlahakis, C and Asaki, Y and others},
  journal={The Astrophysical journal letters},
  volume={808},
  number={1},
  pages={L3},
  year={2015},
  publisher={IOP Publishing}
}

@article{wolf2002detecting,
  title={Detecting planets in protoplanetary disks: a prospective study},
  author={Wolf, S and Gueth, F and Henning, Th and Kley, W},
  journal={The Astrophysical Journal},
  volume={566},
  number={2},
  pages={L97},
  year={2002},
  publisher={IOP Publishing}
}

@article{haworth2016grand,
  title={Grand challenges in protoplanetary disc modelling},
  author={Haworth, Thomas J and Ilee, John D and Forgan, Duncan H and Facchini, Stefano and Price, Daniel J and Boneberg, Dominika M and Booth, Richard A and Clarke, Cathie J and Gonzalez, Jean-Fran{\c{c}}ois and Hutchison, Mark A and others},
  journal={Publications of the Astronomical society of Australia},
  volume={33},
  pages={e053},
  year={2016},
  publisher={Cambridge University Press}
}

@article{andrews2018scaling,
  title={Scaling relations associated with millimeter continuum sizes in protoplanetary disks},
  author={Andrews, Sean M and Terrell, Marie and Tripathi, Anjali and Ansdell, Megan and Williams, Jonathan P and Wilner, David J},
  journal={The Astrophysical Journal},
  volume={865},
  number={2},
  pages={157},
  year={2018},
  publisher={IOP Publishing}
}

@article{isella2018signatures,
  title={Signatures of Young Planets in the Continuum Emission from Protostellar Disks},
  author={Isella, Andrea and Turner, Neal J},
  journal={The Astrophysical Journal},
  volume={860},
  number={1},
  pages={27},
  year={2018},
  publisher={IOP Publishing}
}

@article{bae2022structured,
  title={Structured distributions of gas and solids in protoplanetary disks},
  author={Bae, Jaehan and Isella, Andrea and Zhu, Zhaohuan and Martin, Rebecca and Okuzumi, Satoshi and Suriano, Scott},
  journal={arXiv preprint arXiv:2210.13314},
  year={2022}
}

@article{cieza2021ophiuchus,
  title={The Ophiuchus DIsc Survey Employing ALMA (ODISEA)--III. The evolution of substructures in massive discs at 3--5 au resolution},
  author={Cieza, Lucas A and Gonz{\'a}lez-Ruilova, Camilo and Hales, Antonio S and Pinilla, Paola and Ru{\'\i}z-Rodr{\'\i}guez, Dary and Zurlo, Alice and Casassus, Sim{\'o}n and P{\'e}rez, Sebasti{\'a}n and C{\'a}novas, Hector and Arce-Tord, Carla and others},
  journal={Monthly Notices of the Royal Astronomical Society},
  volume={501},
  number={2},
  pages={2934--2953},
  year={2021},
  publisher={Oxford University Press}
}

@article{huang2020large,
  title={Large-scale CO spiral arms and complex kinematics associated with the T Tauri star RU Lup},
  author={Huang, Jane and Andrews, Sean M and {\"O}berg, Karin I and Ansdell, Megan and Benisty, Myriam and Carpenter, John M and Isella, Andrea and P{\'e}rez, Laura M and Ricci, Luca and Williams, Jonathan P and others},
  journal={The Astrophysical Journal},
  volume={898},
  number={2},
  pages={140},
  year={2020},
  publisher={IOP Publishing}
}

@article{fedele2018alma,
  title={ALMA continuum observations of the protoplanetary disk AS 209-Evidence of multiple gaps opened by a single planet},
  author={Fedele, D and Tazzari, M and Booth, R and Testi, L and Clarke, CJ and Pascucci, I and Kospal, A and Semenov, D and Bruderer, S and Henning, Th and others},
  journal={Astronomy \& Astrophysics},
  volume={610},
  pages={A24},
  year={2018},
  publisher={EDP sciences}
}

@article{toci2019long,
  title={Long-lived Dust Rings around HD 169142},
  author={Toci, Claudia and Lodato, Giuseppe and Fedele, Davide and Testi, Leonardo and Pinte, Christophe},
  journal={The Astrophysical Journal Letters},
  volume={888},
  number={1},
  pages={L4},
  year={2019},
  publisher={IOP Publishing}
}

@article{veronesi2019multiwavelength,
  title={Multiwavelength observations of protoplanetary discs as a proxy for the gas disc mass},
  author={Veronesi, B and Lodato, G and Dipierro, G and Ragusa, E and Hall, C and Price, DJ},
  journal={Monthly Notices of the Royal Astronomical Society},
  volume={489},
  number={3},
  pages={3758--3768},
  year={2019},
  publisher={Oxford University Press}
}

@article{veronesi2021dynamical,
  title={A Dynamical Measurement of the Disk Mass in Elias 2--27},
  author={Veronesi, Benedetta and Paneque-Carreno, Teresa and Lodato, Giuseppe and Testi, Leonardo and P{\'e}rez, Laura M and Bertin, Giuseppe and Hall, Cassandra},
  journal={The Astrophysical Journal Letters},
  volume={914},
  number={2},
  pages={L27},
  year={2021},
  publisher={IOP Publishing}
}

@article{teague2018kinematical,
  title={A kinematical detection of two embedded Jupiter-mass planets in HD 163296},
  author={Teague, Richard and Bae, Jaehan and Bergin, Edwin A and Birnstiel, Tilman and Foreman-Mackey, Daniel},
  journal={The Astrophysical Journal Letters},
  volume={860},
  number={1},
  pages={L12},
  year={2018},
  publisher={IOP Publishing}
}

@article{kanagawa2016mass,
  title={Mass constraint for a planet in a protoplanetary disk from the gap width},
  author={Kanagawa, Kazuhiro D and Muto, Takayuki and Tanaka, Hidekazu and Tanigawa, Takayuki and Takeuchi, Taku and Tsukagoshi, Takashi and Momose, Munetake},
  journal={Publications of the Astronomical Society of Japan},
  volume={68},
  number={3},
  pages={43},
  year={2016},
  publisher={Oxford University Press}
}

@article{dipierro2015dust,
  title={Dust trapping by spiral arms in gravitationally unstable protostellar discs},
  author={Dipierro, Giovanni and Pinilla, Paola and Lodato, Giuseppe and Testi, Leonardo},
  journal={Monthly Notices of the Royal Astronomical Society},
  volume={451},
  number={1},
  pages={974--986},
  year={2015},
  publisher={Oxford University Press}
}

@article{lodato2019newborn,
  title={The newborn planet population emerging from ring-like structures in discs},
  author={Lodato, Giuseppe and Dipierro, Giovanni and Ragusa, Enrico and Long, Feng and Herczeg, Gregory J and Pascucci, Ilaria and Pinilla, Paola and Manara, Carlo F and Tazzari, Marco and Liu, Yao and others},
  journal={Monthly Notices of the Royal Astronomical Society},
  volume={486},
  number={1},
  pages={453--461},
  year={2019},
  publisher={Oxford University Press}
}

@article{zhang2018disk,
  title={The disk substructures at high angular resolution project (DSHARP). VII. The planet--disk interactions interpretation},
  author={Zhang, Shangjia and Zhu, Zhaohuan and Huang, Jane and Guzm{\'a}n, Viviana V and Andrews, Sean M and Birnstiel, Tilman and Dullemond, Cornelis P and Carpenter, John M and Isella, Andrea and P{\'e}rez, Laura M and others},
  journal={The Astrophysical journal letters},
  volume={869},
  number={2},
  pages={L47},
  year={2018},
  publisher={IOP Publishing}
}

@article{long2020dual,
  title={Dual-wavelength ALMA observations of dust rings in protoplanetary disks},
  author={Long, Feng and Pinilla, Paola and Herczeg, Gregory J and Andrews, Sean M and Harsono, Daniel and Johnstone, Doug and Ragusa, Enrico and Pascucci, Ilaria and Wilner, David J and Hendler, Nathan and others},
  journal={The Astrophysical Journal},
  volume={898},
  number={1},
  pages={36},
  year={2020},
  publisher={IOP Publishing}
}

@article{auddy2021dpnnet,
  title={DPNNet-2.0. I. Finding hidden planets from simulated images of protoplanetary disk gaps},
  author={Auddy, Sayantan and Dey, Ramit and Lin, Min-Kai and Hall, Cassandra},
  journal={The Astrophysical Journal},
  volume={920},
  number={1},
  pages={3},
  year={2021},
  publisher={IOP Publishing}
}

@article{zhang2022pgnets,
  title={PGNets: planet mass prediction using convolutional neural networks for radio continuum observations of protoplanetary discs},
  author={Zhang, Shangjia and Zhu, Zhaohuan and Kang, Mingon},
  journal={Monthly Notices of the Royal Astronomical Society},
  volume={510},
  number={3},
  pages={4473--4484},
  year={2022},
  publisher={Oxford University Press}
}

@article{telkamp2022machine,
  title={A Machine Learning Framework to Predict Images of Edge-on Protoplanetary Disks},
  author={Telkamp, Zoie and Mart{\'\i}nez-Palomera, Jorge and Duch{\^e}ne, Gaspard and Ashimbekova, Aishabibi and Wolfe, Edward and Angelo, Isabel and Pinte, Christophe},
  journal={The Astrophysical Journal},
  volume={939},
  number={2},
  pages={73},
  year={2022},
  publisher={IOP Publishing}
}

@article{ribas2020modeling,
  title={Modeling protoplanetary disk SEDs with artificial neural networks-Revisiting the viscous disk model and updated disk masses},
  author={Ribas, {\'A}lvaro and Espaillat, Catherine C and Mac{\'\i}as, Enrique and Sarro, Luis M},
  journal={Astronomy \& Astrophysics},
  volume={642},
  pages={A171},
  year={2020},
  publisher={EDP Sciences}
}

@article{ruzza2024dbnets,
  title={DBNets: A publicly available deep learning tool to measure the masses of young planets in dusty protoplanetary discs},
  author={Ruzza, Alessandro and Lodato, Giuseppe and Rosotti, Giovanni Pietro},
  journal={Astronomy \& Astrophysics},
  volume={685},
  pages={A65},
  year={2024},
  publisher={EDP Sciences}
}

@incollection{pinheiro2021variational,
  title={Variational autoencoder},
  author={Pinheiro Cinelli, Lucas and Ara{\'u}jo Marins, Matheus and Barros da Silva, Eduardo Ant{\'u}nio and Lima Netto, S{\'e}rgio},
  booktitle={Variational methods for machine learning with applications to deep networks},
  pages={111--149},
  year={2021},
  publisher={Springer}
}

@article{benitez2016fargo3d,
  title={FARGO3D: a new GPU-oriented MHD code},
  author={Ben{\'\i}tez-Llambay, Pablo and Masset, Fr{\'e}d{\'e}ric S},
  journal={The Astrophysical Journal Supplement Series},
  volume={223},
  number={1},
  pages={11},
  year={2016},
  publisher={IOP Publishing}
}

@article{dullemond2012radmc,
  title={RADMC-3D: A multi-purpose radiative transfer tool},
  author={Dullemond, CP and Juhasz, A and Pohl, A and Sereshti, F and Shetty, R and Peters, T and Commercon, B and Flock, M},
  journal={Astrophysics Source Code Library},
  pages={ascl--1202},
  year={2012}
}

@inproceedings{mahmud2024vae_exoplanet_masses,
  title     = {Using Variational Autoencoding to Infer the Masses of Exoplanets Embedded in the Disks of Gas and Dust Orbiting Young Stars},
  author    = {Mahmud, Sayed Shafaat and Auddy, Sayantan and Dey, Ramit and Turner, Neal and Bary, Jeffrey},
  booktitle = {Proceedings of the Machine Learning and the Physical Sciences Workshop at NeurIPS 2024},
  year      = {2024},
  url       = {https://ml4physicalsciences.github.io/2024/files/NeurIPS_ML4PS_2024_170.pdf}
}

@article{kingma2014adam,
  title={Adam: A method for stochastic optimization},
  author={Kingma, Diederik P and Ba, Jimmy},
  journal={arXiv preprint arXiv:1412.6980},
  year={2014}
}

@article{dong2017mass,
  title={What is the Mass of a Gap-opening Planet?},
  author={Dong, Ruobing and Fung, Jeffrey},
  journal={The Astrophysical Journal},
  volume={835},
  number={2},
  pages={146},
  year={2017},
  publisher={IOP Publishing}
}

@article{perez2019dust,
  title={Dust unveils the formation of a mini-Neptune planet in a protoplanetary ring},
  author={P{\'e}rez, Sebasti{\'a}n and Casassus, Simon and Baruteau, Cl{\'e}ment and Dong, Ruobing and Hales, Antonio and Cieza, Lucas},
  journal={The Astronomical Journal},
  volume={158},
  number={1},
  pages={15},
  year={2019},
  publisher={American Astronomical Society}
}

@article{jin2016modeling,
  title={Modeling dust emission of HL Tau disk based on planet--disk interactions},
  author={Jin, Sheng and Li, Shengtai and Isella, Andrea and Li, Hui and Ji, Jianghui},
  journal={The Astrophysical Journal},
  volume={818},
  number={1},
  pages={76},
  year={2016},
  publisher={IOP Publishing}
}

@article{facchini2020annular,
  title={Annular substructures in the transition disks around LkCa 15 and J1610},
  author={Facchini, S and Benisty, M and Bae, J and Loomis, R and Perez, L and Ansdell, M and Mayama, S and Pinilla, P and Teague, R and Isella, A and others},
  journal={Astronomy \& Astrophysics},
  volume={639},
  pages={A121},
  year={2020},
  publisher={EDP Sciences}
}

@article{izquierdo2022new,
  title={A New Planet Candidate Detected in a Dust Gap of the Disk around HD 163296 through Localized Kinematic Signatures: An Observational Validation of the discminer},
  author={Izquierdo, Andr{\'e}s F and Facchini, Stefano and Rosotti, Giovanni P and van Dishoeck, Ewine F and Testi, Leonardo},
  journal={The Astrophysical Journal},
  volume={928},
  number={1},
  pages={2},
  year={2022},
  publisher={IOP Publishing}
}

@article{isella2016ringed,
  title={Ringed structures of the HD 163296 protoplanetary disk revealed by ALMA},
  author={Isella, Andrea and Guidi, Greta and Testi, Leonardo and Liu, Shangfei and Li, Hui and Li, Shengtai and Weaver, Erik and Boehler, Yann and Carperter, John M and De Gregorio-Monsalvo, Itziar and others},
  journal={Physical Review Letters},
  volume={117},
  number={25},
  pages={251101},
  year={2016},
  publisher={APS}
}

@article{liu2018new,
  title={New Constraints on Turbulence and Embedded Planet Mass in the HD 163296 Disk from Planet--Disk Hydrodynamic Simulations},
  author={Liu, Shang-Fei and Jin, Sheng and Li, Shengtai and Isella, Andrea and Li, Hui},
  journal={The Astrophysical Journal},
  volume={857},
  number={2},
  pages={87},
  year={2018},
  publisher={IOP Publishing}
}

@article{muller2022emerging,
  title={Emerging population of gap-opening planets around type-A stars-Long-term evolution of the forming planets around HD 163296},
  author={M{\"u}ller-Horn, Johanna and Pichierri, Gabriele and Bitsch, Bertram},
  journal={Astronomy \& Astrophysics},
  volume={663},
  pages={A163},
  year={2022},
  publisher={EDP Sciences}
}

@article{bitsch2019formation,
  title={Formation of planetary systems by pebble accretion and migration: growth of gas giants},
  author={Bitsch, Bertram and Izidoro, Andre and Johansen, Anders and Raymond, Sean N and Morbidelli, Alessandro and Lambrechts, Michiel and Jacobson, Seth A},
  journal={Astronomy \& Astrophysics},
  volume={623},
  pages={A88},
  year={2019},
  publisher={EDP Sciences}
}

@article{rodenkirch2021modeling,
  title={Modeling the nonaxisymmetric structure in the HD 163296 disk with planet-disk interaction},
  author={Rodenkirch, Peter J and Rometsch, Thomas and Dullemond, Cornelis P and Weber, Philipp and Kley, Wilhelm},
  journal={Astronomy \& Astrophysics},
  volume={647},
  pages={A174},
  year={2021},
  publisher={EDP Sciences}
}

@article{jiang2024grain,
  title={Grain-size measurements in protoplanetary disks indicate fragile pebbles and low turbulence},
  author={Jiang, Haochang and Mac{\'\i}as, Enrique and Guerra-Alvarado, Osmar M and Carrasco-Gonz{\'a}lez, Carlos},
  journal={Astronomy \& Astrophysics},
  volume={682},
  pages={A32},
  year={2024},
  publisher={EDP Sciences}
}

@article{tilling2012gas,
  title={Gas modelling in the disc of HD 163296},
  author={Tilling, I and Woitke, Peter and Meeus, Gwendolyn and Mora, Alcione and Montesinos, Benjam{\'\i}n and Rivi{\`e}re-Marichalar, Pablo and Eiroa, Carlos and Thi, W-F and Isella, Andrea and Roberge, Aki and others},
  journal={Astronomy \& Astrophysics},
  volume={538},
  pages={A20},
  year={2012},
  publisher={EDP Sciences}
}

@article{drazkowska2019including,
  title={Including dust coagulation in hydrodynamic models of protoplanetary disks: dust evolution in the vicinity of a jupiter-mass planet},
  author={Dr{\c a}{\.z}kowska, Joanna and Li, Shengtai and Birnstiel, Til and Stammler, Sebastian M and Li, Hui},
  journal={The Astrophysical Journal},
  volume={885},
  number={1},
  pages={91},
  year={2019},
  publisher={IOP Publishing}
}

@article{pinte2015dust,
  title={Dust and gas in the disk of HL Tauri: surface density, dust settling, and dust-to-gas ratio},
  author={Pinte, Christophe and Dent, William RF and M{\'e}nard, Francois and Hales, Antonio and Hill, Tracey and Cortes, Paulo and de Gregorio-Monsalvo, Itziar},
  journal={The Astrophysical Journal},
  volume={816},
  number={1},
  pages={25},
  year={2015},
  publisher={IOP Publishing}
}

@article{wu2018physical,
  title={Physical and Chemical Conditions of the Protostellar Envelope and the Protoplanetary Disk in HL Tau},
  author={Wu, Chun-Ju and Hirano, Naomi and Takakuwa, Shigehisa and Yen, Hsi-Wei and Aso, Yusuke},
  journal={The Astrophysical Journal},
  volume={869},
  number={1},
  pages={59},
  year={2018},
  publisher={IOP Publishing}
}

@article{kingma2014autoencoding,
  title={Auto-Encoding Variational Bayes},
  author={Kingma, Diederik P and Welling, Max},
  journal={arXiv preprint arXiv:1312.6114},
  year={2014}
}

@article{doersch2016tutorial,
  title={Tutorial on variational autoencoders},
  author={Doersch, Carl},
  journal={arXiv preprint arXiv:1606.05908},
  year={2016}
}

@article{elbakyan2021gap,
  title={Gap opening by planets in discs with magnetised winds},
  author={Elbakyan, Vardan and Wu, Yinhao and Nayakshin, Sergei and Rosotti, Giovanni},
  journal={Proceedings of the International Astronomical Union},
  volume={17},
  number={S370},
  pages={194--195},
  year={2021},
  publisher={Cambridge University Press}
}

@article{gonzalez2012planet,
  title={Planet gaps in the dust layer of 3D protoplanetary disks-II. Observability with ALMA},
  author={Gonzalez, J-F and Pinte, Christophe and Maddison, Sarah T and M{\'e}nard, Fran{\c{c}}ois and Fouchet, Laure},
  journal={Astronomy \& Astrophysics},
  volume={547},
  pages={A58},
  year={2012},
  publisher={EDP Sciences}
}

@article{flock2015gaps,
  title={Gaps, rings, and non-axisymmetric structures in protoplanetary disks-from simulations to alma observations},
  author={Flock, Mario and Ruge, Jan Philipp and Dzyurkevich, Natalia and Henning, Th and Klahr, H and Wolf, S},
  journal={Astronomy \& Astrophysics},
  volume={574},
  pages={A68},
  year={2015},
  publisher={EDP Sciences}
}

@article{okuzumi2016sintering,
  title={Sintering-induced dust ring formation in protoplanetary disks: Application to the HL tau disk},
  author={Okuzumi, Satoshi and Momose, Munetake and Sirono, Sin-iti and Kobayashi, Hiroshi and Tanaka, Hidekazu},
  journal={The Astrophysical Journal},
  volume={821},
  number={2},
  pages={82},
  year={2016},
  publisher={IOP Publishing}
}

@article{suriano2019formation,
  title={The formation of rings and gaps in wind-launching non-ideal MHD discs: three-dimensional simulations},
  author={Suriano, Scott S and Li, Zhi-Yun and Krasnopolsky, Ruben and Suzuki, Takeru K and Shang, Hsien},
  journal={Monthly Notices of the Royal Astronomical Society},
  volume={484},
  number={1},
  pages={107--124},
  year={2019},
  publisher={Oxford University Press}
}

@INPROCEEDINGS{fis14,
       author = {{Fischer}, D.~A. and {Howard}, A.~W. and {Laughlin}, G.~P. and
         {Macintosh}, B. and {Mahadevan}, S. and {Sahlmann}, J. and {Yee}, J.~C.},
        title = "{Exoplanet Detection Techniques}",
    booktitle = {Protostars and Planets VI},
         year = 2014,
       editor = {{Beuther}, Henrik and {Klessen}, Ralf S. and {Dullemond}, Cornelis P. and
         {Henning}, Thomas},
        month = jan,
        pages = {715},
          doi = {10.2458/azu_uapress_9780816531240-ch031},
archivePrefix = {arXiv},
       eprint = {1505.06869},
 primaryClass = {astro-ph.EP},
       adsurl = {https://ui.adsabs.harvard.edu/abs/2014prpl.conf..715F}
}

@ARTICLE{lee18,
       author = {{Lee}, Chien-Hsiu},
        title = "{Exoplanets: Past, Present, and Future}",
      journal = {Galaxies},
         year = 2018,
        month = apr,
       volume = {6},
       number = {2},
        pages = {51},
          doi = {10.3390/galaxies6020051},
archivePrefix = {arXiv},
       eprint = {1804.08907},
 primaryClass = {astro-ph.EP},
       adsurl = {https://ui.adsabs.harvard.edu/abs/2018Galax...6...51L}
}

@article{rosotti2023empirical,
  title={Empirical constraints on turbulence in proto-planetary discs},
  author={Rosotti, Giovanni P},
  journal={New Astronomy Reviews},
  volume={96},
  pages={101674},
  year={2023},
  publisher={Elsevier}
}

@article{dominik2024bouncing,
  title={The bouncing barrier revisited: Impact on key planet formation processes and observational signatures},
  author={Dominik, Carsten and Dullemond, CP},
  journal={Astronomy \& Astrophysics},
  volume={682},
  pages={A144},
  year={2024},
  publisher={EDP Sciences}
}

@article{kanagawa2015mass,
  title={Mass estimates of a giant planet in a protoplanetary disk from the gap structures},
  author={Kanagawa, Kazuhiro D and Muto, Takayuki and Tanaka, Hidekazu and Tanigawa, Takayuki and Takeuchi, Taku and Tsukagoshi, Takashi and Momose, Munetake},
  journal={The Astrophysical Journal Letters},
  volume={806},
  number={1},
  pages={L15},
  year={2015},
  publisher={IOP Publishing}
}

@article{de2006comparative,
  title={A comparative study of disc--planet interaction},
  author={de Val-Borro, Miguel and Edgar, RG and Artymowicz, P and Ciecielag, P ea and Cresswell, P and D'Angelo, Gennaro and Delgado-Donate, EJ and Dirksen, G and Fromang, S{\'e}bastien and Gawryszczak, A and others},
  journal={Monthly Notices of the Royal Astronomical Society},
  volume={370},
  number={2},
  pages={529--558},
  year={2006},
  publisher={Blackwell Publishing Ltd Oxford, UK}
}

@article{shakura197313,
  title={13. BLACK HOLES IN BINARY SYSTEMS: OBSERVATIONAL APPEARANCES},
  author={Shakura, NI and Sunyaev, RA},
  journal={X-and Gamma-Ray Astronomy},
  number={55},
  pages={155},
  year={1973},
  publisher={Springer}
}

@article{jacquet2011linear,
  title={On linear dust--gas streaming instabilities in protoplanetary discs},
  author={Jacquet, Emmanuel and Balbus, Steven and Latter, Henrik},
  journal={Monthly Notices of the Royal Astronomical Society},
  volume={415},
  number={4},
  pages={3591--3598},
  year={2011},
  publisher={Blackwell Publishing Ltd Oxford, UK}
}

@article{bjorkman2001radiative,
  title={Radiative equilibrium and temperature correction in Monte Carlo radiation transfer},
  author={Bjorkman, Jon E and Wood, Kenneth},
  journal={The Astrophysical Journal},
  volume={554},
  number={1},
  pages={615},
  year={2001},
  publisher={IOP Publishing}
}

@ARTICLE{fromang2009,
       author = {{Fromang}, S. and {Nelson}, R.~P.},
        title = "{Global MHD simulations of stratified and turbulent protoplanetary discs. II. Dust settling}",
      journal = {\aap},
         year = 2009,
        month = mar,
       volume = {496},
       number = {3},
        pages = {597-608},
          doi = {10.1051/0004-6361/200811220},
archivePrefix = {arXiv},
       eprint = {0901.4434},
 primaryClass = {astro-ph.EP},
       adsurl = {https://ui.adsabs.harvard.edu/abs/2009A&A...496..597F}
}

@article{ziampras2025dusty,
  title={Dusty substructures induced by planets in ALMA discs: how dust growth and dynamics changes the picture},
  author={Ziampras, Alexandros and Sudarshan, Prakruti and Dullemond, Cornelis P and Flock, Mario and Berta, Vittoria and Nelson, Richard P and Mignone, Andrea},
  journal={Monthly Notices of the Royal Astronomical Society},
  volume={536},
  number={4},
  pages={3322--3337},
  year={2025},
  publisher={Oxford University Press}
}

@article{o2015introduction,
  title={An introduction to convolutional neural networks},
  author={O'shea, Keiron and Nash, Ryan},
  journal={arXiv preprint arXiv:1511.08458},
  year={2015}
}

@article{agarap2018deep,
  title={Deep learning using rectified linear units (relu)},
  author={Agarap, Abien Fred},
  journal={arXiv preprint arXiv:1803.08375},
  year={2018}
}

@article{stammler2022dustpy,
  title={DustPy: a python package for dust evolution in protoplanetary disks},
  author={Stammler, Sebastian M and Birnstiel, Tilman},
  journal={The Astrophysical Journal},
  volume={935},
  number={1},
  pages={35},
  year={2022},
  publisher={IOP Publishing}
}

@article{dorschner1995steps,
  title={Steps toward interstellar silicate mineralogy. II. Study of Mg-Fe-silicate glasses of variable composition.},
  author={Dorschner, J and Begemann, B and Henning, Th and Jaeger, C and Mutschke, H},
  journal={Astronomy and Astrophysics, v. 300, p. 503},
  volume={300},
  pages={503},
  year={1995}
}

@article{penny2019predictions,
  title={Predictions of the WFIRST microlensing survey. I. Bound planet detection rates},
  author={Penny, Matthew T and Gaudi, B Scott and Kerins, Eamonn and Rattenbury, Nicholas J and Mao, Shude and Robin, Annie C and Novati, Sebastiano Calchi},
  journal={The Astrophysical Journal Supplement Series},
  volume={241},
  number={1},
  pages={3},
  year={2019},
  publisher={IOP Publishing}
}

@article{welch1947generalization,
  title={The generalization of ‘STUDENT'S’problem when several different population varlances are involved},
  author={Welch, Bernard L},
  journal={Biometrika},
  volume={34},
  number={1-2},
  pages={28--35},
  year={1947},
  publisher={Oxford University Press}
}

@article{akeson2013nasa,
  title={The NASA exoplanet archive: data and tools for exoplanet research},
  author={Akeson, RL and Chen, X and Ciardi, D and Crane, M and Good, J and Harbut, M and Jackson, E and Kane, SR and Laity, AC and Leifer, S and others},
  journal={Publications of the Astronomical Society of the Pacific},
  volume={125},
  number={930},
  pages={989},
  year={2013},
  publisher={IOP Publishing},
  doi = {10.1086/672273}
}

@article{papamakarios2021normalizing,
  title={Normalizing flows for probabilistic modeling and inference},
  author={Papamakarios, George and Nalisnick, Eric and Rezende, Danilo Jimenez and Mohamed, Shakir and Lakshminarayanan, Balaji},
  journal={Journal of Machine Learning Research},
  volume={22},
  number={57},
  pages={1--64},
  year={2021}
}

@article{sharma2017activation,
  title={Activation functions in neural networks},
  author={Sharma, Sagar and Sharma, Simone and Athaiya, Anidhya},
  journal={Towards Data Sci},
  volume={6},
  number={12},
  pages={310--316},
  year={2017}
}

@article{johansen2009zonal,
  title={Zonal flows and long-lived axisymmetric pressure bumps in magnetorotational turbulence},
  author={Johansen, Anders and Youdin, Andrew and Klahr, Hubert},
  journal={The Astrophysical Journal},
  volume={697},
  number={2},
  pages={1269},
  year={2009},
  publisher={IOP Publishing}
}

@article{dittrich2013gravoturbulent,
  title={Gravoturbulent planetesimal formation: the positive effect of long-lived zonal flows},
  author={Dittrich, Karsten and Klahr, Hubert and Johansen, Anders},
  journal={The Astrophysical Journal},
  volume={763},
  number={2},
  pages={117},
  year={2013},
  publisher={IOP Publishing}
}

@article{pinte2019kinematic,
  title={Kinematic detection of a planet carving a gap in a protoplanetary disk},
  author={Pinte, C and van Der Plas, G and M{\'e}nard, F and Price, DJ and Christiaens, Valentin and Hill, T and Mentiplay, D and Ginski, C and Choquet, E and Boehler, Y and others},
  journal={Nature Astronomy},
  volume={3},
  number={12},
  pages={1109--1114},
  year={2019},
  publisher={Nature Publishing Group UK London}
}

@article{mckay2000comparison,
  title={A comparison of three methods for selecting values of input variables in the analysis of output from a computer code},
  author={McKay, Michael D and Beckman, Richard J and Conover, William J},
  journal={Technometrics},
  volume={42},
  number={1},
  pages={55--61},
  year={2000},
  publisher={Taylor \& Francis}
}

@article{youdin2011formation,
  title={On the formation of planetesimals via secular gravitational instabilities with turbulent stirring},
  author={Youdin, Andrew N},
  journal={The Astrophysical Journal},
  volume={731},
  number={2},
  pages={99},
  year={2011},
  publisher={IOP Publishing}
}

@article{takahashi2014two,
  title={Two-component secular gravitational instability in a protoplanetary disk: a possible mechanism for creating ring-like structures},
  author={Takahashi, Sanemichi Z and Inutsuka, Shu-ichiro},
  journal={The Astrophysical Journal},
  volume={794},
  number={1},
  pages={55},
  year={2014},
  publisher={IOP Publishing}
}

@article{zhang2019systematic,
  title={Systematic variations of CO gas abundance with radius in gas-rich protoplanetary disks},
  author={Zhang, Ke and Bergin, Edwin A and Schwarz, Kamber and Krijt, Sebastiaan and Ciesla, Fred},
  journal={The Astrophysical Journal},
  volume={883},
  number={1},
  pages={98},
  year={2019},
  publisher={IOP Publishing}
}

@article{pinilla2017dust,
  title={Dust density distribution and imaging analysis of different ice lines in protoplanetary disks},
  author={Pinilla, P and Pohl, A and Stammler, SM and Birnstiel, T},
  journal={The Astrophysical Journal},
  volume={845},
  number={1},
  pages={68},
  year={2017},
  publisher={IOP Publishing}
}

@article{gonzalez2017self,
  title={Self-induced dust traps: overcoming planet formation barriers},
  author={Gonzalez, J-F and Laibe, Guillaume and Maddison, Sarah T},
  journal={Monthly Notices of the Royal Astronomical Society},
  volume={467},
  number={2},
  pages={1984--1996},
  year={2017},
  publisher={Oxford University Press}
}

@article{barge2017gaps,
  title={Gaps and rings carved by vortices in protoplanetary dust},
  author={Barge, Pierre and Ricci, Luca and Carilli, Christopher Luke and Previn-Ratnasingam, Rathish},
  journal={Astronomy \& Astrophysics},
  volume={605},
  pages={A122},
  year={2017},
  publisher={EDP Sciences}
}

@article{paszke2019pytorch,
  title={Pytorch: An imperative style, high-performance deep learning library},
  author={Paszke, Adam and Gross, Sam and Massa, Francisco and Lerer, Adam and Bradbury, James and Chanan, Gregory and Killeen, Trevor and Lin, Zeming and Gimelshein, Natalia and Antiga, Luca and others},
  journal={Advances in neural information processing systems},
  volume={32},
  year={2019}
}

@article{andrews2018disk,
  title={The disk substructures at high angular resolution project (DSHARP). I. Motivation, sample, calibration, and overview},
  author={Andrews, Sean M and Huang, Jane and P{\'e}rez, Laura M and Isella, Andrea and Dullemond, Cornelis P and Kurtovic, Nicol{\'a}s T and Guzm{\'a}n, Viviana V and Carpenter, John M and Wilner, David J and Zhang, Shangjia and others},
  journal={The Astrophysical Journal Letters},
  volume={869},
  number={2},
  pages={L41},
  year={2018},
  publisher={IOP Publishing}
}

@article{gabbard2022bayesian,
  title={Bayesian parameter estimation using conditional variational autoencoders for gravitational-wave astronomy},
  author={Gabbard, Hunter and Messenger, Chris and Heng, Ik Siong and Tonolini, Francesco and Murray-Smith, Roderick},
  journal={Nature Physics},
  volume={18},
  number={1},
  pages={112--117},
  year={2022},
  publisher={Nature Publishing Group UK London}
}

@article{spindler2021astrovader,
  title={AstroVaDEr: astronomical variational deep embedder for unsupervised morphological classification of galaxies and synthetic image generation},
  author={Spindler, Ashley and Geach, James E and Smith, Michael J},
  journal={Monthly Notices of the Royal Astronomical Society},
  volume={502},
  number={1},
  pages={985--1007},
  year={2021},
  publisher={Oxford University Press}
}

@article{grassi2022reducing,
  title={Reducing the complexity of chemical networks via interpretable autoencoders},
  author={Grassi, Tommaso and Nauman, F and Ramsey, JP and Bovino, S and Picogna, G and Ercolano, B},
  journal={Astronomy \& Astrophysics},
  volume={668},
  pages={A139},
  year={2022},
  publisher={EDP Sciences}
}

@article{gagliano2023physics,
  title={A Physics-Informed Variational Autoencoder for Rapid Galaxy Inference and Anomaly Detection},
  author={Gagliano, Alexander and Villar, V Ashley},
  journal={arXiv preprint arXiv:2312.16687},
  year={2023}
}

@article{portillo2020dimensionality,
  title={Dimensionality reduction of SDSS spectra with variational autoencoders},
  author={Portillo, Stephen KN and Parejko, John K and Vergara, Jorge R and Connolly, Andrew J},
  journal={The Astronomical Journal},
  volume={160},
  number={1},
  pages={45},
  year={2020},
  publisher={IOP Publishing}
}

@article{rafikov2017protoplanetary,
  title={Protoplanetary disks as (possibly) viscous disks},
  author={Rafikov, Roman R},
  journal={The Astrophysical Journal},
  volume={837},
  number={2},
  pages={163},
  year={2017},
  publisher={IOP Publishing}
}

@article{hughes2011empirical,
  title={Empirical constraints on turbulence in protoplanetary accretion disks},
  author={Hughes, A Meredith and Wilner, David J and Andrews, Sean M and Qi, Chunhua and Hogerheijde, Michiel R},
  journal={The Astrophysical Journal},
  volume={727},
  number={2},
  pages={85},
  year={2011},
  publisher={IOP Publishing}
}

@article{andrews2009protoplanetary,
  title={Protoplanetary disk structures in Ophiuchus},
  author={Andrews, Sean M and Wilner, DJ and Hughes, AM and Qi, Chunhua and Dullemond, CP},
  journal={The Astrophysical Journal},
  volume={700},
  number={2},
  pages={1502},
  year={2009},
  publisher={IOP Publishing}
}

@article{birnstiel2012simple,
  title={A simple model for the evolution of the dust population in protoplanetary disks},
  author={Birnstiel, T and Klahr, H and Ercolano, B},
  journal={Astronomy \& Astrophysics},
  volume={539},
  pages={A148},
  year={2012},
  publisher={EDP Sciences}
}

@article{birnstiel2016dust,
  title={Dust evolution and the formation of planetesimals},
  author={Birnstiel, T and Fang, M and Johansen, A},
  journal={Space Science Reviews},
  volume={205},
  number={1},
  pages={41--75},
  year={2016},
  publisher={Springer}
}

@article{quanz2013gaps,
  title={Gaps in the HD 169142 protoplanetary disk revealed by polarimetric imaging: signs of ongoing planet formation?},
  author={Quanz, Sascha P and Avenhaus, Henning and Buenzli, Esther and Garufi, Antonio and Schmid, Hans Martin and Wolf, Sebastian},
  journal={The Astrophysical Journal Letters},
  volume={766},
  number={1},
  pages={L2},
  year={2013},
  publisher={IOP Publishing}
}

@article{winters2003gap,
  title={Gap formation by planets in turbulent protostellar disks},
  author={Winters, Wayne F and Balbus, Steven A and Hawley, John F},
  journal={The Astrophysical Journal},
  volume={589},
  number={1},
  pages={543},
  year={2003},
  publisher={IOP Publishing}
}

@article{zhu2019inclined,
  title={Inclined massive planets in a protoplanetary disc: gap opening, disc breaking, and observational signatures},
  author={Zhu, Zhaohuan},
  journal={Monthly Notices of the Royal Astronomical Society},
  volume={483},
  number={3},
  pages={4221--4241},
  year={2019},
  publisher={Oxford University Press}
}

@article{pinilla2015sequential,
  title={Sequential planet formation in the HD 100546 protoplanetary disk?},
  author={Pinilla, P and Birnstiel, T and Walsh, C},
  journal={Astronomy \& Astrophysics},
  volume={580},
  pages={A105},
  year={2015},
  publisher={EDP Sciences}
}

@article{akiyama2016planetary,
  title={Planetary system formation in the protoplanetary disk around HL Tauri},
  author={Akiyama, Eiji and Hasegawa, Yasuhiro and Hayashi, Masahiko and Iguchi, Satoru},
  journal={The Astrophysical Journal},
  volume={818},
  number={2},
  pages={158},
  year={2016},
  publisher={IOP Publishing}
}

@article{ruzza2025dbnets2,
  title   = {DBNets2.0: Simulation-based inference for planet-induced dust substructures in protoplanetary discs},
  author  = {Ruzza, A. and Lodato, G. and Rosotti, G. P. and Armitage, P. J.},
  journal = {Astronomy \& Astrophysics},
  year    = {2025},
  volume  = {700},
  pages   = {A190},
  month   = aug,
  url     = {https://doi.org/10.1051/0004-6361/202554401}
}

@article{pinilla2012ring,
  title={Ring shaped dust accumulation in transition disks},
  author={Pinilla, Paola and Benisty, Myriam and Birnstiel, Tilman},
  journal={Astronomy \& Astrophysics},
  volume={545},
  pages={A81},
  year={2012},
  publisher={EDP Sciences}
}

@article{dullemond2018disk,
  title={The disk substructures at high angular resolution project (DSHARP). VI. Dust trapping in thin-ringed protoplanetary disks},
  author={Dullemond, Cornelis P and Birnstiel, Tilman and Huang, Jane and Kurtovic, Nicol{\'a}s T and Andrews, Sean M and Guzm{\'a}n, Viviana V and P{\'e}rez, Laura M and Isella, Andrea and Zhu, Zhaohuan and Benisty, Myriam and others},
  journal={The Astrophysical Journal Letters},
  volume={869},
  number={2},
  pages={L46},
  year={2018},
  publisher={IOP Publishing}
}

@article{mao2023ppdonet,
  title={Ppdonet: Deep operator networks for fast prediction of steady-state solutions in disk--planet systems},
  author={Mao, Shunyuan and Dong, Ruobing and Lu, Lu and Yi, Kwang Moo and Wang, Sifan and Perdikaris, Paris},
  journal={The Astrophysical Journal Letters},
  volume={950},
  number={2},
  pages={L12},
  year={2023},
  publisher={IOP Publishing}
}

@article{mao2024disk2planet,
  title={Disk2planet: A robust and automated machine learning tool for parameter inference in disk--planet systems},
  author={Mao, Shunyuan and Dong, Ruobing and Yi, Kwang Moo and Lu, Lu and Wang, Sifan and Perdikaris, Paris},
  journal={The Astrophysical Journal},
  volume={976},
  number={2},
  pages={200},
  year={2024},
  publisher={IOP Publishing}
}

@article{wang2004image,
  title={Image quality assessment: from error visibility to structural similarity},
  author={Wang, Zhou and Bovik, Alan C and Sheikh, Hamid R and Simoncelli, Eero P},
  journal={IEEE transactions on image processing},
  volume={13},
  number={4},
  pages={600--612},
  year={2004},
  publisher={IEEE}
}

@article{andrews2016ringed,
  title={Ringed substructure and a gap at 1 au in the nearest protoplanetary disk},
  author={Andrews, Sean M and Wilner, David J and Zhu, Zhaohuan and Birnstiel, Tilman and Carpenter, John M and P{\'e}rez, Laura M and Bai, Xue-Ning and {\"O}berg, Karin I and Hughes, A Meredith and Isella, Andrea and others},
  journal={The Astrophysical Journal Letters},
  volume={820},
  number={2},
  pages={L40},
  year={2016},
  publisher={IOP Publishing}
}

@article{long2018gaps,
  title={Gaps and Rings in an ALMA Survey of Disks in the Taurus Star-forming Region},
  author={Long, Feng and Pinilla, Paola and Herczeg, Gregory J and Harsono, Daniel and Dipierro, Giovanni and Pascucci, Ilaria and Hendler, Nathan and Tazzari, Marco and Ragusa, Enrico and Salyk, Colette and others},
  journal={The Astrophysical Journal},
  volume={869},
  number={1},
  pages={17},
  year={2018},
  publisher={IOP Publishing}
}

@article{marino2021constraining,
  title={Constraining planetesimal stirring: how sharp are debris disc edges?},
  author={Marino, Sebastian},
  journal={Monthly Notices of the Royal Astronomical Society},
  volume={503},
  number={4},
  pages={5100--5114},
  year={2021},
  publisher={Oxford University Press}
}

@article{tamayo2015dynamical,
  title={Dynamical stability of imaged planetary systems in formation: Application to HL tau},
  author={Tamayo, Daniel and Triaud, Amaury HMJ and Menou, Kristen and Rein, Hanno},
  journal={The Astrophysical Journal},
  volume={805},
  number={2},
  pages={100},
  year={2015},
  publisher={IOP Publishing}
}

@article{simbulan2017connecting,
  title={Connecting HL Tau to the observed exoplanet sample},
  author={Simbulan, Christopher and Tamayo, Daniel and Petrovich, Cristobal and Rein, Hanno and Murray, Norman},
  journal={Monthly Notices of the Royal Astronomical Society},
  volume={469},
  number={3},
  pages={3337--3346},
  year={2017},
  publisher={Oxford University Press}
}

@article{price2022astropy,
  title={The Astropy Project: sustaining and growing a community-oriented open-source project and the latest major release (v5. 0) of the core package},
  author={Price-Whelan, Adrian M and Lim, Pey Lian and Earl, Nicholas and Starkman, Nathaniel and Bradley, Larry and Shupe, David L and Patil, Aarya A and Corrales, Lia and Brasseur, CE and N{\"o}the, Maximilian and others},
  journal={The Astrophysical Journal},
  volume={935},
  number={2},
  pages={167},
  year={2022},
  publisher={IOP Publishing}
}

@article{price2018astropy,
  title={The astropy project: Building an open-science project and status of the v2. 0 core package},
  author={Price-Whelan, Adrian M and Sip{\H{o}}cz, BM and G{\"u}nther, HM and Lim, PL and Crawford, SM and Conseil, S and Shupe, DL and Craig, MW and Dencheva, N and Ginsburg, Adam and others},
  journal={The Astronomical Journal},
  volume={156},
  number={3},
  pages={123},
  year={2018},
  publisher={IOP Publishing}
}

@misc{auddy_sayantan_2022_7332423,
  author       = {Auddy Sayantan},
  title        = {DPNNet-RT},
  month        = nov,
  year         = 2022,
  publisher    = {Zenodo},
  doi          = {10.5281/zenodo.7332423},
  url          = {https://doi.org/10.5281/zenodo.7332423},
}
\bibliographystyle{aasjournal}



\end{document}